\documentclass[secnumarabic,12pt,tightenlines,eqsecnum,floats,aps,amsmath,amssymb,nofootinbib,superscriptaddress,showpacs,prd]{revtex4-1}

\usepackage{amsmath,amsthm,amssymb,amsfonts}
\usepackage{graphicx}
\usepackage{palatino}
\usepackage{supertabular}
\usepackage[mathcal]{euler}

\newcommand\T{\rule{0pt}{2.6ex}}
\newcommand\B{\rule[-1.7ex]{0pt}{0pt}}

\numberwithin{equation}{section}

\begin{document}

\title{Detailed black hole state counting in loop quantum gravity}

\author{Ivan  \surname{Agullo}}
\affiliation{Physics Department, University of Wisconsin-Milwaukee, P.O.Box 413, Milwaukee, WI
53201 USA}

\author{J. Fernando \surname{Barbero G.}}
\affiliation{Instituto de
Estructura de la Materia, CSIC, Serrano 123, 28006 Madrid, Spain}

\author{Enrique F. \surname{Borja}}
 \affiliation{Institute for Theoretical Physics III, University of Erlangen-N\"{u}rnberg, Erlangen, Germany}

\author{Jacobo  \surname{Diaz-Polo}}
\affiliation{Institute for Gravitation and the Cosmos \& Physics Department,
Penn State, University Park, PA 16802-6300, USA}

\author{Eduardo J. \surname{S. Villase\~nor}}
\affiliation{Instituto Gregorio
Mill\'an, Grupo de Modelizaci\'on y Simulaci\'on Num\'erica,
Universidad Carlos III de Madrid, Avda. de la Universidad 30, 28911
Legan\'es, Spain} \affiliation{Instituto de Estructura de la
Materia, CSIC, Serrano 123, 28006 Madrid, Spain}

\date{July 2, 2010}

\begin{abstract}
We give a complete and detailed description of the computation of black hole entropy in loop quantum gravity by employing the most recently introduced number-theoretic and combinatorial methods. The use of these techniques allows us to perform a detailed analysis of the precise structure of the entropy spectrum for small black holes, showing some relevant features that were not discernible in previous computations. The ability to manipulate and understand the spectrum up to the level of detail that we describe in the paper is a crucial step towards obtaining the behavior of entropy in the asymptotic (large horizon area) regime.
\end{abstract}


\maketitle

\tableofcontents

\section{Introduction} The identification of the microscopic degrees of freedom responsible for the black hole entropy is one of the key results expected from any candidate quantum gravity theory. In this respect loop quantum gravity (LQG) \cite{Thiemann:2007zz,Rovelli:2004tv,Ashtekar:2004eh} can claim a reasonable success because it accounts for the black holes degrees of freedom in a beautiful and mathematically appealing way \cite{Ashtekar:1997yu,Ashtekar:2000eq}. Black hole degrees of freedom are described in this framework as Chern-Simons (CS) states residing on the ``surface'' of a black hole modeled as an isolated horizon \cite{Ashtekar:2004cn}. This leads to an interesting interplay among quantum geometry, Chern-Simons theory, and statistical mechanics. When the states describing a black hole are counted and this number is related to the black hole horizon area $a$, the resulting entropy is found to be proportional to $a$. By taking advantage of the free Immirzi parameter $\gamma$ that labels inequivalent quantum sectors of general relativity in the connection formulation, it is actually possible to make the proportionality factor between the entropy and the area equal to $1/4$, thus reproducing the Bekenstein-Hawking area law. The choice of  $\gamma$, initially made in the study of spherical non-rotating black holes, is universal in the sense that it works also for other types of black holes (i.e. rotating or coupled to some matter fields). As of today, there are no exceptions to this universality (see, however, \cite{Jacobson:2007uj} for some interesting suggestions about the role of $\gamma$). The available framework also predicts the first subleading corrections to the Bekenstein-Hawking law which turn out to be logarithmic with area. These corrections are generically independent of $\gamma$ and in qualitative agreement with the ones obtained by using completely different approaches \cite{Carlip:2000nv}.

In addition to the successful derivation of the entropy-area law, there have been some recent results concerning small black holes \cite{Corichi:2006bs,Corichi:2006wn,Corichi:2007zz}. In particular, a persistent ``periodicity'' has been found in the so called black hole degeneracy spectrum when plotted as a function of the area.\footnote{Here and in the following, \textit{periodicity} actually means a modulation with a regular period of some growing magnitude.} The most degenerate quantum configurations accumulate around certain evenly spaced values of area, with a much lower degeneracy  in region between those values. This produces an effectively equidistant area spectrum, despite the fact that the area spectrum in LQG is not equidistant. This phenomenon makes contact in a non trivial way with the evenly spaced black hole horizon area spectrum predicted by Bekenstein and Mukhanov in \cite{Bekenstein:1974jk,Bekenstein:1995ju} under quite general conditions. The periodicity in the degeneracy spectrum leads to a striking staircase behavior when the entropy is plotted as a function of the area.

In a recent paper \cite{Agullo:2008yv} we have proposed a novel way to understand the black hole degeneracy spectrum by relying on number-theoretic and combinatorial methods. The main goal of that paper was to confirm the original results obtained in \cite{Corichi:2006bs,Corichi:2006wn,Corichi:2007zz} and extend them by using an improved computer algorithm based on the new mathematical understanding of the problem. Once this goal has been achieved, the real challenge is trying to see if macroscopic black holes display the same periodicity in the entropy as the microscopic ones. To this end, one has to find appropriate exact expressions for the entropy as a function of the area, which are suitable to derive its asymptotic behavior. A key step in this direction is to codify the solution to the combinatorial problem of computing the number of relevant microstates  as the coefficients of a formal power series expansion of the so called generating function (see \cite{Sahlmann:2007jt,Sahlmann:2007zp} for a pioneering suggestion in this direction). In many cases this formal expansion converges in non empty open disks in the complex plane and admits an analytic extension allowing the coefficients to be written as contour integrals whose asymptotic behavior can then be studied by more or less standard methods. The relevant generating functions for our problem were obtained and used to derive the expression of the entropy as a function of the area in \cite{BarberoG.:2008ue, G.:2008mj}.

The study of the asymptotics of black hole entropy has been attempted before \cite{Meissner:2004ju}. In this remarkable paper, Meissner gave a clever method to sidestep many of the fine-grained issues that we have looked at with our number-theoretic methods. There he obtained a closed expression for the entropy as a function of the area in the form of an inverse Laplace (actually Laplace-Fourier) transform. This poses some interesting questions about the compatibility of his approach and ours: Are they compatible? Is it really necessary to resort to the number theoretical methods of \cite{Agullo:2008yv,BarberoG.:2008ue}? These questions have been answered in \cite{G.:2008mj} (and the present work offers more evidence supporting the need to use our methods to satisfactorily understand several important issues related to black hole entropy). A first relevant point to highlight is the difficulty in confirming the structure seen in \cite{Corichi:2006wn} by directly computing Meissner's integrals due to the subtly oscillatory nature of their integrands. To this end, the number theoretical and combinatorial methods of \cite{Agullo:2008yv} are clearly better suited. The issue is then to show that they provide the same answer as Meissner's approach. This has been done in \cite{G.:2008mj} where the integral expressions of \cite{Meissner:2004ju} have been obtained from the generating functions introduced in \cite{BarberoG.:2008ue}. Another important issue that has been considered in \cite{G.:2008mj} is the analytic structure of the integrand of the inverse Laplace transform that gives the entropy. The presence or absence of the sought for periodic structure for macroscopic black holes depends on the location of the poles of the integrand. The results presented in \cite{G.:2008mj} do not prove that the periodicity exists for large black holes, but certainly do not exclude this possibility owing to the subtle behavior of the real parts of the poles. An alternative way to understand the periodicity of the entropy (and justify why it should be present) has been given in \cite{Agullo:2008eg}. We want to mention here that some non-trivial ``intermediate'' regimes may be significant in the behavior of the entropy because it is not inconceivable that the sizes of real astrophysical black holes are actually outside the large-area asymptotic regime. This would mean that a non-trivial behavior could be displayed by these objects even if the asymptotics of the entropy is more or less trivial (i.e. just linear). Finally, the number-theoretic methods that we propose lead to a clear identification of the role played by the different sources of degeneracy in the black hole spectrum and how they explain the observed non-trivial behavior of black hole entropy in LQG.

The purposes of this paper are manifold. First of all, we give a detailed and expanded derivation of the number-theoretic methods of \cite{Agullo:2008yv} and show how they can be effectively used to understand in full detail many issues related to the behavior of black hole entropy. To avoid confusion, we consider separately the computation of the entropy in the Ashtekar-Baez-Corichi-Krasnov (ABCK) framework \cite{Ashtekar:2000eq}, as reinterpreted by Domagala and Lewandowski (DL) \cite{Domagala:2004jt}, and other countings proposed in the literature \cite{Ghosh:2006ph,Kaul:1998xv}. We also consider the novel proposal of \cite{Engle:2009vc} where the authors study black hole entropy from first principles without using any internal gauge fixing of the $SU(2)$ symmetry on the horizon. The reason to consider these other points of view is to illustrate the flexibility of our approach. Our second goal is to derive and discuss the black hole generating functions that codify all the information about black hole entropy as a function of the horizon area. We will show how they can be immediately applied to derive the asymptotic behavior of the black hole degeneracy spectrum for some important subsets of the area spectrum (a discussion of how they can be used to get an integral representation for the black hole entropy appears in \cite{G.:2008mj}). The third goal of the paper is to provide a detailed account of how the periodic pattern in the black hole degeneracy spectrum arises. We will take advantage of the number-theoretical methods described in the beginning of the paper to study the interplay between the different contributions of the different combinatorial factors to the entropy and also to show what the most relevant area configurations are. We will also provide a simple heuristic argument that complements the one given in \cite{Agullo:2008eg}. Another important result in this respect will be to show that the interesting behavior of the black hole entropy is present even without taking into account the so-called ``projection constraint'' (though in this case the logarithmic correction to the entropy is absent). Finally we give a number-theoretic prescription to label and classify the substructures (``bands'') that appear in the black hole degeneracy spectrum. Summarizing, the main new results presented in the paper are:

\begin{itemize}
\item We provide several independent methods to count the configurations selected by the projection constraint, in particular, we explain how the counting can be made by using auxiliary quantum Hamiltonians for spin systems. We also give a group-theoretic treatment that can be used for all the different countings discussed in the literature.
\item We explain how to obtain the asymptotic behavior of the black hole degeneracy spectrum for subfamilies of area eigenvalues characterized by a set of square free integers. We use then this method to show how the band structure of the degeneracy spectrum originates.
\item We discuss the relative importance of the different sources of degeneracy. In particular we disentangle the contributions coming from the projection constraint from the remaining ones.
\item We provide a new argument that explains how the configurations contributing to the bands in the degeneracy spectrum can be labeled by a simple function $P$ of the spin labels at the punctures on the black hole horizon. We do this by using a straightforward continuum approximation.
\item We introduce a new type of generating function, defined with the help of $P$, that allows us to isolate the configurations contributing to specific bands in the degeneracy spectrum.
\item We give the full details regarding the degeneracy spectrum and entropy for the black holes with the smallest possible areas.
\end{itemize}

The paper is organized as follows. After this introduction we devote the next section (section \ref{sect:Isolated}) to a quick discussion of how black holes are modeled with the help of isolated horizons. The aim of this section is to provide a frame to assess and compare the different proposals, in particular those suggesting to employ a $SU(2)$ Chern-Simons model to describe the quantum degrees of freedom on the horizon. Section \ref{sect:ABCK} describes in detail the methods used to count the number of states relevant for the computation of the entropy  the ABCK prescription according to DL. In this section we introduce several precise definitions (black hole configurations, degeneracy spectrum, etc.) and solve the relevant number-theoretic and combinatorial problems step by step. Section \ref{sect:Other} will discuss the application of our methods to other countings and proposals, in particular, there is a subsection devoted to the Ghosh-Mitra counting \cite{Ghosh:2006ph} and another to the $SU(2)$-black entropy proposal of Engle, Noui and Perez \cite{Engle:2009vc}. In the process we will describe several ways to solve the so called projection constraint. One of them is interesting because it offers some tantalizing hints about the possible connection of the problem of computing black hole entropy with conformal field theory techniques \cite{Agullo:2009zt}. Another one makes use of a simple type of generating function (a fact that originally suggested that it might be possible to find generating functions for the exact black hole entropy \cite{BarberoG.:2008ue}). A very short summary of the different schemes is presented in section \ref{sect:Summary}. Section \ref{sect:Detailed} provides a detailed account of the features of the black hole degeneracy spectrum, how it arises, what the most relevant area eigenvalues are, and the mathematical classification of the characteristic bands that it shows. We give here an argument, alternative to the one given in \cite{Agullo:2008eg}, that explains how the periodicity of the black hole degeneracy spectrum arises and we also provide a generating function that selects the configurations that define the bands. In particular, the analysis presented in this section contains new important information about the detailed features of the entropy spectrum, and introduces new techniques for the study of its periodic structures. All this information is valuable for the purpose of obtaining the asymptotic behavior of the entropy in the limit of very large (astrophysical) black holes, which is the main open problem in the number theoretic approach that we are following. Section \ref{sect:Conclusions} is devoted to our conclusions, comments, and suggests directions for future work. We end the paper with several appendices. The first one (Appendix \ref{App:A}) provides a pedagogical derivation of the generating functions used in the paper. Appendix \ref{App:B} gives a complete and explicit computation of the black hole degeneracy spectrum and the entropy for \textit{all} the small black holes with areas smaller than  $18\times 4\pi\gamma\ell_P^2$. (Throughout the paper we will denote by $\ell_P$   the Planck length.) Finally, Appendix \ref{App:C} gives a unified, group-theoretic treatment to solve the so called projection constraint in any of the counting proposals discussed in the paper.

\section{\label{sect:Isolated}Isolated horizons and black hole entropy}

The purpose of this section is to give a short history of the study of black holes in the loop quantum gravity framework and, in particular, of the entropy computations leading to the Bekenstein-Hawking law. Our main goal here is to provide the reader with the background information necessary to establish the connections between the different approaches and proposals, assess their degree of rigor and state of development, and compare their relative merits. Several excellent reviews on the subject are available (see for example \cite{Ashtekar:2004eh,Corichi:2009wn}).

The successful derivation of the Bekenstein-Hawking law is a key stepping stone in the quest to arrive at a working theory of quantum gravity. This explains why the main contenders in this area of fundamental physics --string theory and loop quantum gravity-- have struggled to derive this result within the respective frameworks.\footnote{An interesting discussion of the relative merits and problems of both approaches can be found in \cite{Hor}.} In both approaches, the first problem that must be faced is the appropriate description of a quantum black hole or, at least, a physical approximation to it. The history of the study of black hole entropy in LQG can be divided in two different periods separated by the landmark paper of Ashtekar, Baez, Corichi and Krasnov \cite{Ashtekar:1997yu}. The main importance of this work lies in the fact that it proposes the key idea of describing black holes by using isolated horizons. The reason why this is so important is because a ``reduction'' of general relativity consisting of spacetimes admitting isolated horizons as inner boundaries can be described within the Hamiltonian formalism. This idea is similar in spirit, but not equal, to the study of the quantization of the symmetry reductions of general relativity that gives rise to the popular mini and midisuperspace models. For both types of systems a subset of the possible gravitational field configurations is chosen by imposing some restriction on the configuration space of metrics. In the mini and midisuperspace models this is accomplished by restricting the allowed configurations to metrics satisfying some kind of symmetry principle whereas in the case of black holes the requirement is that the allowed metrics must have an isolated horizon.

During the period preceding the appearance of \cite{Ashtekar:1997yu}, several authors made interesting suggestions that have played a key role in the development of the standard ABCK model. In fact, it is fair to say that many of the ideas that have been instrumental in the currently accepted framework made their appearance during those years. We want to mention explicitly the work of Smolin in \cite{Smolin:1995vq}, where he discusses, among other things, the importance of considering the quantization of the gravitational field in spacetimes with inner boundaries of finite area and the important role of the Chern-Simons theory and quantum groups. After this paper, it is mandatory to highlight the works of Krasnov \cite{Krasnov:1996tb,Krasnov:1996wc} and Rovelli \cite{Rovelli:1996dv,Rovelli:1996ti}. The first author gives several important insights related, in particular, to the issue of the distinguishable character of the spin network labels on the horizon, the types of spin networks that are relevant to describe black holes and the concrete role of the CS theory to describe horizon degrees of freedom.\footnote{It is interesting to mention at this point that the original proposal made use of a $SU(2)$-CS model.} Rovelli, on his part, made a direct counting and used simple combinatorial methods to arrive at an approximate Bekenstein-Hawking law (in the sense that the linear growth of the entropy as a function of the area was found but the proportionality coefficient was not fixed). He also insisted on the role played by the distinguishability of punctures on the black hole horizon.

The currently accepted treatment of the problem is given in \cite{Ashtekar:1997yu,Ashtekar:2000eq} (see also the review by Corichi \cite{Corichi:2009wn}). The starting point is the Hamiltonian description of spacetimes with isolated horizons in terms of Ashtekar variables. A crucial issue is the treatment of the inner boundary and the consequences that its introduction has for the final Hamiltonian description of the model. The following facts are particularly relevant.

\begin{itemize}
\item The isolated horizon condition translates itself into a matching boundary condition involving the pullback of the Ashtekar connection and the tetrads to the spherical slices of the inner boundary.

\item The inner horizon boundary conditions force the $SU(2)$ connection $A^\imath\,_\jmath$ to be reducible on the boundary \cite{Ashtekar:1997yu} (here the indices  $\imath$ and $\jmath$ are $\mathfrak{su}(2)$ indices). This means that there exists an internal vector field $r^\imath$, defined on the spherical section of the isolated horizon, satisfying the condition $d_Ar^\imath=dr^\imath+A^\imath\,_\jmath \,r^\jmath =0$. A choice of this internal vector [that amounts to a partial gauge fixing of the $SU(2)$ symmetry] is used to explicitly take into account the reducibility of the connection on the boundary. The $U(1)$ invariance on the isolated horizon is guaranteed by the $r^\imath$ projection of the matching conditions.

\item The symplectic structure consists on a volume part (the same as in the usual models without boundary) and a boundary part corresponding to a $U(1)$-CS theory. This fact strongly suggests that the Hilbert space appropriate to describe the system is a tensor product of a $U(1)$ Chern-Simons Hilbert space and a LQG volume Hilbert space.

\item In the quantum formalism the $r^\imath$-projected matching condition is quantized and used to select the physical quantum states. The solutions to this condition automatically satisfy the Gauss law. This means that it is treated as a first class constraint as far as quantization is concerned.
\end{itemize}

The reducibility condition of the connections on the horizon is a somehow unexpected feature of the present scheme. Several authors have explored the physical consequences of heuristically forgetting about this fact and retain a $SU(2)$-CS model on the horizon \cite{Kaul:1998xv,Kaul:2000kf}, whereas other authors have addressed the development of a $SU(2)$ invariant formalism from first principles \cite{Engle:2009vc,Engle:2010kt}. The possibility of using a $SU(2)$ description is an interesting topic. In fact there is a recently published but old proposal by Krasnov and Rovelli \cite{Krasnov:2009pd} to describe quantum black holes --without the use of any semiclassical approximations-- that lead to results very similar to the ones presented in \cite{Engle:2009vc,Engle:2010kt}.

Within all these models it is possible to obtain the Bekenstein-Hawking law by appropriately adjusting $\gamma$. There are also logarithmic corrections to the entropy that play an important role because they are specific predictions of the different proposals that can be used, in principle, to choose among them. The combinatorial problems associated with the computation of the entropy for all these schemes are rather similar in nature and can be approached with the techniques that we describe in this paper. In that respect, our methods are quite robust and well adapted to the nature of the combinatorial problems that appear, so we expect them to play a significant role also in the understanding of dynamical aspects of black hole physics such as Hawking radiation or black hole evaporation.

\section{\label{sect:ABCK}The ABCK quantum isolated horizon and the DL counting }

The study of black holes within LQG \cite{Ashtekar:2000eq} makes use of the \textit{isolated horizon} concept (see \cite{Ashtekar:2004cn} and references therein). In this framework the horizon is introduced as an inner boundary of the classical spacetime manifold. Several conditions are imposed on it to guarantee that the relevant physical features of a black hole are captured, in particular its thermodynamical behavior. Spacetimes of this class admit 3-dimensional, spatial, partial Cauchy surfaces. Each of them is  bounded by a topological 2-sphere, that we will refer to as the horizon. A Hamiltonian description for this sector of GR is available and is the starting point for canonical quantization \cite{Ashtekar:1999wa}. The details of the canonical description strongly suggest that the appropriate Hilbert space  should be  built as a tensor product of bulk and horizon Hilbert spaces,  $\mathcal{H}^\kappa_{\rm Kin}=\mathcal{H}^\kappa_{\rm Hor}\otimes\mathcal{H}_{\rm Bul}$. In this approach, it is natural to adapt the bulk Hilbert space $\mathcal{H}_{\rm Bul}$ to the presence of an inner boundary. As usual in LQG, it is convenient to use a spin network basis\footnote{A spin network is an oriented graph embedded in a spatial section of the spacetime, carrying a $SU(2)$ irreducible representation labeled by a spin number $j$ on each of its edges, and a gauge invariant operator (intertwiner) linking incoming and outgoing representations at each vertex \cite{Ashtekar:2004eh}.}. A particular bulk spin network can pierce the horizon or not. If it does,  it will do so at a finite number of points that we will refer to as \textit{punctures}. These punctures have a distinguishable character, as a consequence of the action of general diffeomorphisms over the horizon surface \cite{Ashtekar:2000eq}. The punctures carry two quantum numbers $(j,m)$. The first one is just the spin $j$-label of the edge that defines it. These spin quantum numbers allow us, in particular, to calculate the horizon area according to the standard prescription of quantum geometry. The other quantum numbers are spin components, $m$, defined with the help of the preferred $\mathfrak{su}(2)$-internal vector field $r^\imath$ on the horizon (see section \ref{sect:Isolated}). These labels play an important role in the implementation of the quantum boundary conditions. Explicitly, the bulk Hilbert space ${\cal H}_{\rm Bul}$ is spanned by states of the form
\begin{equation}
|(0),\cdots\rangle_{\rm Bul},\ldots,
|(m_1,j_1,\ldots,m_N,j_N),\cdots\rangle_{\rm Bul}\,,\ldots\label{bulkstates}
\end{equation}
where the half integers $j_I\in\frac{1}{2}\mathbb{N}$ label irreducible representations of
$SU(2)$ and
\begin{equation}
m_I \in \{-j_I,\,-j_I+1,\ldots,j_I-1,j_I\}
\end{equation}
are spin components in the direction of $r^\imath$. The numbers $m_I$ and $j_I$ represent the quantum degrees of freedom of the bulk geometry ``close'' to the horizon, and the ``$\cdots$'' in
the bulk state $|(m_1,j_1,\ldots,m_N,j_N),\cdots\rangle_{\rm Bul}$  refer to bulk degrees of freedom away from the horizon. Finally $|(0),\cdots\rangle_{\rm Bul}$ denotes the states corresponding to spin networks that do not pierce the horizon.

The surface horizon Hilbert space ${\cal H}^\kappa_{\rm Hor}$ is described by a $U(1)$ Chern-Simons theory. The level $\kappa\in \mathbb{N}$ of this quantum CS-theory gives rise to a prequantized value $a_\kappa=4\pi\gamma\ell^2_P \kappa$ for the area of the isolated horizon. In $\mathcal{H}^\kappa_{\rm Hor}$,  the $U(1)$-CS basis states over the punctured sphere $|(c_1,\ldots,c_N)\rangle^\kappa_{\rm Hor}$ are characterized by arbitrarily long (ordered) sequences $(c_I)_{I=1}^N=(c_1,\ldots,c_N)$ of non-zero congruence classes of integers modulo $\kappa$. Each of the $c_I$ can be thought of as an integer number belonging to $\{1,2,\ldots, \kappa-1\}$ that  labels the quantized deficit angle $4\pi c_I/\kappa$ associated with the $I$-th puncture. If the isolated horizon has a spherical topology, there is an additional restriction over the total curvature, which translates into a condition
$$\sum_I c_I=0 \quad  (\mathrm{mod} \,\, \kappa),$$
for the  congruence classes appearing in a given sequence. This condition  can be interpreted  as the quantum equivalent of the Gauss-Bonnet theorem. These $c_I$ labels turn out to be related, through the quantized isolated horizon boundary conditions, to the label $m_I$ in the $j_I$ representation of the spin network edge, $|(j_I,m_I)_{I=1}^N,\cdots\rangle_{\rm Bul}\in \mathcal{H}_{\rm Bul}$,  piercing at the corresponding puncture, according to
$$c_I=-2m_I \quad (\mathrm{mod}\,\,\kappa)\ ,$$
with $m_I\in\left\{-j_I, -j_I+1,\ldots,j_I\right\}$. This restriction on the form of the basis states $|(c_I)\rangle_{\rm Hor}^\kappa\otimes |(j_I,m_I),\cdots\rangle_{\rm Bul}$ of $\mathcal{H}^\kappa_{\rm Kin}$ is the quantum counterpart of the isolated horizon boundary condition.

The area induced on the horizon by a given bulk state is the eigenvalue of the standard area operator
\begin{eqnarray}
\label{area} a^{{\scriptscriptstyle \rm LQG}}(j_I,m_I)=8\pi\gamma\ell_P^2\sum_I{\sqrt{j_I(j_I+1)}}\,,
 \end{eqnarray}
given in terms of the $j_I$-labels of the punctures. For a fixed value of the black hole area, there are different combinations of horizon labels compatible with it (i.e. the area eigenvalues are degenerate). The quantum states of the horizon belonging to $\mathcal{H}_{\rm Hor}^\kappa$ and ``compatible'' with the area (\ref{area}) are precisely the ones responsible for the black hole entropy.

Some important comments, relevant to the definition of the entropy, are in order now:
\begin{itemize}
\item \textbf{Compatibility of areas:} Two different types of areas associated with the horizon are relevant here. The prequantized area $a_\kappa$ necessary for the quantization of the CS theory and (\ref{area}). The $a_\kappa$
    does not belong to the spectrum of the area operator, i.e.
$$
a_\kappa=4\pi\gamma\ell^2_P \kappa\neq a^{{\scriptscriptstyle \rm LQG}}(j_I,m_I)=8\pi\gamma\ell_P^2\sum_I{\sqrt{j_I(j_I+1)}}\,.
$$
This forces us to introduce a suitable, and physically sensible, notion of compatibility.\footnote{In this respect we want to point out that there is another possible choice \cite{FernandoBarbero:2009ai} for the horizon area operator in LQG corresponding to an evenly spaced area spectrum. In this case the Bekenstein-Mukhanov \cite{Bekenstein:1974jk,Bekenstein:1995ju} scheme is realized from the start and there is no need to introduce an area interval at this stage.}

\item  \textbf{Entropy:} In Quantum Statistical Mechanics the standard definition of the entropy in the microcanonical ensemble makes use of an energy interval \cite{G}. In vacuum general relativity, the non-rotating and neutral black holes are fully characterized by their areas. Therefore, it is convenient to use this geometrical feature in the discussion of their statistical mechanics. This is explained at length in \cite{Krasnov:1996tb,Krasnov:1997yt}. Here, then, we introduce an area interval $[a_\kappa-\delta,a_\kappa+\delta]$ in order to define an appropriate statistical ensemble. This helps to solve the problem mentioned in the previous item because, once the prequantized value of the area is fixed, there are always bulk quantum states with area eigenvalues belonging to $[a_\kappa-\delta,a_\kappa+\delta]$ for reasonable choices of $\delta$. The entropy is obtained by tracing out the bulk degrees of freedom to get a density matrix that describes a maximal entropy mixture of surface states with area eigenvalues in the previous interval. The value of the entropy is computed by counting the number of allowed sequences $(c_I)$ of non-zero elements of $\mathbb{Z}_\kappa$ satisfying $c_1+\cdots+c_N=0$, such that $c_I=-2m_I\, (\mathrm{mod}\, \kappa)$ for some permissible spin components $(m_I)$. Here permissible means that there exists a sequence of non-vanishing spins $(j_I)$ such that each $m_I$ is a spin component of $j_I$ and
\begin{eqnarray}
a_\kappa-\delta\leq a^{{\scriptscriptstyle \rm LQG}}(j_I,m_I)=8\pi\gamma\ell_P^2\sum_I\sqrt{j_I(j_I+1)}\leq a_\kappa+\delta.
\end{eqnarray}
The counting of \textit{horizon} $c$-labels is equivalent to the determination of the dimension of the Hilbert subspace of ${\cal H}^{\kappa}_{\rm Hor}$ that represents the black hole degrees of freedom.

\item \textbf{Thermodynamical limit:}  In the thermodynamical limit, for standard statistical mechanical systems, the choice of energy interval introduced in the definition of the microcanonical ensemble is irrelevant \cite{G}, i.e. it is equivalent to consider $[E-\Delta, E+\Delta]$, or $[E-\Delta, E]$, or $[E, E+\Delta]$, or even $[E_0,E]$, where $E_0$ is the minimum  energy of the system. Moreover, only in the thermodynamical limit the entropy is a smooth function of the energy with the possible exception of points related to phase transitions. This smoothness is crucial to define the derivatives appearing in the definition of thermodynamical magnitudes. In the present case, one should follow a similar path (with the area playing the role of the energy).
\end{itemize}

The possibility of taking an appropriate area interval has been used to simplify the actual computation of the entropy. By using an interval of the form $[0,a_\kappa]$, Domagala and Lewandowski \cite{Domagala:2004jt} proved   that the  black hole entropy can be obtained according to the following prescription involving only the $m_I$ \textit{bulk} labels:

\bigskip

\textit{
The entropy $S_\leq(a_\kappa)$ of a quantum horizon of classical area $a_\kappa$, according to Quantum Geometry
and the Ashtekar-Baez-Corichi-Krasnov  framework, is
$$S_\leq(a_\kappa) = \log (1+N_\leq(a_\kappa))\,,$$
where $N_\leq(a_\kappa)$ is  the number of all the finite, arbitrarily long, sequences $(m_1,\ldots,m_N)$ of non-zero half integers, such that the following equality and inequality are satisfied:
$$\sum_{I=1}^N m_I=0, \quad \sum_{I=1}^N\sqrt{|m_I|(|m_I|+1)}\leq \frac{a_\kappa}{8\pi\gamma\ell_P^2}.$$
The extra term 1 above comes from the trivial sequence.}

\bigskip

\noindent Notice that, in the ABCK prescription, the entropy $S_\leq(a)$ is defined \textit{only} for area values of the form $a=a_\kappa$. However, we will extend the definition to arbitrary values of $a\in[0,\infty)$ by just requiring that
$$
\sum_{I=1}^N\sqrt{|m_I|(|m_I|+1)}\leq \frac{a}{8\pi\gamma\ell_P^2}\,.
$$
We introduce this extension because we think that the detailed form of the area spectrum predicted by loop quantum gravity may play a prominent role in gravitational systems for which area is an important observable quantity. In the following, unless stated otherwise, we will write areas in units of $4\pi\gamma\ell_P^2$.

\bigskip

The rest of this section is devoted to describing methods based on number theory that are useful to understand the structure of the black hole degeneracy spectrum and entropy. The counting of the allowed sequences according to the previous definition is conveniently performed in four successive steps:
\begin{enumerate}
\item[Step 1.] Fix a value for the area $a$ and obtain all the possible choices for the half integers $|m_I|$ compatible with the area, i.e. satisfying
    $$
    \sum_{I=1}^N\sqrt{|m_I|(|m_I|+1)}= \frac{a}{2}.
    $$
    Notice that, at this point, we are considering the possible choices of absolute values of the spin components as the elements of a multiset (and hence there is no ordering of the labels). In other words, at this stage we only want to know how many times the spin component $1/2$ appears, how many $1$'s appear, and so on.
\item[Step 2.] Count the different ways in which the previous multisets can be reordered.
\item[Step 3.] Count all the different ways of introducing signs in the sequences of positive half-integers $|m_I|$ obtained in the previous step in such a way that the projection constraint $\sum_{I=1}^N m_I=0$ is satisfied.
\item[Step 4.] Repeat the same procedure for all the area eigenvalues smaller than $a$ and add the number of sequences obtained in each case.
\end{enumerate}

\subsection{{\label{subPell}}Step 1: The Pell equation}

The first step can be thought of, in fact, as a characterization of the part of the spectrum of the area operator relevant to the computation of black hole entropy (regarding, in  particular, the degeneracy of the different area eigenvalues). Let us start by introducing the positive integer variables $k_I:=2|m_I|$ and write
\begin{eqnarray}
2\sum_{I=1}^N\sqrt{|m_I|(|m_I|+1)}=a \Leftrightarrow\sum_{I=1}^N\sqrt{(k_I+1)^2-1}=a\Leftrightarrow\sum_{k=1}^{k_{\rm max}}N_k\sqrt{(k+1)^2-1}=a\hspace{9mm}\label{eqkn_k}
\end{eqnarray}
where the non-negative integers $N_k$ (that will be allowed to be zero) in the last sum tell us the number of times that the label $k/2\in\mathbb{N}/2$ appears in the sequence $(|m_I|)=(k_I/2)$. Also, we denote as $k_{\rm max}=k_{\rm max}(a)$ the maximum value of the positive integer $k$ compatible with the area $a$. At this point we are interested in finding out the multisets mentioned above, i.e. all the sets of pairs $\{(k,N_k):k\in\mathbb{N},\,N_k\in\mathbb{N}\cup\{0\}\}$ such that equation (\ref{eqkn_k}) is satisfied. Notice that in the description provided by the multiset, we can restrict ourselves to list only the values of $k$ that do actually appear (i.e. those for which $N_k\neq 0$). We want to point out now a simple but important fact. By using the prime factor decomposition of $k(k+2)$, it is always possible to write $\sqrt{(k+1)^2-1}$ as the product of an integer times the square root of a square-free positive number $p_i$. Square-free numbers are integers that are not divisible by non-trivial square numbers (that we enumerate as $p_1=2$, $p_2=3$, $p_3=5$, and so on). Then, equation (\ref{eqkn_k}) tells us that the area eigenvalue $a$ must be an integer linear combination of square roots of squarefree numbers, and have the form
$$
a=\sum_{i=1}^{i_{\rm max}} q_i \sqrt{p_i},\quad q_i\in\mathbb{N}\cup\{0\}\,,
$$
or else equation (\ref{eqkn_k}) cannot be satisfied. This leads then to the following condition
\begin{equation}
\sum_{k=1}^{k_{\rm max}}N_k\sqrt{(k+1)^2-1}=\sum_{i=1}^{i_{\rm max}} q_i\sqrt{p_i}\,.
\label{eqdiof1}
\end{equation}
As a preliminary step to solve this equation, we want to separately consider each of the square-free numbers $p_i$ and find out the possible values of $k$ such that $\sqrt{(k+1)^2-1}$ is an integer multiple of $\sqrt{p_i}$, i.e. we first solve the equations
$$
\sqrt{(k_i+1)^2-1}=y_i\sqrt{p_i}
$$
in the two unknowns $k_i$ and $y_i$ (here the label $i$ refers to the square-free number $p_i$). These are equivalent to
\begin{equation}
x_i^2-p_i y_i^2=1\,,
\label{Pell}
\end{equation}
where we have written $x_i:=k_i+1$. For each fixed square-free integer $p_i$,  (\ref{Pell}) is the well known Pell equation \cite{Burton}, whose general solution can be found in the following way. Obtain first a so called \textit{fundamental solution} $(x_1^i,y_1^i)$ with the smallest positive value of $x$ by using continued fractions as explained in \cite{Burton}. Once this solution is known the (infinitely many) remaining ones are given by $(x_i,y_i):=(x_\alpha^i,y_\alpha^i),$ $\alpha\in\mathbb{N}$,  where
\begin{eqnarray*}
&&x^i_{\alpha}=\frac{1}{2}
\left[(x^i_1+y^i_1\sqrt{p_i})^{\alpha}+(x^i_1-y^i_1\sqrt{p_i})^{\alpha}\right]\,,\\
&&y^i_{\alpha}=\frac{1}{2\sqrt{p_i}}
\left[(x^i_1+y^i_1\sqrt{p_i})^{\alpha}-(x^i_1-y^i_1\sqrt{p_i})^{\alpha}\right]\,,
\end{eqnarray*}
from which we get the solutions to the original equation that we will label as $(k_\alpha^i,y_\alpha^i)$. For instance, for the first square-free $p_1=2$, the previous sequence $(k^1_\alpha,y^1_\alpha)$ starts as $(2,2),(16,12),(98,70),(576,408),\ldots$ (see section \ref{subsect:disentangling} for more details).

Once we know the values of $k^i_{\alpha}$ that can contribute when a given $p_i$ appears in the value of the area, equation (\ref{eqdiof1}) can be written as
$$
\sum_{k=1}^{k_{\rm max}}N_k\sqrt{(k+1)^2-1}=\sum_{i=1}^{i_{\rm max}}\sum_{{\alpha}=1}^\infty N_{k_{\alpha}^i}y_{\alpha}^i\sqrt{p_i}=\sum_{i=1}^{i_{\rm max}} q_i\sqrt{p_i}\,.
$$
Now we make use of the fact that the square roots of the square-free numbers are linearly independent over the rationals (and hence also over the integers) to show that the previous equation is equivalent to the following system of $i_{\rm max}$  uncoupled, linear, diophantine equations
\begin{equation}
\sum_{{\alpha}=1}^\infty y_{\alpha}^i N_{k_{\alpha}^i}=q_i,\quad i=1,\ldots, i_{\rm max}.\label{diofN}
\end{equation}
Several comments are in order now. The first is that, for a fixed value of the area $a$ (necessarily an integer linear combination of $\sqrt{p_i}$ with a finite number of coefficients $q_i$) only a finite number of labels $k_\alpha^i$ come into play. Second, it may happen that some of these equations admit no solutions, in that case $a$ does not belong to the spectrum of the area operator (this happens, for instance, for $a=\sqrt{2}$). Finally, when they can be solved, their solution tell us exactly what the allowed values for $k$ are and the number of times $N_k$ that they appear. This construction identifies the set of allowed \textit{configurations} $\mathcal{C}(a)$ consisting of all multisets  $c=\{(k_\alpha^i,N_{k_\alpha^i})\}$ associated with a value of the area $a=\sum_{i}q_i\sqrt{p_i}$.

\subsection{Step 2: The reordering degeneracy ($r$-degeneracy)}

Up to this point we have found the number of all the possible choices for the absolute values of the spin components $|m|=k/2$ compatible with a given value of the area $a$ together with their multiplicities $N_k$. Each of the configurations $c\in\mathcal{C}(a)$ can be represented in the following schematic form
$$
c=\Big(\underbrace{\frac{1}{2},\ldots,\frac{1}{2}}_{N_1},\ldots, \underbrace{\frac{k}{2},\ldots,\frac{k}{2}}_{N_k},\ldots, \underbrace{\frac{k_{\rm max}}{2},\ldots,\frac{k_{\rm max}}{2}}_{N_{k_{\rm max}}}\Big)\,,
$$
where, in the previous representation, if a particular $N_k$ is zero then there are no $k/2$ terms. The number of different sequences $(k_I/2)$ obtained from each configuration $c$ by reordering its elements is given by the multinomial coefficient
\begin{equation}
d_r(c):=\frac{\big(\sum_{k=1}^{k_{\rm max}} N_k\big)!}{\prod_{k=1}^{k_{\rm max}} N_k!}.\label{rdeg}
\end{equation}
In the following we will refer to $d_r(c)$ as the $r$-degeneracy of the configuration $c$. We will also define the $r$-degeneracy associated with a given value of the area $a=\sum_{i}q_i\sqrt{p_i}$ as
\begin{equation}
D_r(a)=\sum_{c\in\mathcal{C}(a)}
d_r(c)\,.
\end{equation}
As we will see, this quantity plays a central role in the appearance of the black hole entropy structure found in \cite{Corichi:2006wn}.

\subsection{\label{sub413} Step 3: Solving the projection constraint ($m$-degeneracy)}

Once we have identified all the possible sequences $(|m_I|)=(k_I/2)$ of positive half-integers satisfying the area condition, we have now to introduce \textit{signs} in each $m_I$ and find out how many of the resulting sequences $(m_I)$ satisfy the so-called projection constraint
\begin{equation}
\sum_{I=1}^N m_I=0\,.
\label{proyconst}
\end{equation}
There are several approaches to solve this problem that we will describe here. The reasons to look at these different ways to solve the projection constraint are the following. First, some of them are specially suited to be used in the computer algorithms that we have employed in the actual black hole entropy computations. Second, the solution in terms of generating functions is a preliminary step towards the obtention of the black hole generating functions that we define below. Finally, some of the solutions suggest intriguing connections with other interesting problems (such as conformal field theories as we will mention briefly in Appendix \ref{App:C}). Let us consider them one by one.

\bigskip

\paragraph*{The partition problem:}
The problem of finding all the possible different ways to ``sprinkle'' the signs on a fixed sequence $(k_I/2)$ has already been considered in the literature. In fact, it has a proper name: \textit{the partition problem}, that can be stated as follows. Given a sequence $(k_I/2)=(k_1/2,k_2/2,\ldots,k_N/2)$  of $N$ real numbers (positive half-integers in our case) find all the different partitions of  $\{1,2,\ldots, N\}=\mathcal{N}_+ \cup\mathcal{N}_-$ such that
 $$
\sum_{I\in \mathcal{N}_+} k_I -\sum_{I\in \mathcal{N}_-} k_I=0\,.
 $$
Here we will solve the following slightly different problem: given $M\in \mathbb{Z}/2$, find out the \textit{number} of different ways to partition $\{1,2,\ldots, N\}$ in such a way that the following condition holds
\begin{eqnarray}
\sum_{I\in \mathcal{N}_+} k_I -\sum_{I\in \mathcal{N}_-} k_I=2M\,.\label{2M}
\end{eqnarray}
The answer to this question is known and can be found, for example, in \cite{DeRaedt}. It is given by
\begin{eqnarray}
\label{DLcounting}
\frac{2^N}{L}\sum_{\ell=0}^{L-1}e^{-4\pi i \ell M/L}\prod_{I=1}^{N}\cos(2\pi \ell k_I/L)=
\frac{2^N}{L}\sum_{\ell=0}^{L-1}e^{-4\pi i \ell M/L}\prod_{I=1}^{N}\cos(4\pi \ell |m_I|/L)\hspace*{5mm}
\end{eqnarray}
where $L$ is a conveniently chosen integer. As the result is independent of this choice (as long as $L$ is big enough, see below) we will take $L:=1+2M+\sum_{I=1}^N k_I$.

The formula (\ref{DLcounting}) is obtained in the following way \cite{DeRaedt}. For a given sequence of positive integers $(k_1,k_2,\ldots,k_N)\in \mathbb{N}^N$ and $M\in \mathbb{Z}/2$, let us define the auxiliary Hamiltonian operator  $H:\mathbb{C}^{2\otimes N}\rightarrow\mathbb{C}^{2\otimes N}$ as
$$
H=2M-\sum_{I=1}^N k_I \sigma_I^{(3)}\,,
$$
where the operator $\sigma_I^{(3)}=1\otimes \cdots \otimes1\otimes \sigma^{(3)}\otimes 1\otimes\cdots\otimes 1$ acts  as the $\sigma^{(3)}$ Pauli matrix on the $I^{\rm th}$ $\mathbb{C}^2$-factor of $\mathbb{C}^{2\otimes N}$ and trivially on the others.
This operator satisfies
\begin{eqnarray*}
\frac{1}{L}\sum_{\ell =0 }^{L-1}\mathrm{Tr}\Big(\exp(-2\pi i\ell H/L)\Big)&=&\frac{1}{L}\sum_{\ell =0 }^{L-1}\sum_{\{\sigma_I=\pm1\}} \exp\Big(\frac{2\pi i \ell}{L}\Big(\sum_{I=1}^N k_I\sigma_I-2M\Big)\Big).
\end{eqnarray*}
When $\sum_{I=1}^N k_I\sigma_I-2M=0$ we have
$$\frac{1}{L}\sum_{\ell =0 }^{L-1} \exp\Big(\frac{2\pi i \ell}{L}\Big(\sum_{I=1}^N k_I\sigma_I-2M\Big)\Big)=1.$$ On the other hand, when $\sum_{I=1}^N k_I\sigma_I-2M\neq0$ we can write
\begin{eqnarray*}
\sum_{\ell =0 }^{L-1} \exp\Big(\frac{2\pi i \ell}{L}\Big(\sum_{I=1}^N k_I\sigma_I-2M\Big)\Big)= \frac{1-\exp\Big(2\pi i \Big(\sum_{I=1}^N k_I\sigma_I-2M\Big)\Big)}{1-\exp\Big(2\pi i \Big(\sum_{I=1}^N k_I\sigma_I-2M\Big)/L\Big)}=0\,.
\end{eqnarray*}
In this last equation we have used the fact that $\sum_{I=1}^N k_I\sigma_I-2M\in \mathbb{Z}$ and  $L=1+2M+\sum_{I=1}^N k_I$ to guarantee that the denominator in the previous expression never vanishes. This way we conclude that
\begin{eqnarray}
\frac{1}{L}\sum_{\ell =0 }^{L-1}\mathrm{Tr}\Big(\exp(-2\pi i\ell H/L)\Big)= \sum_{\{\sigma_I=\pm1\}} \delta\big(2M,\sum_{I}k_I\sigma_I\big)\,,\label{suma1}
\end{eqnarray}
i.e. the trace above counts the partitions such that (\ref{2M}) holds.
The left hand side of equation (\ref{suma1}) can be explicitly computed as
\begin{eqnarray}
\frac{1}{L}\sum_{\ell =0 }^{L-1}\mathrm{Tr}\Big(\exp(-2\pi i\ell H/L)\Big)&=&\frac{1}{L}\sum_{\ell =0 }^{L-1}e^{-4\pi i \ell M/L}\prod_{I=1}^N \mathrm{Tr}\left(\exp(2\pi i \ell k_I\sigma^{(3)}/L)\right)\nonumber\\
&=&\frac{2^N}{L}\sum_{\ell =0 }^{L-1}e^{-4\pi i \ell M/L}\prod_{I=1}^N \cos(2\pi  \ell k_I/L)\,.\label{traza}
\end{eqnarray}
By comparing (\ref{suma1}) and (\ref{traza}) it is clear that (\ref{DLcounting}) provides the number of solutions to the partition problem. It is important to notice that equation (\ref{suma1}) implies that expression (\ref{DLcounting}) is zero
whenever no solutions to the projection constraint can be found.

\bigskip

\paragraph*{Generating functions:} Another method to solve the projection constraint (\ref{2M}) is based on the use of a suitable generating function. Let us consider  an ordered, finite, sequence $(k_I/2)$ of positive half integers of the form $k/2$  with multiplicities given by $N_k>0$, and take the following function of the variable $z$ (a Laurent polynomial)
\begin{equation}
\prod_{I=1}^N(z^{k_I}+z^{-k_I})=\prod_{k}(z^{k}+z^{-k})^{N_k}.
\label{gf_proy}
\end{equation}
By expanding it in powers of $z$, it is easy to see that the coefficient of the power $z^{2M},$
 \begin{eqnarray*}
[z^{2M}]\prod_{k}(z^{k}+z^{-k})^{N_k}\,,
\end{eqnarray*}
is \textit{precisely} the number of different ways to distribute signs among the elements of the multiset  in such a way that the sum of all the elements $k/2$ of the sequence equals $M$.
In particular, if we look for the constant, i.e. $[z^0]$, term in (\ref{gf_proy}) we get the number of solutions for the projection constraint that can be built from a given sequence.

A convenient way to extract this type of information is by using Cauchy's theorem. This allows us to extract these coefficients (for a given $M$) by computing the contour integral
\begin{equation}
[z^{2M}]\prod_{k}(z^{k}+z^{-k})^{N_k}=\frac{1}{2\pi i}\oint_C \frac{\mathrm{d}z}{z^{2M+1}}\prod_k(z^{k}+z^{-k})^{N_k}\,,
\label{integral_cont}
\end{equation}
where $C$ is an index-one curve surrounding the origin. By choosing for it a unit circumference, the previous integral can be written in the following useful alternative form
\begin{eqnarray}
[z^{2M}]\prod_{k}(z^{k}+z^{-k})^{N_k}&=&\frac{2^{N-1}}{\pi}\int_{-\pi}^\pi \cos (2M\theta) \prod_k\cos^{N_k}(k\theta)\,\mathrm{d}\theta
\label{integral_cont2}\\
&=&\frac{2^{N-1}}{\pi}\int_{-\pi}^\pi \cos (2M\theta) \prod_{I=1}^N\cos(k_I\theta)\,\mathrm{d}\theta\,,\nonumber
\end{eqnarray}
where $N=\sum_k N_k$ denotes the number of elements in the sequence $(k_I/2)$. It is important to point out that (\ref{integral_cont2}) and (\ref{DLcounting}) provide the same answer to the counting problem. The advantage of this solution is that it is specially appropriate to find generating functions for the black hole entropy as will be shown later. In Appendix \ref{App:C} we provide a different procedure to solve the same problem based on elegant group theoretical methods [compare equations (\ref{integral_cont2}) and (\ref{dlmdeg})].

\bigskip

\paragraph*{The definition of $m$-degeneracy:} Given a value of the area $a=\sum_{i}q_i\sqrt{p_i}$, we will define the $m$-degeneracy $d_m^{\rm DL}(c)$ of the configuration $c\in \mathcal{C}(a)$ as
\begin{eqnarray*}
d_m^{\rm DL}(c)&:=&\frac{2^{N-1}}{\pi}\int_{-\pi}^\pi  \prod_{i}\prod_{\alpha}\cos^{N_{k^i_\alpha}}(k_\alpha^i \theta)\,\mathrm{d}\theta\\
&=&\frac{2^N}{L}\sum_{\ell=0}^{L-1}\prod_{i}\prod_\alpha \cos^{N_{k^i_\alpha}}(2\pi \ell k_\alpha^i/L)
\end{eqnarray*}
where $N=N(c)=\sum_{i}\sum_\alpha N_{k^i_\alpha}$ and $L=L(c)=1+\sum_{i}\sum_\alpha k_\alpha^iN_{k^i_\alpha}$.

\begin{figure}[htbp]
\includegraphics[width=16.5cm]{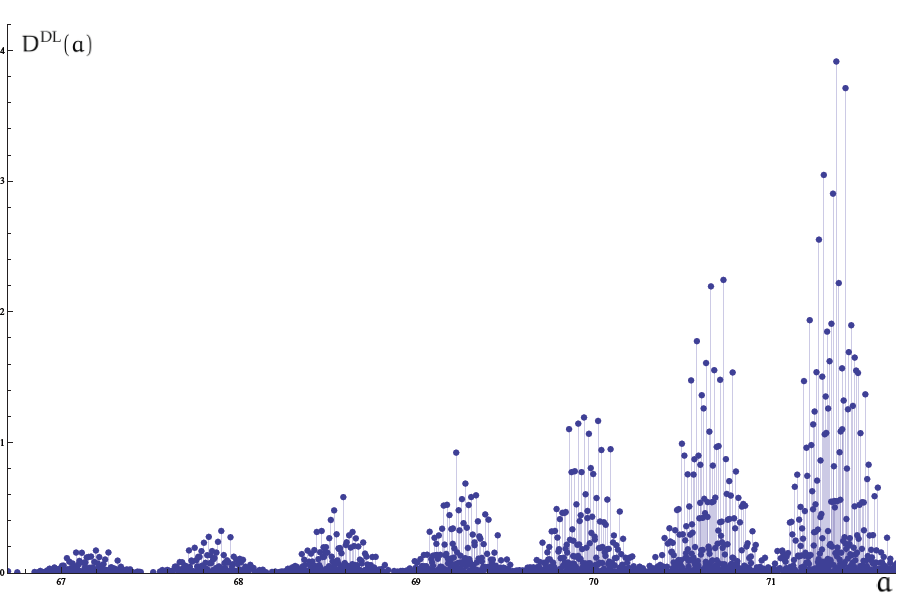}
\caption{Plot of the black hole degeneracy spectrum $D^{\rm DL}$ (in units of $10^{19}$) in terms of the area (in units of $4\pi\gamma\ell^2_P$) for a range of area values. The periodicity can be traced all the way back to the smaller values of the area.} \label{Fig:DDL}
\end{figure}

\paragraph*{Black hole degeneracy spectrum:}
Once we have closed expressions for the $r$ and $m$ degeneracies, we can define the \textit{black hole degeneracy} associated with a given value of the area $a=\sum_{i}q_i\sqrt{p_i}$ as
\begin{equation}
D^{\rm DL}(a)=\sum_{c\in \mathcal{C}(a)} d_r(c) d^{\rm DL}_m(c)\,. \label{DDL}
\end{equation}
We will refer in the following to $D^{\rm DL}(a)$ as the \textit{black hole degeneracy spectrum} for the DL counting. According to the previous discussion, for every value of the area $a$,  $D^{\rm DL}(a)$ gives  the number of sequences of non-zero half-integers satisfying the two conditions
$$
\sum_{I=1}^N m_I=0, \quad \sum_{I=1}^N\sqrt{|m_I|(|m_I|+1)}=\frac{a}{2}.
$$
The procedure described above can be efficiently implemented in a computer to
explicitly obtain the black hole degeneracy in terms of the area. In
order to see the fundamental structure of the degeneracy, we can plot
the black hole degeneracy $D^{DL}(a)$ versus $a$. This is shown in
Fig. \ref{Fig:DDL}. We can see that the result obtained in \cite{Corichi:2006wn}
by a brute force analysis is reproduced. Specifically, we can see
that the number of sequences $D^{DL}(a)$ is distributed forming a
``band structure'' in terms of the area. Peaks of degeneracy appear
in an evenly spaced fashion, interspaced with regions where the degeneracy is several orders of magnitude smaller. This gives rise to an effectively equidistant spectrum \cite{Corichi:2006wn}.

In the process of understanding the structure of the black hole degeneracy spectrum (\ref{DDL}) shown in Fig. \ref{Fig:DDL}, the $m$ and $r$ degeneracies play different roles. In order to disentangle them it will be useful to consider an auxiliary description where the projection constraint is ignored. This means that, once we have identified all the possible sequences $(|m_I|)=(k_I/2)$ of positive half-integers satisfying the area condition, we introduce \textit{signs} in each $m_I$ without any additional restriction. For this auxiliary problem, the $m$-degeneracy of a configuration is given by
$$
d_*^{\rm DL}(c)=\prod_{i}\prod_{\alpha} 2^{N_{k^i_\alpha}}=2^{N(c)}
$$
and the black hole degeneracy becomes
\begin{equation}
D_*^{\rm DL}(a)=\sum_{c\in \mathcal{C}(a)} d_r(c) d_*^{\rm DL}(c)\,.
\label{deg}
\end{equation}
We will sometimes refer to $D_*^{\rm DL}(a)$ as the ``black hole degeneracy without projection constraint''.

\subsection{\label{subsect:adding}Step 4: Adding up}

In order to take into account the inequality appearing in the DL-prescription for the black hole entropy, one can repeat the previous procedure for each of the relevant area eigenvalues $a'\leq a$ and add up the resulting black hole degeneracies $D^{\rm DL}(a')$. For a hypothetical equally spaced area spectrum this task could be easily accomplished by using generating functions (as has been done, for example, in \cite{Sahlmann:2007jt,FernandoBarbero:2009ai}). However, for the general case that we are considering here, one has to resort to more complicated methods based on the use of functional equations or a combination of generating functions and integral transforms \cite{Meissner:2004ju,G.:2008mj}. It should be pointed out that, as the area spectrum is a countable set, it is possible in principle to build the sequence $(D^{\rm DL}(a^{{\scriptscriptstyle \rm LQG}}_n))$ consisting of the values of the black hole degeneracy corresponding to the $n^{th}$ area eigenvalue (ordered in such a way that $a^{{\scriptscriptstyle \rm LQG}}_n<a^{{\scriptscriptstyle \rm LQG}}_{n+1}$), write a generating function for it, and perform the sum by the same method used in \cite{Sahlmann:2007jt,Sahlmann:2007zp}. In practice, however, this is a difficult problem due to the (current) lack of a simple enough algorithm to obtain the value of the $n^{\rm th}$ area eigenvalue $a^{{\scriptscriptstyle \rm LQG}}_n$ in terms of $n$. This is the reason why we are forced to use Laplace transform methods to complete this last step.

In order to fix ideas, let us consider  the sequence $(a^{{\scriptscriptstyle \rm LQG}}_n)$  of  eigenvalues  of the (relevant sector of the) area operator and some number sequence $(\beta_n)$ related to them (for example the black hole degeneracies corresponding to the areas $a^{{\scriptscriptstyle \rm LQG}}_n$). We want to solve the following problem: given an area $a$, compute the sum
$$
\sum_{\{n\,:\, a^{{\scriptscriptstyle \rm LQG}}_n\leq a\}} \beta_n\,.
$$
The solution when $a$ does not belong to the area spectrum is trivially given by
\begin{eqnarray}
\sum_{\{n\,:\, a^{{\scriptscriptstyle \rm LQG}}_n\leq a\}} \beta_n=\int_0^a   \sum_{n \in \mathbb{N}} \beta_n \delta(a'-a^{{\scriptscriptstyle \rm LQG}}_n) \,\mathrm{d}a' \,.\label{suma}
\end{eqnarray}
We will take into account now the following two facts\footnote{Here and in the following $\mathcal{L}(F(a),s)$ denotes the Laplace transform, expressed in
the variable $s$, of the function $F(a)$. Also $\mathcal{L}^{-1}(f(s),a)$ denotes the inverse Laplace transform of the function $f(s)$ in terms of the variable $a$.}
\begin{enumerate}
\item $\mathcal{L}(\delta(a-\alpha),s)=e^{-\alpha s}$, for $\alpha \geq 0$.
\item If $f(s)=\mathcal{L}(F(a), s)$, then the Laplace transform of $\int_0^a F(a')\, da'$ is simply $s^{-1}f(s)$.
\end{enumerate}
These properties allow us to rewrite (\ref{suma}) in the form
\begin{eqnarray}
\sum_{\{n\,:\, a^{{\scriptscriptstyle \rm LQG}}_n\leq a\}} \beta_n=\mathcal{L}^{-1}\left(s^{-1}\sum_{n \in \mathbb{N}} \beta_n \exp(-a^{{\scriptscriptstyle \rm LQG}}_n s) , a \right)\,.\label{a}
\end{eqnarray}
For the cases in which $\beta_n = D^{\rm DL} (a^{{\scriptscriptstyle \rm LQG}}_n)$ or $\beta_n = D^{\rm DL}_* (a^{{\scriptscriptstyle \rm LQG}}_n)$ the sum \begin{eqnarray}
\sum_{n \in \mathbb{N}} \beta_n \exp(-a^{{\scriptscriptstyle \rm LQG}}_n s)=:P(s)\label{P}
\end{eqnarray}
can be conveniently obtained from the generating functions introduced in Appendix \ref{App:A}.
The idea is to take first the generating function given in \cite{BarberoG.:2008ue} (see Appendix \ref{App:A} for details)
\begin{eqnarray}
G^{\rm DL}(z,x_1,x_2,\dots)&=&\left(\displaystyle 1-\sum_{i=1}^\infty\sum_{\alpha=1}^\infty
\Big( z^{k^i_\alpha}+z^{-k_\alpha^i}\Big) x_i^{y^i_\alpha}\right)^{-1}\label{GDL}
\end{eqnarray}
or, in the case of ignoring the projection constraint (by putting $z=1$),
\begin{eqnarray}
G^{\rm DL}_*(x_1,x_2,\dots)&=&G^{\rm DL}(1,x_1,x_2,\dots)=\left(\displaystyle 1-2\sum_{i=1}^\infty\sum_{\alpha=1}^\infty x_i^{y^i_\alpha}\right)^{-1}\label{GDL1}\,,
\end{eqnarray}
and perform the substitutions  $x_1=e^{-s\sqrt{p_1}}$, $x_2=e^{-s\sqrt{p_2}}$, and so on.
It is important to point out that, in order to deal with the projection constraint, we need to introduce an extra variable $z$. This means that, in this case, we will not directly get the function $P(s)$ appearing in (\ref{P}) but rather a function of $P(s,z)$ such that the inverse Laplace transform of $s^{-1}P(s,z)$ in the variable $s$ is a Laurent polynomial in $z$, with area dependent coefficients, whose constant term gives the desired sum. In practice this requires the computation of a contour integral in $z$ or, equivalently, an inverse Fourier transform in the additional variable $\omega$ defined by $z=e^{i\omega}$ \cite{G.:2008mj}. By following this procedure we find
\begin{eqnarray*}
P^{\rm DL}(s,\omega):=G^{\rm DL}(e^{i\omega};e^{-s\sqrt{p_1}},e^{-s\sqrt{p_2}},\dots)&=&\Big(\displaystyle 1-\sum_{i=1}^\infty\sum_{\alpha=1}^\infty (e^{i\omega k^i_\alpha} +e^{-i\omega k^i_\alpha}) e^{-sy^i_\alpha\sqrt{p_i}}\Big)^{-1}\,.
\end{eqnarray*}
The exponentials $e^{-sy^i_\alpha\sqrt{p_i}}$ appearing in this function can be simplified if we use the Pell equations (\ref{Pell}), $y^i_\alpha\sqrt{p_i}=\sqrt{k^i_\alpha(k^i_\alpha+2)}$, to  get
$$
P^{\rm DL}(s,\omega)=\Big(\displaystyle 1-2\sum_{i=1}^\infty\sum_{\alpha=1}^\infty  e^{-s\sqrt{k^i_\alpha(k^i_\alpha+2)}}\cos(\omega k^i_\alpha)\Big)^{-1}\!\!=\!\!\Big(\displaystyle 1-2\sum_{k=1}^\infty  e^{-s\sqrt{k(k+2)}}\cos\omega k\Big)^{-1}\,.
$$
Now, by using (\ref{a}) we obtain the following expression for the entropy as an inverse Laplace transform (and an additional inverse Fourier transform to deal with the projection constraint)
\begin{equation}
\exp S_\leq(a)=\frac{1}{(2\pi)^2 i}\int_{0}^{2\pi} \int_{x_0-i\infty}^{x_0+i\infty}
s^{-1}\Big(\displaystyle 1-2\sum_{k=1}^\infty e^{-s\sqrt{k(k+2)}}\cos \omega k\Big)^{-1} e^{as}\,\mathrm{d}s \,\mathrm{d}\omega\,,\label{numint}
\end{equation}
where $x_0$ is a real number larger than the real part of all the singularities of the integrand in the previous expression. This expression is valid for those area values $a\geq0$ that do not belong to the spectrum of the LQG area operator and, in particular, it gives the exact value of $S_\leq(a_\kappa)$ for the CS prequantized values of the area. For areas of the form $a_n^{{\scriptscriptstyle \rm LQG}}$ the previous formula gives the arithmetic mean of the left and right limits when $a\rightarrow a_n^{{\scriptscriptstyle \rm LQG} \,\pm}$.

\begin{figure}[htbp]
\includegraphics[width=16.5cm]{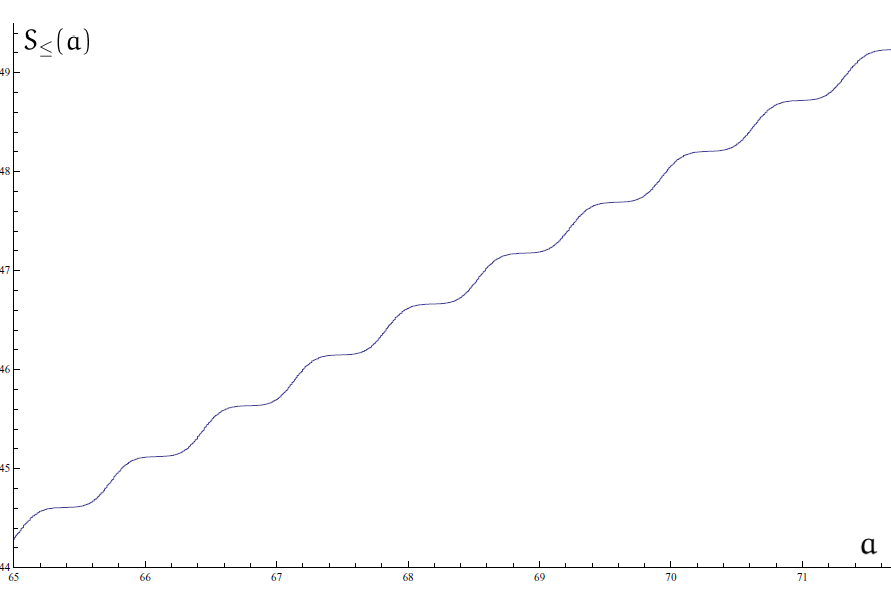}
\caption{Plot of $S_{\leq}(a)$ in terms of the area (in units of $4\pi\gamma\ell^2_P$).
The computation of $S_{\leq}(a)$ has been done by using the algorithm based
in number-theoretical method discussed in the paper. } \label{Fig:SDL}
\end{figure}

The expression (\ref{numint}) can be used to study the asymptotic behavior of the entropy, whose exponential growth  as a function of the area is explained by the presence of the pole in the integrand of (\ref{numint}) with the largest real part \cite{Meissner:2004ju}. This pole determines, in particular, the value of the Immirzi parameter $\gamma=0.237\ldots$ that must be chosen in order to reproduce the Bekenstein-Hawking law. An additional logarithmic correction of the form
$$
-\frac{1}{2}\log(a/\ell^2_P)
$$
can also be derived from (\ref{numint}). When the projection constraint is ignored, the entropy is given by
\begin{equation}
\exp S^*_\leq(a)=\frac{1}{2\pi i} \int_{x_0-i\infty}^{x_0+i\infty}
s^{-1}\Big(\displaystyle 1-2\sum_{k=1}^\infty e^{-s\sqrt{k(k+2)}}\Big)^{-1} e^{as}\,\mathrm{d}s\,, \label{numint1}
\end{equation}
from which it is easy to see that the linear behavior of entropy with area and the value of $\gamma$ are unaltered. However, there are no logarithmic corrections in this case.

It must be pointed out that, from a purely numerical point of view, (\ref{numint}) and (\ref{numint1}) are rather bad because of the inherent difficulties associated with numerically computing improper integrals. Nonetheless, it can be seen that the formula (\ref{numint}) gives the exact result for the lowest values of the area spectrum  by comparing to a direct numerical computation (that is impractical to extend for larger areas). Entropy computations are carried out in practice by using the exact values for the black hole degeneracy spectrum obtained by using the exact combinatorial methods described above and adding up. A sample result for $S_{\leq}(a)$ is shown in Fig. \ref{Fig:SDL}, which displays a characteristic step modulation superimposed to a linear growth. This behavior may be present for large areas as a consequence of the fact that the analytic structure of the integrand in (\ref{numint}) is rather complicated and, in particular, the (real) pole that determines the growth of the entropy and fixes the value of $\gamma$ is an accumulation point for the real parts of the poles in the integrand \cite{G.:2008mj}.

\section{\label{sect:Other}Other approaches}

In the preceding sections we have discussed the computation of the entropy within the framework developed in \cite{Ashtekar:1997yu,Ashtekar:2000eq} and expanded in \cite{Domagala:2004jt,Meissner:2004ju}. These papers have motivated other authors to propose modifications in which some elements differ from the original approach. In particular, Ghosh and Mitra have put forward a modified counting scheme where the set of spin labels of horizon states is expanded to include $j_I$ and $m_I$ labels \cite{Ghosh:2006ph}. There is another series of papers by Kaul and Majumdar that explore the possibility of using a $SU(2)$-CS theory to describe the horizon degrees of freedom \cite{Kaul:1998xv,Kaul:2000kf}. We think that all these proposals have, at least, a heuristic value. Finally, a concrete way to show that the horizon degrees of freedom are indeed accounted for by a $SU(2)$-CS theory is given in \cite{Engle:2009vc,Engle:2010kt}. As these schemes differ in their predictions for $\gamma$ and/or the logarithmic corrections for the entropy, it should be possible in principle to falsify some (or all) of them on physical grounds. Here we take a pragmatic approach motivated by our desire to show that all these models can be rather easily handled by the techniques discussed in the present paper. This shows that our methods are a sharp tool that can be used to derive definite predictions within a broad class of LQG inspired models. The result of the analysis of the different schemes can be summarized by saying that they all lead to the Bekenstein-Hawking law (for different choices of $\gamma$), they predict similar logarithmic corrections (though their coefficients differ among the proposals) and all of them display the interesting substructure found in \cite{Corichi:2006bs,Corichi:2006wn}.

\subsection{\label{subsectGM}The Ghosh-Mitra counting}

We will start by considering the alternative proposal put forward by Ghosh and Mitra (GM) in \cite{Ghosh:2006ph} and showing how it can be handled by the same methods that we have used for the ABCK approach.  We will then analyze the results and compare both models. It is important to highlight at this point that, as far as the computation of the degeneracy spectrum is concerned, both countings can be formulated in very similar terms as done in \cite{Agullo:2008yv}, where they were presented in such a way that the only differences between them could be traced back to the form of the projection constraint.

In section \ref{sect:ABCK} we have seen that the horizon states in the ABCK framework are described by $U(1)$-CS labels that can be identified (by using the quantum boundary condition and introducing an area interval) with lists of spin components ($m_I$-labels) that can be conveniently used to determine the entropy \cite{Domagala:2004jt}. We want to insist at this point on the equivalence of the original ABCK entropy definition and the prescription provided in \cite{Domagala:2004jt} to compute the entropy by employing only $m_I$-labels.
Punctures carrying $m_I$-labels are also characterized by the $SU(2)$ irreducible representation $j_I$ associated with the edges piercing the horizon. These play a role in the ABCK framework because the \textit{compatible} $m_I$-lists that are counted in the DL prescription must be such that they correspond to lists of $j_I$-labels leading to an area eigenvalue in the prescribed area interval.  The GM counting, however, takes the $j$'s as ``independent'' labels. Although the GM approach has not been derived from first principles, according to some authors ``[it] is not an unreasonable proposal'' \cite{Thiemann:2007zz}; in fact,  we will show that it leads to predictions that are quite close to the ones derived by Engle, Noui and Perez in \cite{Engle:2009vc} (and in qualitative agreement with the ABCK approach).

If we follow this approach, the prescription to obtain the entropy can be stated as follows:

\bigskip

\textit{
The entropy $S_\leq^{\rm GM}(a)$ of a quantum horizon of the classical area $a$, according to the GM prescription, is defined by  computing
$$S_\leq^{\rm GM}(a) = \log (1+N_\leq^{\rm GM}(a))\,,$$
where $N_\leq^{\rm GM}(a)$ is the number of all the finite, arbitrarily long, sequences $((j_1,m_1),\ldots,(j_N,m_N))$ of ordered pairs of non-zero half integers $j_I$ and  spin components $m_I\in\left\{-j_I,-j_I+1,\ldots,j_I\right\}$, satisfying
$$2\sum_{I=1}^N\sqrt{j_I(j_I+1)}\leq a\quad\textrm{ and }\quad \sum_I{m_I}=0.$$
}

\noindent Here we want to comment on the type of area interval used in the previous definition. Whereas the mathematical  details of the DL prescription require that the area interval must have the form $[0,a]$, there is no restriction in the GM approach on the form of the interval. By invoking the thermodynamical limit, one would expect that the width of the area interval is irrelevant. As far as the applicability of our method is concerned, it is equally easy to handle an interval of the type $[a-\delta,a+\delta]$, as originally proposed by Ghosh and Mitra, than an interval such as $[0,a]$, because the associated counting can be obtained by subtracting the results for $[0,a+\delta]$ and $[0,a-\delta]$.

\bigskip

The  combinatorial problems associated with the computation of $S_\leq^{\rm GM}(a)$ can be solved by essentially following the same steps as in the previous section. These are now:

\begin{enumerate}
\item[Step 1.] Fix a value for the area $a$ and obtain all the possible choices for the half integers $j_I$ compatible with the area, i.e. satisfying
    $$
    \sum_{I=1}^N\sqrt{j_I(j_I+1)}=\frac{a}{2}.
    $$
    At this point we do not look yet at the ordering of the labels. Notice that this step is \textit{precisely equivalent} to the first step in the DM counting, if one now writes $j_I=k_I/2$ instead of $|m_I|=k_I/2$ for the corresponding labels of each configuration $c\in\mathcal{C}(a)$.
\item[Step 2.] Count the different ways that we can reorder each of the previous multisets.
\item[Step 3.] For each of the configurations corresponding to the first two steps (ordered sequences of non-zero half-integers), we have to find the possible choices for the $m_I$-labels that constitute the second element of the pairs that define the configurations is this case. Notice that one can think of this step as a generalization of the third step in the DL setting. The difference is that there we only had to worry about the signs whereas here each $m_I$ is a spin component taking values in $\left\{-j_I,-j_I+1,\ldots,j_I\right\}$.
\item[Step 4.] Repeat the same procedure for all the area eigenvalues in the interval $[0,a]$ and add the number of sequences $((j_I,m_I))$ obtained for each value of the area.
\end{enumerate}

\bigskip

\noindent \textbf{Step 1: The Pell equation again.}

\bigskip

The first step now is identical to the one described for the DL counting. The only thing that has to be remembered is the different meaning of the $k/2$ labels obtained in the GM case.

\bigskip

\noindent \textbf{Step 2: The reordering degeneracy ($r$-degeneracy)}

\bigskip

This is again the same as before. The number of different reorderings is the one obtained in the previous section. An interesting comment that must be made here is that, as we will show later, the origin of the non-trivial structure in the black hole entropy and the degeneracy spectrum comes from this reordering degeneracy. Hence, we expect to have the same type of qualitative behavior both for the DL and the GM counting.

\bigskip

\noindent \textbf{Step 3: Solving the projection constraint ($m$-degeneracy)}

\bigskip

Up to this point, we have just determined the possible values of $j_I=k_I/2$ in the sequences $((j_I,m_I))$. In this counting procedure ``$m$-degeneracy'' will refer to the number of different ways to assign one of the $2j_I +1$ values of $m_I\in\{-j_I, -j_I+1,\ldots,j_I\}$ to each of the  $j_I$ labels. This assignment must be subject to the restriction given by the projection constraint  $\sum_I m_I=0$. This problem can be solved in several ways. Some of them are analogous to the ones used in the study of the DL counting. However there is an additional method --relying on the use of fusion matrices-- that suggests an intriguing connection with conformal field theory and the $SU(2)$ proposals that we will discuss later in subsection \ref{sect:SU2}.

\bigskip

\paragraph*{The partition problem:}   The first approach to solving this problem is based on the one used in \cite{DeRaedt} for the resolution of the partition problem as described in \ref{sub413}.
Given the sequence $(j_1,\ldots,j_N)$, $j_I\in \mathbb{N}/2$, we want to find the number of sequences $(m_I)$ satisfying  $\sum_{I=1}^N m_I=M$ and $m_I\in\{-j_I,-j_I+1,\ldots,j_I\}$. This number can be computed by introducing, as before, an appropriate Hamiltonian for an auxiliary system of spins. In this case we just have to substitute the multiples of the Pauli matrices $k_I\sigma_I^{(3)}$ by the operator $S_I^{(3)}=1\otimes 1\cdots \otimes 1\otimes s_I^{(3)}\otimes1\otimes \cdots \otimes 1$, where  $s_I^{(3)}:\mathbb{C}^{2j_I+1}\rightarrow \mathbb{C}^{2j_I+1}$ are third-spin component matrices associated with a $j_I$-spin. We consider then the Hamiltonian $H:\mathbb{C}^{\otimes(2j_I+1)}\rightarrow \mathbb{C}^{\otimes(2j_I+1)}$
$$
H=M-\sum_{I=1}^{N}S_I^{(3)}\,.
$$
Proceeding as above, we see that
\begin{eqnarray*}
\frac{1}{L}\sum_{\ell=0}^{L-1} \mathrm{Tr} \exp\left(-4\pi i\ell H/L \right) &=&\frac{1}{L}\left(\prod_{I=1}^N(2j_I+1)+\sum_{\ell=1}^{L-1}e^{-4\pi i \ell M/L}\prod_{I=1}^N\frac{\sin \big(2\pi(2j_I+1)\ell/L\big)}{\sin (2\pi \ell/L)}
\right)\\
 &=&\sum_{\{m_I=-j_I,\ldots,j_I\}}\delta(M,\sum_I m_I)\,,
\end{eqnarray*}
where $L$ is an integer that can be fixed as $L=1+2M+\sum_{I=1}^Nk_I=1+2M+2\sum_I j_I$ when  $2M+2\sum_I j_I$ is even and $L=2+2M+2\sum_I j_I$ when $2M+2\sum_I j_I$ is odd (in order to prevent the vanishing of the denominators). We conclude that the number that we are looking for is
\begin{eqnarray*}
\frac{1}{L}\left(\prod_{I=1}^N(2j_I+1)+\sum_{\ell=1}^{L-1}e^{-4\pi i \ell M/L}\prod_{I=1}^N\frac{\sin \big(2\pi(2j_I+1)\ell/L\big)}{\sin (2\pi \ell/L)}
\right)\,.
\end{eqnarray*}

\bigskip

\paragraph*{Fusion matrices:} The second approach relies on techniques used in the context of conformal field theories \cite{DiFrancesco:1997nk} and, in fact, is suggestive of a deep connection between them and the problem of computing black hole entropy in LQG \cite{Agullo:2009zt}. The key insight now is to realize that the problem of determining the number of solutions to the projection constraint $\sum_Im_I=0$ is equivalent to counting the number of $SU(2)$ irreducible representations, including multiplicities, that appear in the tensor product $\bigotimes_{I=1}^N[j_I]$. In the following we will denote the $SU(2)$ irreducible representation corresponding to spin $j_I$ as $[j_I]=[k_I/2]$.

We start by writing the tensor product of two $SU(2)$ representations as
$$
\left[\frac{k_1}{2}\right]\otimes\left[\frac{k_2}{2}\right]
=\bigoplus_{k_3=0}^\infty\mathcal{N}^{k_3}_{k_1k_2}\left[\frac{k_3}{2}\right]\
,
$$
where the integers $\mathcal{N}^{k_3}_{k_1k_2}$, called
\textit{fusion numbers} \cite{DiFrancesco:1997nk},  give us the number of
times that the representation $[k_3/2]$ appears in the
tensor product of $[k_1/2]$ and $[k_2/2]$. For each $k\in
\mathbb{N}\cup\{0\}$, we introduce then the infinite \textit{fusion
matrices} $(C_k)_{k_1k_2}:=\mathcal{N}^{k_2}_{k_1k}$, where $k_1$,
$k_2\in \mathbb{N}\cup\{0\}$. These satisfy the recursion relation
\begin{equation}
C_{k+2}=X C_{k+1}-C_k,\quad k=0,1,\ldots\label{recurrencia}
\end{equation}
where we use the notation $X:=C_1$. Explicitly,
$X_{k_1k_2}=\delta_{k_1,k_2-1}+\delta_{k_1,k_2+1}$, which shows that
$X$ is a so-called infinite Toeplitz matrix \cite{Bottcher}. The solution to
(\ref{recurrencia}), with initial conditions $C_0=I$ and $C_1=X$,
is
$$
C_k=U_k(X/2),\quad k=0,1,\ldots
$$
where the $U_k$ are the Chebyshev polynomials of the second kind. The
tensor product of any number of representations can be
decomposed as a direct sum of irreducible representations by
multiplying the fusion matrices defined above. This way we get
$$
\left[\frac{k_1}{2}\right]\otimes \left[\frac{k_2}{2}\right]
\otimes\cdots \otimes
\left[\frac{k_N}{2}\right]=\bigoplus_{k=0}^\infty(C_{k_2}C_{k_3}\cdots
C_{k_N})_{k_1 k}\left[\frac{k}{2}\right].
$$
Notice that the product of matrices
$$C_{k_2}C_{k_3}\cdots C_{k_N}=U_{k_2}(X/2)U_{k_3}(X/2)\cdots U_{k_N}(X/2)$$
is a polynomial in $X$.  The total number of
representations giving the solution to the combinatorial
problem that we are trying to solve now is
\begin{eqnarray}
\sum_{k=0}^\infty(C_{k_2}C_{k_3}\cdots C_{k_N})_{k_1 k}=\sum_{k=0}^\infty(U_{k_2}(X/2)U_{k_3}(X/2)\cdots U_{k_N}(X/2))_{k_1 k}\,,
\label{prodC}
\end{eqnarray}
i.e. the sum of the (finite number of non zero) elements in
the $k_1$ row of the infinite matrix $C_{k_2}C_{k_3}\cdots C_{k_N}$. An
integral representation for this sum can be obtained by introducing, as in \cite{Bottcher},
a resolution of the identity for the Toeplitz matrix  $X$  and the
identity $U_k(\cos\theta)=\sin[(k+1)\theta]/\sin\theta$ for the Chebyshev polynomials.
Equation (\ref{prodC}) can be equivalently written as
\begin{equation}
\hspace{-3mm}\sum_{k=0}^\infty(C_{k_2}C_{k_3}\cdots C_{k_N})_{k_1 k}\!=\!\frac{2}{\pi} \int_0^\pi\!\!
\cos\frac{\theta}{2}\left(\cos\frac{\theta}{2}-\cos\big(\sum_{I=1}^N k_I+\frac{3}{2}\big)\theta\right)
\prod_{I=1}^N\frac{\sin(k_I+1)\theta}{\sin\theta} \, \mathrm{d}\theta \,.
\label{integral}
\end{equation}

\bigskip

\paragraph*{Generating functions:} The third way to solve the projection constraint for the GM counting makes use of generating functions. The idea is to take now
$$
\prod_{k}\left(\sum_{n=0}^{k}z^{k-2n}\right)^{N_k}=\prod_{k}\left(\frac{z^{k+1}-z^{-k-1}}{z-z^{-1}}\right)^{N_k}\,.
$$
This function is similar to the one corresponding to the DL counting (\ref{gf_proy}). The number of solutions to the projection constraint $\sum_Im_I=M$ is given now by the coefficient of $z^{2M}$ in the previous expression and can be extracted as before by using Cauchy's theorem. The relevant coefficients for a given $2M$ are given now by
\begin{equation}
[z^{2M}]\prod_{k}\left(\frac{z^{k+1}-z^{-k-1}}{z-z^{-1}}\right)^{N_k}=\frac{1}{2\pi i}\oint_C \frac{\mathrm{d}z}{z^{2M+1}}\prod_{k}\left(\frac{z^{k+1}-z^{-k-1}}{z-z^{-1}}\right)^{N_k}
\label{integral_cont_GM}
\end{equation}
where $C$ is, again, an index-one curve surrounding the origin. By choosing for it a unit circumference, the previous integral can be written in the following useful alternative form
\begin{equation}
[z^{2M}]\prod_{k}\left(\frac{z^{k+1}-z^{-k-1}}{z-z^{-1}}\right)^{N_k}=\frac{1}{2\pi}\int_{-\pi}^\pi \cos (2M \theta) \prod_{k}\frac{\sin^{N_k}(k+1)\theta}{\sin^{N_k}\theta}\, \mathrm{d}\theta\,.
\label{integral_cont2_GM}
\end{equation}
In Appendix \ref{App:C} we give yet another procedure to solve the projection constraint based on group theoretical methods [see equation (\ref{mcaracteres}) and compare equations (\ref{integral_cont2_GM}) and (\ref{mdeg1})].

\bigskip

\noindent \textbf{GM black hole degeneracy spectrum.}

\bigskip

As we did in the DL case, given an area $a=\sum_{i}q_i\sqrt{p_i}$ we can define the $m$-degeneracy $d_m^{\rm GM}(c)$ of the configuration\footnote{The definition of the set of configurations $\mathcal{C}(a)$ here is the same as in the DL scheme and relies on the solutions to the Pell equations as before.} $c\in \mathcal{C}(a)$ as
\begin{eqnarray*}
d_m^{\rm GM}(c)&:=& \frac{1}{L} \left( \prod_{i}\prod_\alpha(k^i_\alpha+1)^{N_{k^i_\alpha}}
+\sum_{\ell=1}^{L-1}\prod_{i}\prod_\alpha\frac{\sin^{N_{k^i_\alpha}} \big(2\pi(k^i_\alpha+1)\ell/L\big)}{\sin^{N_{k^i_\alpha}} (2\pi \ell/L)}\right)
\\&=&
\frac{2}{\pi} \int_0^\pi
\cos\frac{\theta}{2}\left(\cos\frac{\theta}{2}-\cos\big(\sum_{i}\sum_\alpha k_\alpha^i N_{k_\alpha^i}+\frac{3}{2}\big)\theta\right)
\prod_{i}\prod_\alpha \frac{\sin^{N_{k_\alpha^i}}(k_\alpha^i+1)\theta}{\sin^{N_{k_\alpha^i}}\theta} \, \mathrm{d}\theta
\\
&=&\frac{1}{2\pi}\int_{-\pi}^\pi  \prod_{i}\prod_\alpha \frac{\sin^{N_{k^i_\alpha}}(k^i_\alpha+1)\theta}{\sin^{N_{k^i_\alpha}}\theta}\, \mathrm{d}\theta\,,
\end{eqnarray*}
where the integer $L=L(c)$ has been defined above.

We will also define the \textit{black hole degeneracy} associated with a given value of the area $a=\sum_{i}q_i\sqrt{p_i}$ as
$$
D^{\rm GM}(a)=\sum_{c\in \mathcal{C}(a)} d_r(c) d^{\rm GM}_m(c)\,,
$$
where $d_r(c)$ was given in (\ref{rdeg}). In the following we will refer to $D^{\rm GM}(a)$ as the \textit{black hole degeneracy spectrum} for the GM counting. According to the previous discussion, for every value of the area $a$,  it gives  the number of sequences $((j_I,m_I))$,  where $j_I\in N/2$ and $m_I\in \mathbb{Z}/2$, satisfying the conditions
$$
2\sum_{I=1}^N\sqrt{j_I(j_I+1)}=a\,,\quad m_I\in\{j_I-n \,:\, n=0,\ldots,2j_I \}\,,\quad \sum_{I=1}^N m_I=0\,.
$$

\bigskip

\noindent \textbf{Step 4: Adding up.}

\bigskip

In the previous paragraphs we have given a procedure to compute the black hole degeneracy spectrum for the GM counting. As in the DL counting, we have to add the degeneracies corresponding to all the possible values of the area spectrum in a certain interval and, again, use integral transform techniques to deal with the difficult combinatorial problem of summing for all the relevant values of the area. To this end we proceed as in subsection \ref{subsect:adding}. Let us consider the generating function for the black hole degeneracy $D^{{\rm GM}}(a)$ (see Appendix \ref{App:A})
\begin{equation}
G^{\rm GM}(z,x_1,x_2,\dots)=\left(\displaystyle 1-\sum_{i=1}^\infty\sum_{m=1}^\infty \Big( \frac{z^{k^i_m+1}-z^{-k^i_m-1}}{z-z^{-1}}\Big) x_i^{y^i_m}\right)^{-1}\,.\label{GGM}
\end{equation}
and perform the substitutions $z=e^{i\omega}$, $x_1=e^{-s\sqrt{p_1}}$, $x_2=e^{-s\sqrt{p_2}}\,,\ldots,$ to get
\begin{eqnarray}
\exp S^{\rm GM}_\leq (a) =\frac{1}{(2\pi)^2 i} \int_0^{2\pi}\int_{x_0-i\infty}^{x_0+i\infty}s^{-1}\Bigg(\displaystyle 1-\sum_{k=1}^\infty \frac{\sin (k+1)\omega}{\sin \omega} e^{-s\sqrt{k(k+2)}}\Bigg)^{-1} e^{as}\,\mathrm{d}s \,\mathrm{d}\omega\,.\hspace{5mm}\label{numintGM}
\end{eqnarray}
This expression corresponds to an area interval of the form $[0,a]$ in the definition of the entropy. As explained before, the values for an area interval $[a-\delta,a+\delta]$ can be easily obtained from it. The behavior of $S^{\rm GM}_\leq (a)$ as a function of area is the same as for the DL counting, namely a linear dependence  and a logarithmic correction with $-1/2$ coefficient. The Bekenstein-Hawking entropy-area relation is recovered now by taking  $\gamma=0.274\ldots$.

As in the case of the DL counting, it is possible to define a simplified combinatorial problem by ignoring the projection constraint and considering all the $2j+1$ choices of $m$-labels \cite{Meissner:2004ju,Ghosh:2006ph}. This can be easily done just by replacing $\sin (k+1)\omega/\sin \omega$ by $k+1$ in equation (\ref{numintGM}).

\subsection{\label{sect:SU2}$SU(2)$ black hole entropy}

The initial proposals of Smolin and Krasnov \cite{Smolin:1995vq,Krasnov:1996wc} to study black hole entropy in LQG (predating the more detailed ABCK approach) suggest that the horizon degrees of freedom could be accounted for by a $SU(2)$-CS theory. One of the non-trivial points of the ABCK treatment is the claim that the horizon degrees of freedom are, in fact, described by a $U(1)$-CS model. This has been justified in this framework by a combination of ideas involving the analysis of gauge transformations on the isolated horizon and Hamiltonian methods. However, the definition of an isolated horizon does not require the introduction of any internal symmetry so, on the face of it, no restriction on the internal symmetry on the horizon is expected when connection variables are used \cite{Engle:2009vc}. The subtleties associated with this issue have led some authors to explore the possible consequences of ascribing the black hole degrees of freedom (and, hence, the entropy) to a $SU(2)$-CS theory. In particular, Kaul and Majumdar \cite{Kaul:1998xv,Kaul:2000kf} have argued that the Bekenstein-Hawking area law can be explained in a such scheme and have also proposed logarithmic corrections to the black hole entropy. The entropy computations carried out in \cite{Kaul:1998xv,Kaul:2000kf} rely on well-established conformal field theory methods \cite{DiFrancesco:1997nk} and, hence, will not be reviewed here. The main problem with this approach is that the $SU(2)$-CS model on the horizon is really put in by hand and not derived from the quantization of the relevant sector of general relativity. Although it is definitely interesting to explore alternative approaches it stands to reason that any model aspiring to provide a fundamental physical description of black holes should be solidly rooted in theory.

Other authors have attempted to see if a $SU(2)$-CS description can be derived from first principles within LQG. In particular Engle, Noui and Perez (ENP) have proposed a covariant Hamiltonian scheme leading to a pure $SU(2)$ formulation \cite{Engle:2009vc} (see also \cite{Engle:2010kt}). These results are not compatible with the ABCK model but lead to a remarkable entropy definition in the sense that it gives the same Immirzi parameter predicted by Ghosh and Mitra and the same logarithmic correction proposed by Kaul and Majumdar.

The entropy in the ENP model is obtained according to the following prescription:
\bigskip

\textit{
The entropy $S_\leq^{\rm ENP}(a)$ of a quantum horizon of the classical area $a_\kappa=4\pi\gamma\ell^2_P\kappa$ (when $\gamma\leq\sqrt{3}$) is defined as
$$S^{\rm ENP}_\leq(a) = \log (1+N^{\rm ENP}_\leq(a))\,,$$
where $N^{\rm ENP}_\leq(a)$ is the number of all the finite, arbitrarily long, sequences $(j_1,\ldots,j_N)$ of non-zero half integers $j_I$ satisfying
$$8\pi\gamma\ell^2_P\sum_{I=1}^N\sqrt{j_I(j_I+1)}\leq a_\kappa$$
and counted with a multiplicity given by the dimension of the invariant subspace $\mathrm{Inv}(\otimes_I[j_I])$.}

\bigskip

The last condition in the ENP entropy prescription plays the role of the projection constraint characteristic of the other approaches. Actually, if instead of using $\mathrm{Inv}(\otimes_I[j_I])$ each sequence is counted with a multiplicity given by the number of irreducible representations –-taking into account multiplicities–- that appear in the direct sum decomposition of the tensor product $\otimes_I [j_I]$ we recover \textit{precisely} the GM counting. Hence, the number of configurations accounting for the entropy for a given area in the ENP approach is slightly lower than the one in the GM counting. It only changes the logarithmic correction (see below).

We discuss now in an schematic way how the ENP entropy is computed in our scheme.
\begin{enumerate}
\item[Step 1.] Fix a value for the area $a$ and obtain all the possible choices (without ordering)  for the half integers $j_I$ compatible with the area, i.e. satisfying
    $$
    \sum_{I=1}^N\sqrt{j_I(j_I+1)}= \frac{a}{2}.
    $$
\item[Step 2.] Count the different ways to reorder each of the previous multisets.
\item[Step 3.] Determine the dimension of the invariant subspace $\mathrm{Inv}(\otimes_I[j_I])$ associated with each sequence $(j_I)$.
\item[Step 4.] Repeat the same procedure for all the area eigenvalues smaller than $a$ and add the numbers obtained in each case.
\end{enumerate}

\begin{figure}[htbp]
\begin{center}
\includegraphics[width=16.5cm]{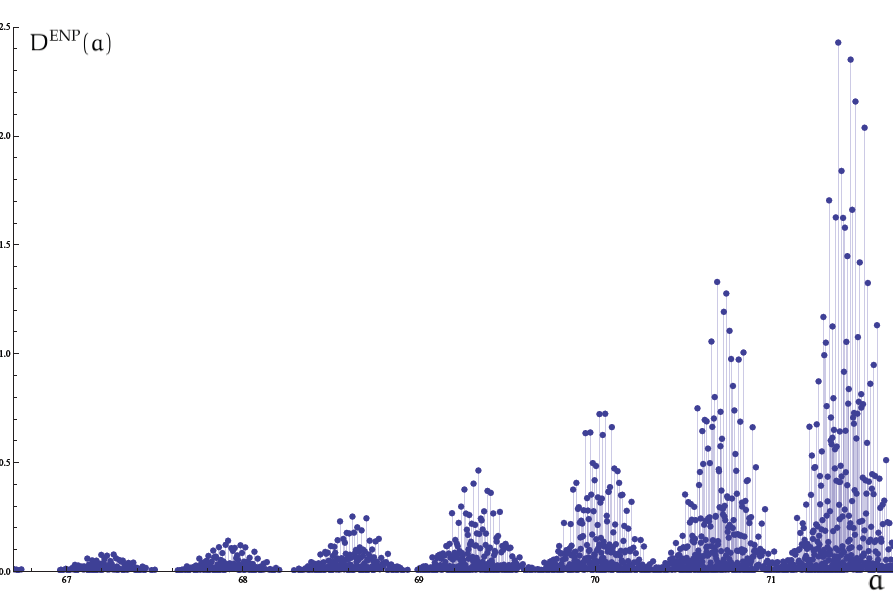}
\caption{Plot of the black hole degeneracy spectrum $D^{\rm ENP}$ (in units of $10^{21}$) in terms of the area (in units of $4\pi\gamma\ell^2_P$) for a range of area values. } \label{Fig:DENP}
\end{center}
\end{figure}

\begin{figure}[htbp]
\begin{center}
\includegraphics[width=16.5cm]{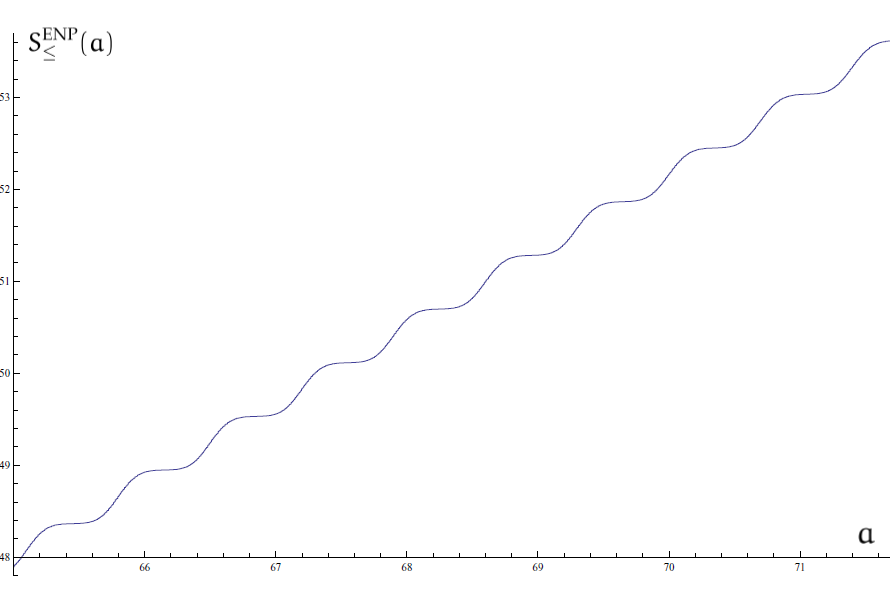}
\caption{Plot of $S^{\rm ENP}_{\leq}(a)$ in terms of the area (in units of $4\pi\gamma\ell^2_P$). This plot is essentially the same for $S^{\rm GM}_{\leq}(a)$.}
\label{Fig:SENP}
\end{center}
\end{figure}

The only difference with the GM case is in step 3. In practice \cite{Agullo:2009eq}, this means that it suffices to replace the $m$-degeneracy $d_m^{\rm GM}(c)$  by (see equation (\ref{enp_c}) in Appendix \ref{App:C})
\begin{eqnarray*}
d^{\rm ENP}_m(c)&=&-\frac{1}{2\pi i}\oint_C\frac{\mathrm{d}z}{z}\frac{(z-z^{-1})^2}{2}\prod_{i}\prod_\alpha \frac{(z^{k^i_\alpha+1}-z^{-k^i_\alpha-1})^{N_{k^i_\alpha}}}{(z-z^{-1})^{N_{k^i_\alpha}}}\\&=&
\frac{1}{\pi}\int_0^{2\pi}  \sin^2\theta \prod_{i}\prod_\alpha \frac{\sin^{N_{k^i_\alpha}}(k^i_\alpha+1)\theta}{\sin^{N_{k^i_\alpha}}\theta}\, \mathrm{d}\theta\,.
\end{eqnarray*}
We will also define the \textit{black hole degeneracy} associated with a given value of the area $a=\sum_{i}q_i\sqrt{p_i}$ as
$$
D^{\rm ENP}(a)=\sum_{c\in \mathcal{C}(a)} d_r(c) d^{\rm ENP}_m(c)\,,
$$
where $d_r(c)$ is given by (\ref{rdeg}). We will refer to $D^{\rm ENP}(a)$ as the \textit{black hole degeneracy spectrum} as before. The entropy can be computed by summing up black hole degeneracies. In practice, this can be done by considering the generating function  \cite{Agullo:2009eq}
\begin{equation}
G^{\rm ENP}(z,x_1,x_2,\dots)=-\frac{(z-z^{-1})^2}{2}\left(\displaystyle 1-\sum_{i=1}^\infty\sum_{m=1}^\infty \Big( \frac{z^{k^i_m+1}-z^{-k^i_m-1}}{z-z^{-1}}\Big) x_i^{y^i_m}\right)^{-1}\label{GGM}
\end{equation}
and performing  the substitutions $z=e^{i\omega}$, $x_1=e^{-s\sqrt{p_1}}$, $x_2=e^{-s\sqrt{p_2}},\ldots,$  to get
\begin{eqnarray*}
\exp S^{\rm ENP}_\leq(a) =\frac{2}{(2\pi)^2 i} \int_0^{2\pi}\int_{x_0-i\infty}^{x_0+i\infty}s^{-1}\sin^2\omega \, \Big(\displaystyle 1-\sum_{k=1}^\infty \frac{\sin (k+1)\omega}{\sin \omega} e^{-s\sqrt{k(k+2)}}\Big)^{-1} e^{as}\,\mathrm{d}s \,\mathrm{d}\omega\,.\label{numintSU2}
\end{eqnarray*}
This expression leads us to fix $\gamma=0.274\ldots$ and gives the logarithmic correction
$$
-\frac{3}{2}\log(a/\ell^2_P)\,.
$$
In Figs. \ref{Fig:DENP} and \ref{Fig:SENP} we plot $D^{\rm ENP}(a)$ and $S^{\rm ENP}_\leq(a)$ as functions of the area. Notice that these figures are similar to Figs. \ref{Fig:DDL} and \ref{Fig:SDL}.

\section{\label{sect:Summary}Summary of the different schemes}

The behavior of the entropy as a function of the area, obtained by using the different countings considered in the previous sections, is compatible with the Bekenstein-Hawking law for appropriate choices of the Immirzi parameter. The qualitative behavior for the different models is shown in Figs.  \ref{Fig:SDL} and \ref{Fig:SENP} (notice that the GM and ENP countings differ only in the logarithmic corrections and, hence, the plots of the entropy in these two cases are essentially indistinguishable). These figures show that the entropy grows in steps of a characteristic width $\Delta a$ that appears to be the same for all the countings (when measured in units of $4\pi\gamma \ell^2_P$), i.e.
$$
\frac{\chi}{4 \pi}:=\frac{(\Delta a)_{\rm DL}}{4\pi \gamma_{\rm DL}\ell_P^2}=\frac{(\Delta a)_{\rm GM}}{4\pi \gamma_{\rm GM}\ell_P^2}=\frac{(\Delta a)_{\rm ENP}}{4\pi \gamma_{\rm ENP}\ell_P^2}= \frac{8.78\cdots}{4\pi} = 0.69\cdots
$$
There are also subdominant logarithmic corrections that differ from model to model. These results are summarized in the following table.

\bigskip

\begin{center}
\begin{tabular}{|c|c|c|c|}
  \hline
  \phantom{...}\textbf{Approach}\phantom{...} \T\B& $ \gamma$ \T\B&\phantom{...}\textbf{ Logarithmic correction}\phantom{...}  \T\B&\phantom{......} $\Delta a$ \phantom{......} \T\B\\  \hline  \hline
ABCK-DL & $\,\,\gamma_{\rm DL}=0.237\cdots$ & $-1/2\log(a/\ell_P^2)$ & $\chi \gamma_{\rm DL}\ell_P^2$\\  \hline
  GM &  $\,\,\gamma_{\rm GM}=0.274\cdots$ & $-1/2\log(a/\ell_P^2)$ &  $\chi\gamma_{\rm GM}\ell_P^2$\\  \hline
  ENP &  $\gamma_{\rm ENP}=\gamma_{\rm GM}$  \phantom{....}& $-3/2\log(a/\ell_P^2)$ & $\chi\gamma_{\rm ENP}\ell_P^2$\\  \hline
  \hline
\end{tabular}
\end{center}

\bigskip

A comment regarding $\Delta a$ is in order now. In the previous sections, we have measured areas in units of $4\pi\gamma\ell^2_P$. This choice of units is, in principle, model-dependent because the value of $\gamma$ must be chosen differently in order to arrive at the Bekenstein-Hawking law. This means that area steps in the entropy have different (but related) sizes for the different countings and hence the actual value of $\gamma$ leaves a characteristic signature in the behavior of the entropy.  A final remark regarding $\chi$ is the conjecture appearing in \cite{Agullo:2008eg} stating that $\chi= 8\log 3$.

\section{\label{sect:Detailed}Detailed analysis of the entropy microstructure}

The main goal of this section is to use the techniques presented in
the preceding parts of the paper to obtain a detailed understanding of the behavior of the entropy. The ``modular'' nature of the number-theoretic approach that we have followed will allow us to ascribe some of the interesting behaviors that we want to understand (in particular the staircase growth of the entropy) to some of the specific steps that we have followed. This analysis is valuable on its own but can provide, in addition, important information to tackle the difficult problem of fully understanding the asymptotic behavior of the entropy in the large area limit.

We are going to separate the analysis in different parts. On the one hand, we are going to study the relative importance of each of the possible values of the $k$-labels in the different configurations. On the other hand, we are going to analyze the contribution of each type of degeneracy, namely the $r$-degeneracy and  the $m$-degeneracy, and their influence in the observed patterns.

From a practical point of view it is better to focus on the behavior of the degeneracy spectrum $D(a)$ instead of the entropy. In the following we will concentrate on the DL-counting because the arguments can be copied for all the other models that we have discussed in the paper.

\subsection{\label{subsect:disentangling}Disentangling the area spectrum with the help of squarefree integers.}

The methods developed in section \ref{sect:ABCK} will allow us to analyze in detail the structure of the black hole degeneracy spectrum and the entropy. The area spectrum is given by certain integer linear combinations of square roots of squarefree numbers $a=\sum_i q_i \sqrt{p_i}$. The solutions to the diophantine equations described above (subsection \ref{subPell}) tell us how to identify the configurations $\mathcal{C}(a)$ associated with each value of the area. One of the important features of our formalism is the central role of the squarefree numbers in the classification of the points in the area spectrum. In many instances, it is much more convenient to use them instead of spin labels. Of course, once a squarefree is fixed, there is an infinite number of possible spin labels given by the solutions to the Pell equation (\ref{Pell}). The squarefree numbers and the associated spin labels $k/2$ are given in table \ref{tablasquarefree}. The $i$ label appearing in the first column identifies the elements of the sequence of squarefree numbers listed in increasing order ($p_1=2$, $p_2=3$, $p_3=5,\ldots$). The squarefree numbers themselves are shown in the second column. They are listed according to their order of appearance in the successive eigenvalues of the area spectrum. The spin labels in the column $k^i_1$ are associated with the fundamental solution of the corresponding Pell equation. The spins in this column grow in units of $1/2$. This pattern is only interrupted by the appearance of ``secondary'' solutions in $k^i_2,k^i_3,\ldots$ (for example $k/2=3$ appears as the second solution to the Pell equation for $p_2=3$, i.e. in the $k^i_2$ column).
\begin{center}
\begin{table}[h!b!p!]
\caption{Pell equation: area spectrum and squarefrees}
\begin{tabular}{|c|c||c|c|c|c||c|}
  \hline
 $i$ \T\B& $\sqrt{p_i}$ \T\B& \phantom{...}$k^i_1/2$\phantom{...} \T\B&\phantom{..} $k^i_2/2$ \phantom{..}\T\B& $k^i_3/2$ \T\B& \phantom{...} $\cdots$\phantom{...}\T\B& $s_i$\T\B\\\hline\hline
  2  & $\sqrt{3}$ & 1/2 & 3 & 25/2 & $\cdots$ &0.4526794\\\hline
  1   &  $\sqrt{2}$ & 1 & 8 & 49& $\cdots$ &0.2545910\\\hline
10  &  $\sqrt{15}$ & 3/2 & 15 & 243/2 & $\cdots$&0.1808782\\\hline
4   &  $\sqrt{6}$  & 2 & 24 & 242& $\cdots$ &0.1418796\\\hline
22  &  $\sqrt{35}$ & 5/2 & 35 &  845/2& $\cdots$ &0.1172453\\\hline
5 &   $\sqrt{7}$  & 7/2 & 63 & 2023/2 & $\cdots$&0.0873322\\\hline
3  &  $\sqrt{5}$  & 4 & 80 & 1444 & $\cdots$&0.0774971\\\hline
7  &  $\sqrt{11}$  & 9/2 & 99 & 3969/2& $\cdots$ &0.0696641\\\hline
18  &  $\sqrt{30}$  & 5 & 120 & 2645& $\cdots$ &0.0632754\\\hline
88 &   $\sqrt{143}$  & 11/2 & 143 & 6875/2& $\cdots$ &0.0579639\\\hline
27 & $\sqrt{42}$   & 6 & 168 & 4374 & $\cdots$ &0.0534775\\\hline
119 &   $\sqrt{195}$  & 13/2 & 195 & 10933/2 & $\cdots$ &0.0496373\\\hline
9 &  $\sqrt{14}$  & 7 & 224 & 6727& $\cdots$ &0.0463128\\\hline
156  & $\sqrt{255}$ &  15/2  & 255 & 16335/2& $\cdots$ &0.0434066\\\hline
198 &   $\sqrt{323}$ & 17/2 & 323 &  23273/2 & $\cdots$&0.0385677\\\hline
$\vdots$ &   $\vdots$ & $\vdots$ & $\vdots$ &  $\vdots$ & $\ddots$&$\vdots$\\\hline
  \hline
\end{tabular}
 \label{tablasquarefree}
\end{table}
\end{center}

As we know form previous analyses \cite{Ghosh:2006ph,Domagala:2004jt}, the smaller
the $k$-label is, the higher its contribution to the most degenerate
states. This is the reason why for many practical purposes we can concentrate on the study of the lowest squarefree numbers. For example, if we want to consider only spin labels up to $5/2$, we only have to take the squarefree numbers $3, 2, 15, 6,$ and $35$.  The fact that the very first value of $k$, (corresponding to $\sqrt{3}$) is $1/2$ justifies the well known observation that punctures with spin labels equal to $1/2$ give the most significant contribution to the entropy.

\begin{figure}[htbp1]
\vspace*{-12mm}\includegraphics[width=16.3cm]{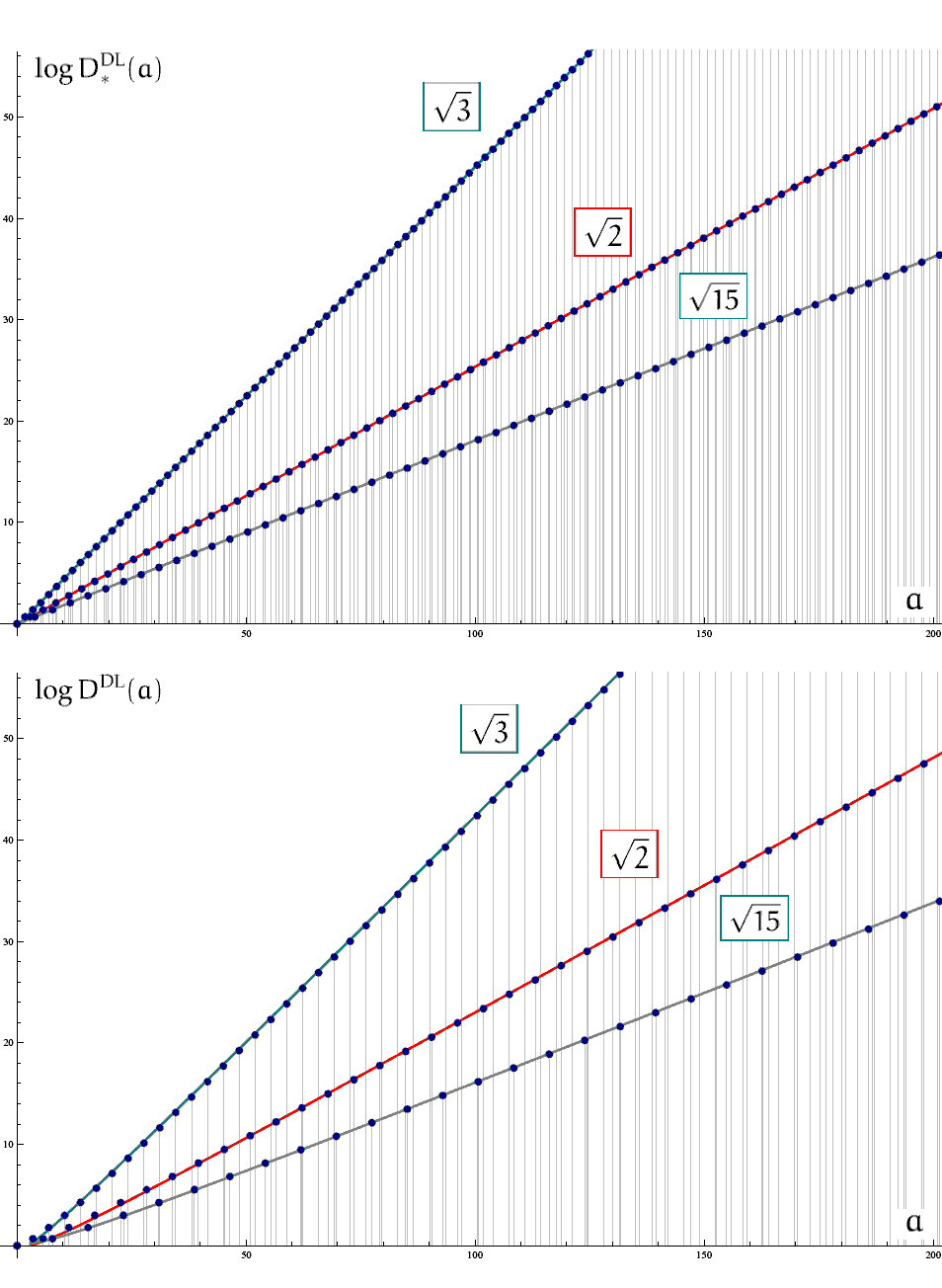}
\vspace*{-5mm}\caption{Plot of  $\log D_*^{\rm DL}(a)$ and $\log D^{\rm DL}(a)$ in terms of the area (in units of $4\pi\gamma\ell^2_P$). The figure shows the values associated with area eigenvalues multiples of $\sqrt{3}$, $\sqrt{2}$ and $\sqrt{15}$ respectively. The solid lines correspond to the asymptotic approximations given by straight lines with the slopes $s_2$, $s_1$ and $s_{10}$ appearing in table \ref{tablasquarefree}. When the projection constraint is taken into account (lower plot) the asymptotic approximations have also a logarithmic correction $-(\log a)/2$.}
\label{Fig:onerootLP}
\end{figure}

Subfamilies in the area spectrum consisting of area eigenvalues that are multiples of the square root of a \textit{single} square free number $\sqrt{p_i}$ are specially easy to analyze. Figure  \ref{Fig:onerootLP} shows the behavior of $\log D_*^{\rm DL}(a)$ and $\log D^{\rm DL}(a)$ in three different cases given by the subfamilies of the area spectrum that can be written as integer multiples of $\sqrt{3}$, $\sqrt{2}$ and  $\sqrt{15}$. As we can see each of these subfamilies is characterized by a different growth rate. The logarithm of the black hole degeneracy spectrum grows linearly with the area with a slope $s_i$ that can be determined numerically for each squarefree number $p_i$ by finding the unique real solution to the equation \cite{G.:2008mj}
\begin{eqnarray}
1-2\sum_{\alpha=1}^\infty e^{-s_i\sqrt{k^i_\alpha(k^i_\alpha+2)}}=0
\end{eqnarray}
in the unknown $s_i$.
The corresponding values are given in the last column of table \ref{tablasquarefree}. As we can see they decrease monotonically for each successive $p_i$. In addition to the linear growth there is a logarithmic correction of the form $-(\log a)/2$ in the physically relevant case for which the projection constraint is taken into account. It is important to point out here that the actual growth rate of each family can only be determined once the value of the Immirzi parameter $\gamma$ has been fixed by enforcing the Bekenstein-Hawking area law.

The asymptotic approximation for the subfamilies considered above can be easily obtained from the generating functions discussed in Appendix \ref{App:A}. We will describe next how this can be done for more general subfamilies of points in the area spectrum consisting of integer linear combinations of a fixed number of square roots of squarefree numbers $p_{i_1}$,\ldots, $p_{i_n}$ labeled by the subset $\mathcal{I}=\{i_1,\ldots,i_n\}\subset \mathbb{N}$. The idea is to restrict the full generating function (\ref{GDL}) by eliminating all the variables with the exception of a finite number of them, $(x_{i_1},\ldots,x_{i_n})$, associated with the square free numbers defining the subfamily. This way we get
$$G^{\rm DL}_\mathcal{I}(z,x_{i_1},\ldots,x_{i_n})=\Big(1-\sum_{i\in \mathcal{I}}\sum_{\alpha=1}^\infty (z^{k^i_\alpha}+z^{k^i_\alpha}) x_i^{y^i_\alpha}\Big)^{-1}\,.$$
By using the Laplace transform methods described in \cite{G.:2008mj} we can easily find the asymptotics corresponding to the subfamilies by performing the substitutions $x_i=e^{-s\sqrt{p_i}}$, $z=1$,  and solving the equation
\begin{eqnarray}
1-2\sum_{i\in \mathcal{I}}\sum_{\alpha=1}^\infty e^{-s_\mathcal{I}\sqrt{k^i_\alpha(k^i_\alpha+2)}}=0\label{polo}
\end{eqnarray}
that gives the real pole $s_\mathcal{I}$. This directly provides the slope of the asymptotic linear growth of the degeneracy spectrum associated with the chosen subfamily. The functions of the type
$$
Q_\mathcal{I}(s)= 1-2\sum_{i\in \mathcal{I}}\sum_{\alpha=1}^\infty e^{-s\sqrt{k^i_\alpha(k^i_\alpha+2)}}
$$
have some distinctive properties that justify the previous procedure:
\begin{itemize}
\item They have a single real zero $s_\mathcal{I}$.
\item The remaining complex zeros have real parts that are smaller or equal than $s_\mathcal{I}$.
\item The real parts of these zeros accumulate to the value $s_\mathcal{I}$.
\end{itemize}
These properties explain why the real solution to  equation (\ref{polo}) captures the effective asymptotic behavior of the black hole degeneracy for the chosen subfamily, and leaves room for interesting (i.e. periodic) substructures. Several comments are in order now:
\begin{itemize}
\item The derived growth applies to the ``convex'' envelope associated with the chosen subfamily.
\item There may be additional substructures (in the form of bands for the degeneracy spectrum or steps for the full entropy).
\item There may be subdominant (i.e. logarithmic) terms in the asymptotic expansions. To find these the $z$ variable must be reintroduced.
\item For a given subfamily there are sub-subfamilies (chosen by further restricting the number of squarefrees) with different growth rates.
\end{itemize}

As we have mentioned above, the behavior of the black hole degeneracy spectrum for area eigenvalues that can be written as integer multiples of a single squarefree is shown in Fig. \ref{Fig:onerootLP}. The expected linear growth (with the corresponding logarithmic corrections when the projection constraint is included) can be readily seen.

\begin{figure}[htbp2]
\vspace*{-15mm}\includegraphics[width=16.3cm]{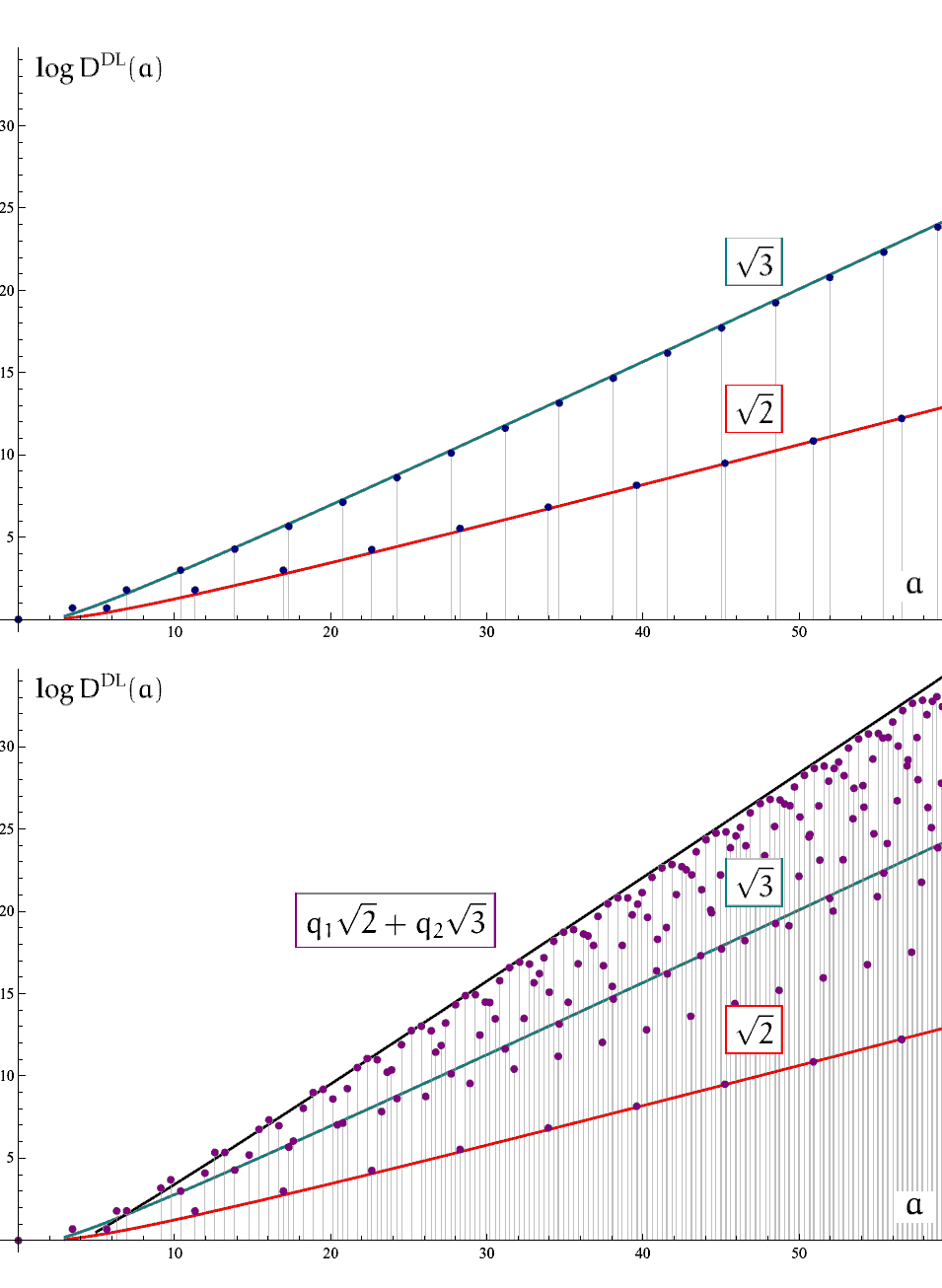}
\vspace*{-5mm}\caption{Plot of $\log D^{\rm DL}(a)$ for area eigenvalues of the form $q_1\sqrt{2}+q_2\sqrt{2}$ with $q_1,q_2\in\mathbb{N}\cup \{0\}$. The solid line with the largest slope $s_{\{1,2\}}=0.645008\cdots$ (determined by equation (\ref{polo}) for $\mathcal{I}=\{1,2\}$) gives the asymptotic approximation for their growth. The other solid lines correspond to the $\sqrt{2}$ and $\sqrt{3}$ subfamilies. As it can be seen the ``envelope'' of the points grows faster than these subfamilies. In all the cases the logarithmic correction $-(\log a)/2$ has been included.}
\label{Fig:2rootLP}
\end{figure}

Figure \ref{Fig:2rootLP} shows the  degeneracy spectrum for the subset of area eigenvalues consisting of integer linear combinations of $\sqrt{2}$ and $\sqrt{3}$ when the projection constraint is included. There are several interesting features that can be seen in this case:
\begin{itemize}
\item The set of points represented in  Fig. \ref{Fig:2rootLP} corresponds to a subset of the area eigenvalues consisting in the union of subfamilies $\mathcal{A}_i(q^0)$ of the form
$$\mathcal{A}_1(q_2^0)=\{ q_1\sqrt{2}+ q_2^0\sqrt{3}\,:\, q_2\in \mathbb{N}\cup \{0\}\}\,,\quad \mathcal{A}_2(q_1^0)=\{q_1^0\sqrt{2}+ q_2\sqrt{3}\,:\, q_2\in \mathbb{N}\cup \{0\}\}$$
where $q_i^0$ are fixed non-negative integers.
\item The asymptotic growth of the degeneracy spectrum within each family $\mathcal{A}_i(q^0)$ is controlled by $s_i$.
\item Notice, however, that the ``envelope'' of the plot grows with a rate given by the $s_\mathcal{I}=0.645008...$ with $\mathcal{I}=\{1,2\}$. This is larger than both $s_1$ and $s_2$. Also notice that some points in each of these families contribute to the envelope that eventually defines the growth of the degeneracy spectrum for intermediate values of the area.
\item Although the band structure of the full degeneracy spectrum is not apparent at this stage, the separation in area of the points that define the envelope is quite close to the actual periodicity of the full spectrum. In fact it is possible to argue that these points somehow define the ``roots'' of the bands.
\end{itemize}

\begin{figure}[htbp2]
\includegraphics[width=16.5cm]{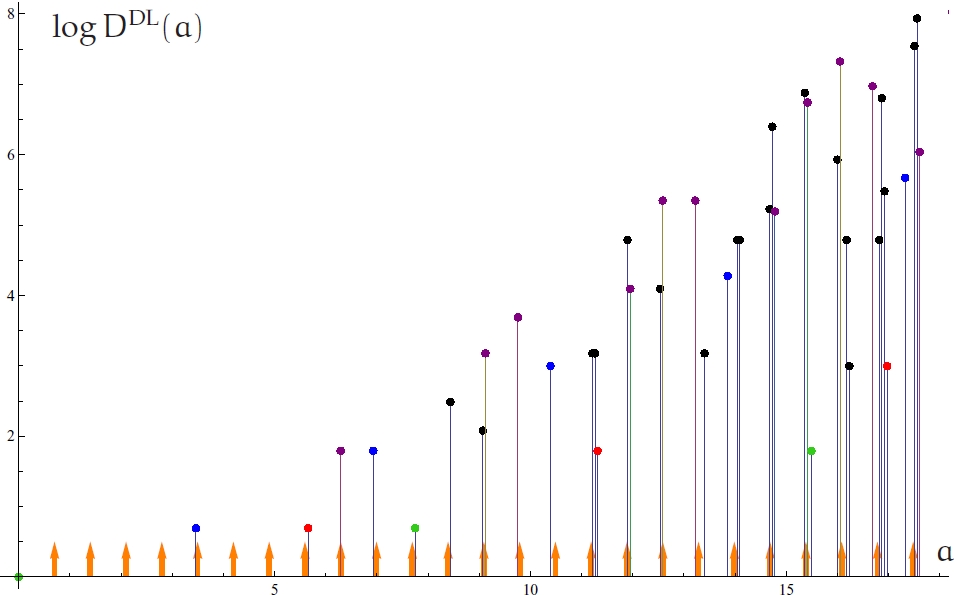}
\caption{Plot of the logarithm of the black hole degeneracy spectrum, $\log D^{\rm DL}(a)$, in terms of the area (in units of $4\pi\gamma\ell^2_P$). The figure shows the values associated with area eigenvalues that can be written as integer linear combinations of $\sqrt{3}$, $\sqrt{2}$ and $\sqrt{15}$. The arrows mark the position of the bands as predicted by equation (\ref{0349P}), taking into account that only even values of $P$ contribute when the projection constraint is considered. Notice that beyond $a=5$ there are bands at each of these positions. Notice also that there are many values of the area spectrum for which $D^{\rm DL}(a)=0$ and, hence, do not contribute to the band structure shown here. The colors indicate the subfamilies to which the different points in the plot belong (blue for $\sqrt{3}$, red for $\sqrt{2}$, green for $\sqrt{15}$, purple for linear combinations of $\sqrt{2}$ and $\sqrt{3}$ and black for linear combinations of $\sqrt{2}$, $\sqrt{3}$ and $\sqrt{15}$).}
\label{Fig:3rootLP}
\end{figure}

\begin{figure}[htbp]
\includegraphics[width=16.5cm]{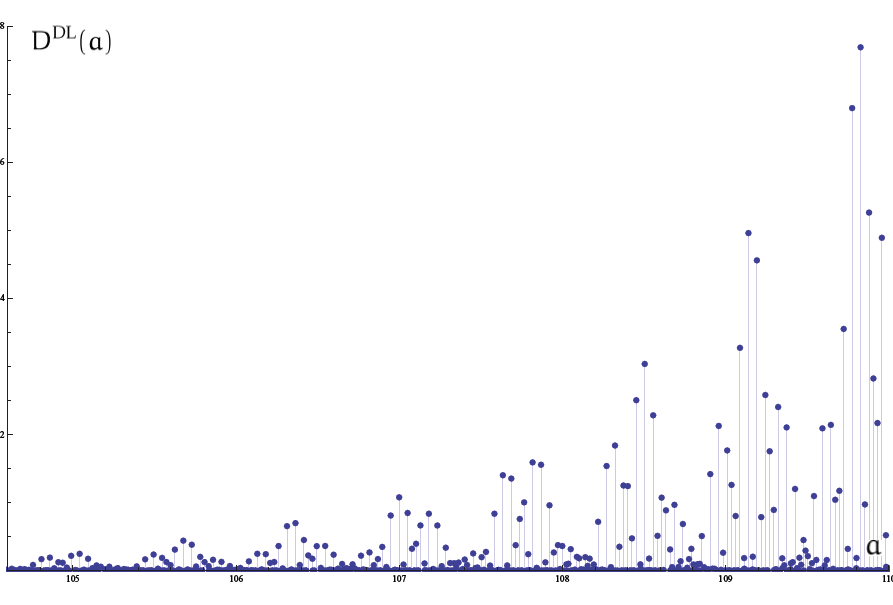}
\caption{Plot of $D^{\rm DL}$ (in units of $10^{30}$) versus area (in units of $4\pi\gamma\ell^2_P$) considering
only areas involving the squarefrees numbers  2, 3, and 15.}
\label{3famt}
\end{figure}

Finally, Fig. \ref{Fig:3rootLP} shows the result of considering area eigenvalues that can be written as integer linear combinations of $\sqrt{2}$, $\sqrt{3}$ and $\sqrt{15}$. The appearance of the band structure in the degeneracy spectrum is evident now. Figure \ref{3famt} shows a detail of the plot for larger values of the area spectrum where the band structure can be clearly seen. The envelope of the graph grows now with a slope given by  $s_\mathcal{I}=0.708187\cdots$ which is actually quite close to the actual value obtained for the full area spectrum $s_{\mathbb{N}}=0.746232\cdots$  Several comments are in order now:

\bigskip

The first is to notice that the band structure in the spectrum appears clearly for area eigenvalues involving a small number of square roots of squarefree integers. In fact, with linear combinations of the first two of them ($\sqrt{3}$ and $\sqrt{2}$) it is not possible to distinguish those structures in the spectrum, but as soon as the first three are taken into account, the bands readily show up as can be seen in Figs. \ref{Fig:3rootLP} and \ref{3famt}.

The second point to make is that, once the first three squarefrees are considered, adding more linear combinations does not significantly change the structure and position of the bands. As one adds new squarefree numbers to the values of the
area spectrum, the only effect that can be observed is that more and more points appear `in' the peaks, but not `out' the peaks, giving as a result a better resolution of the shape, but not modifying any feature of the structure in a significant way.

The third point is that, as soon as the five or six first squarefree numbers (in the order given in table \ref{tablasquarefree}) are considered, the spectrum obtained is almost indistinguishable from the complete one. This means that, with the first five squarefree numbers, we are able to reproduce all the points in the spectrum that have a degeneracy of the same order of magnitude as the maximum. A possible way to quantify this fact is by looking at the asymptotic growth rate of the envelope given by  (\ref{polo}) as a function of the number of squarefrees used. This is shown in table \ref{tablagrowth}.
\begin{table}[h!b!p!]
\caption{Asymptotic growth rate}
\begin{tabular}{|l|c|}
  \hline
  $s_2$ & $0.452679\ldots$ \\ \hline
  $s_{\{1,2\}}$ & $0.645008\ldots$ \\ \hline
  $s_{\{1,2,10\}}$ & $0.708187\ldots$ \\ \hline
  $s_{\{1,2,4,10\}}$ & $0.732215\ldots$ \\ \hline
  $s_{\{1,2,4,10,22\}}$ & $0.742270\ldots$ \\ \hline
  $s_{\{1,2,4,5,10,22\}}$ & $0.744388\ldots$ \\ \hline
  $s_{\{1,2,3,4,5,10,22\}}$ & $0.745368\ldots$ \\ \hline
    $\hspace{1.1cm}\vdots$ & $\vdots$ \\ \hline
  $s_{\mathbb{N}}$ &$0.746232\ldots$ \\
  \hline
\end{tabular}
\label{tablagrowth}
\end{table}
The information that one can extract from the above analysis is
valuable, in particular the fact that the highest degeneracy states
responsible for the observed pattern in the degeneracy spectrum are
(mostly) composed by punctures with $k=1,2,3,4,5,$ and $6$. Moreover,
we know that the area values corresponding to these configurations are
those containing multiples of square roots of the squarefree numbers
$3,2,15,6,$ and $35$. Then, considering only these values of the area, we are
sure that no relevant information is being missed but the calculations are simplified. In fact, the plots
appearing in the present paper corresponding to the largest area
eigenvalues considered have been obtained by using this approximation.

\subsection{Effects of the $r$-degeneracy and  $m$-degeneracy on the entropy microstructure}

Irrespective of the model (DL, GM, ENP), the black hole degeneracy spectrum $D(a)$ has the form
$$
D(a)=\sum_{c\in\mathcal{C}(a)}d_r(c)d_m(c)\,.
$$
There are then two different contributions to the degeneracy of \textit{each} configuration: one, the $r$-degeneracy, is model independent and  comes from the possible reorderings of its labels. The other, the $m$-degeneracy,  originates in the projection constraint (or similar conditions). In this subsection we want to study the relative contributions of each of them and, specifically, try to understand if any of these can explain by itself the observed microstructure of $D(a)$. To this end we will look at some auxiliary objects built out of the $d_r(c)$ and $d_m(c)$, in particular
$$
D_r(a):=\sum_{c\in\mathcal{C}(a)}d_r(c)\,,\quad D_m(a):=\sum_{c\in\mathcal{C}(a)}d_m(c)\,.
$$

\begin{figure}[htbp]
\includegraphics[width=16.5cm]{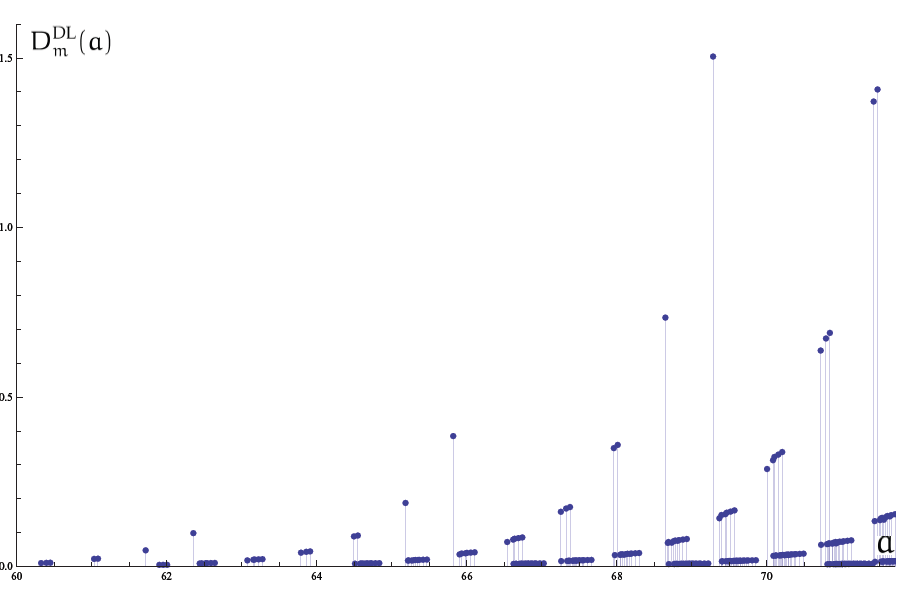}
\caption{Plot of $D_m^{\rm DL}(a):=\sum_{c\in \mathcal{C}(a)} d_m^{\rm DL}(c)$ (in units of $10^{11}$) versus area (in units of $4\pi\gamma\ell^2_P$) }
\label{6famm}
\end{figure}

\begin{figure}[htbp]
\includegraphics[width=16.5cm]{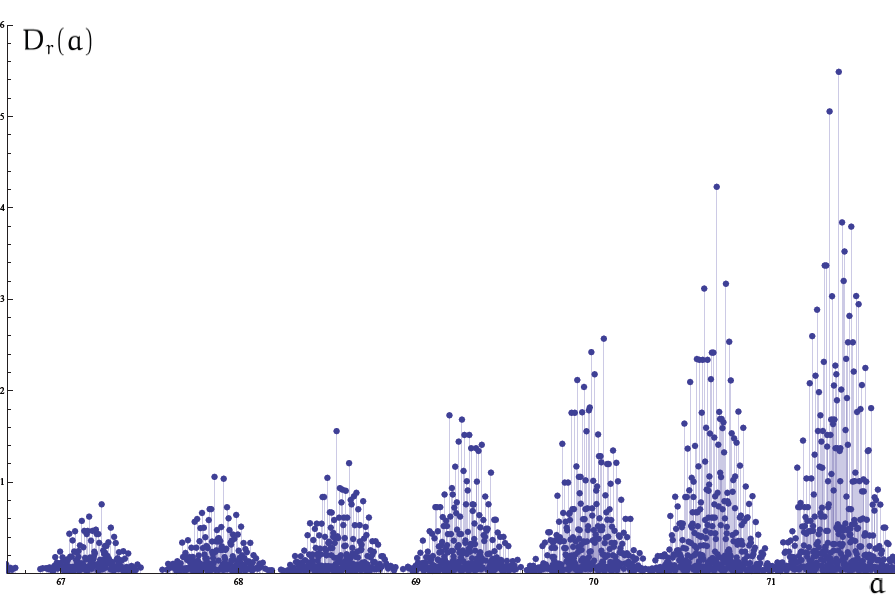}
\caption{Plot of $r$-degeneracy $D_r(a)=\sum_{c\in\mathcal{C}(a)}d_r(c)\delta_m(c)$ (in units of $10^{12}$) versus area $a$ expressed in units of $4\pi\gamma\ell^2_P$. Here $\delta_m(c)$ is one if the configuration $c$ satisfies the projection constraint and zero otherwise. Without this factor the period of the peaks decreases by a factor of two.}
\label{6famr}
\end{figure}

Figure \ref{6famm} shows $D_m(a)$ for the DL case (the other cases behave in a similar way). As we can see, there is some non-trivial substructure in this plot but this does not resemble the band structure that can be seen in the full degeneracy spectrum $D(a)$. On the other hand, the plot of $D_r(a)$ (Fig.\ref{6famr}) readily shows a band structure similar to the one found in $D(a)$. The only significant difference is that the exponential growth of the peaks is less pronounced. Therefore, we can conclude that the origin of the band structure resides in the $r$-degeneracy, which is rooted in the distinguishable character of the punctures. This is a consequence of the action of general diffeomorphisms over the horizon states. The fact that this degeneracy is common for all the countings explains why the band structure appears for all of them.

\subsection{\label{subsect:peaks} The peaks in the degeneracy spectrum}

In order to study the appearance of the peaks in the degeneracy spectrum it is very useful to introduce the following function in the space of black hole configurations \cite{Agullo:2008eg}
$$
P(c):=3K(c)+2N(c)\,,
$$
where, given $c=\{(k,N_k)\}$, we define $K(c):=\sum_k k N_k$ and $N(c):=\sum_k N_k$ (the sums extend to all the $k$-labels appearing in $c$). $P(c)$ is always a natural number, however, it is important to notice that those configurations giving rise to odd values of $P(c)$ correspond to zero degeneracy. This is so because the projection constraint cannot be satisfied if $P$ is odd.

\begin{figure}[htbp]
\includegraphics[width=16.5cm]{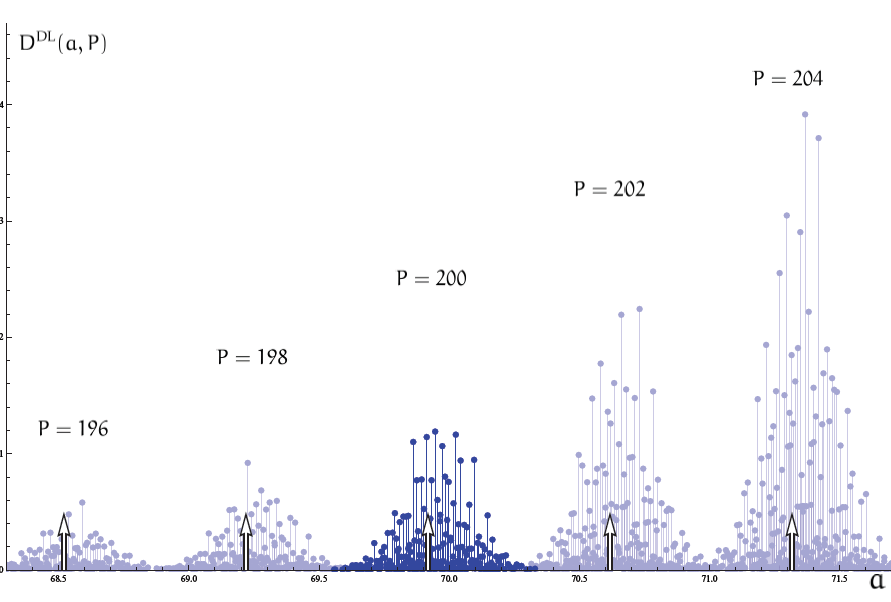}
\caption{This is a new version of Fig. \ref{Fig:DDL} where we have highlighted one of the peaks singled out by the peak counter $P$. We also show the position of the peaks according to formula (\ref{0349P}).
} \label{6famrclr}
\end{figure}

As we justify now, the crucial property of $P(c)$ is that each peak in the degeneracy spectrum is characterized by a single value of $P$ (see Fig. \ref{6famrclr}). This fact can be understood by following a simplified analysis, that we explain next,  that captures the essential features of the problem. To begin with let us consider the function defined by
$$
F^{\rm DL}_*(c):=d_r(c)d^{DL}_*(c)=\frac{(\sum N_k)!}{\prod N_k!}2^{\sum N_k}
$$
in the configuration space $\mathcal{C}=\bigcup_a \mathcal{C}(a)$. Here we are neglecting the projection constraint (though we will restrict ourselves to considering only even values of $P$). Notice, by the way, that the degeneracy given by (\ref{deg}) is just the sum of $F^{\rm DL}_*(c)$ for all the configurations $c\in \mathcal{C}(a)\subset \mathcal{C}$ corresponding to a specific area $a$. Now, it is possible to partition the space of configurations according to the values of $P(c)$. This procedure should be thought of as a coarse grained alternative to the standard approach of describing the degeneracy spectrum in terms of $a$ because, for a fixed value of $P$, there are configurations corresponding to a range of areas. The remarkable fact, as we will see, is that --in the continuous approximation-- there is a single maximum of $F^{\rm DL}_*(c)$ for each value of $P$. Furthermore, the areas associated with the configurations that maximize $F^{\rm DL}_*$ for the different values of $P$ closely match the observed position of the peaks in the degeneracy spectrum.

We show why this is so by considering continuous values for the variables $N_k$ and extending the function $F^{\rm DL}_*$ accordingly. We then use the Lagrange multipliers method to enforce the condition that $P(c)=P$ with $P\in \mathbb{N}$. By taking $\log F^{\rm DL}_*(c)$, using the Stirling approximation, introducing a single Lagrange multiplier $\lambda$, and using the notation $\hat{N}_k:=N_k/(\sum_{k^\prime} N_{k^\prime})$, the extrema are determined by
\begin{equation}
\hat{N}_k=2\exp{\big(-\lambda(3k+2)\big)}\,.\label{Distr}
\end{equation}
Adding the previous expressions in the label $k$ we get the condition
$$
1=\sum_{k=1}^\infty 2 \exp{\big(-\lambda(3k+2)\big)}=\frac{2e^{-2\lambda}}{e^{3\lambda}-1}
$$
that fixes the value of $\lambda=-\log\nu_0$ in terms of the single real solution $\nu_0$ of an auxiliary quintic equation
$$
\quad 2\nu^5+\nu^3-1=0\,.
$$
The solution $\nu_0$ can be written in terms of hypergeometric functions as
\begin{eqnarray*}
\nu_0&=& \, _4F_3\left(\frac{1}{15},\frac{4}{15},\frac{7}{15},\frac{13}{15};\frac{1}{3},\frac{2}{3},\frac{7}{6};-\frac{6250}{27}\right)-\frac{2}{3}\,_4F_3\left(\frac{2}{5},\frac{3}{5},\frac{4}{5},\frac{6}{5};\frac{2}{3},\frac{4}{3},\frac{3}{2};-\frac{6250}{27}\right)\\
&+&\frac{16}{9}\,_4F_3\left(\frac{11}{15},\frac{14}{15},\frac{17}{15},\frac{23}{15};\frac{4}{3},\frac{5}{3},\frac{11}{6};-\frac{6250}{27}\right) \end{eqnarray*}
and it gives $\lambda=0.260847\cdots$

\begin{figure}[htbp]
\includegraphics[width=16.5cm]{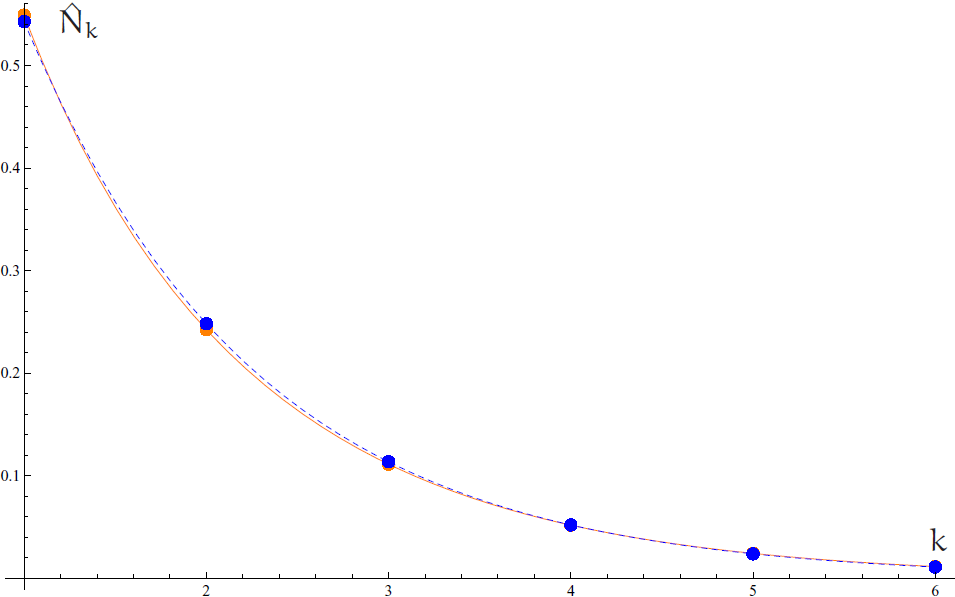}
\caption{Plot of the degeneracy distribution given by formula (\ref{Distr}) (dashed line) compared to the  maximum degeneracy distribution (\ref{mdd}) (solid line).} \label{fig:MDD}
\end{figure}

Introducing back the value of $\lambda$ in (\ref{Distr}), we obtain the distribution for the $\hat{N}_k$. This determines the values of the $N_k$ --maximizing $\log F^{\rm DL}_*$ for a given $P$-- modulo the value of the sum $\sum_k N_k$ that can be obtained by using the constraint
$$
P=3K(c)+2N(c)=\Big(\sum_{k=1}^\infty N_k \Big)\cdot\Big(2+6\sum_{k=1}^\infty k e^{-\lambda (3k+2)}\Big)\Rightarrow \sum_{k=1}^\infty N_k=\frac{P}{2+6\sum_{k=1}^\infty k e^{-\lambda (3k+2)}}\,.
$$
Finally, the area corresponding to the maximum of $F^{\rm DL}_*$ in terms of the peak label $P$ is
\begin{eqnarray}
a(P)=\frac{\sum_{k=1}^\infty e^{-\lambda (3k+2)}\sqrt{k(k+2)}}{1+3\sum_{k=1}^\infty k e^{-\lambda (3k+2)}} \,P= (0.34959\cdots) P\,.
\label{0349P}
\end{eqnarray}
Several comments are in order now. First, it is interesting to compare the distribution given by (\ref{Distr}) with the so called ``maximum degeneracy distribution'' (MDD) given by
\begin{eqnarray}
\hat{N}_k=2\exp{\big(-\lambda_{DL}\sqrt{k(k+2)}\big)}\,,\quad \textrm{where }\,\lambda_{DL}=0.746232\ldots\label{mdd}
\end{eqnarray}
The MDD can be obtained by maximizing $\log F^{\rm DL}_*$ subject to the condition that the \textit{area} $\sum_k N_k\sqrt{k(k+2)}$ is fixed. As can be seen in Fig. \ref{fig:MDD}, they are very close to each other and, in particular, they are almost identical for the most relevant spin labels. A second remark concerns the positions of the $a(P)$. As can be readily seen in (\ref{0349P}) they are integer multiples of the constant $0.34959\cdots$, and in the physically relevant case where the projection constraint is taken into account (even $P$'s), they are multiples of $0.69918\cdots$ These positions are very well correlated with the observed maxima of the degeneracy distribution. It is interesting to remark here that a very similar distribution has been obtained by a completely different method in \cite{Agullo:2008eg}. This compatibility is a non trivial cross check of both approaches. Third, the previous computations have been performed in a continuum approximation that neglects the fact that the values of $N_k$ are integers. In practice, this means that the actual shape of the peaks may be more irregular and secondary maxima may be present.

Once we have checked that $P$ plays an important role in the identification and labeling of the peaks appearing in the degeneracy spectrum it is straightforward to write down generating functions that allow us to select the contribution to the degeneracy spectrum of the configurations associated with a given value of $P$. Following the approach described in Appendix \ref{App:A}, it is possible to take care of the values of $P$ by introducing a new variable $\nu$ in the generating functions. This way we find
\begin{eqnarray}
G^{\rm DL}(\nu;z,x_1,x_2,\ldots)&=&\left(1-\sum_{i=1}^\infty\sum_{\alpha=1}^\infty \nu^{3k^i_\alpha+2}(z^{k^i_\alpha}+z^{-k^i_\alpha}) x_i^{y^i_\alpha}\right)^{-1}\,.\label{Gpicos}
\end{eqnarray}
The coefficient
$$
D^{\rm DL}(a,P):=[z^0][x_1^{q_1}x_2^{q_2}\cdots][\nu^P]G^{\rm DL}(\nu;z,x_1,x_2,\ldots)$$ gives the sum of the degenerations $d_r(c)d_m^{\rm DL}(c)$ of all the configurations with area $a=\sum_i q_i\sqrt{p_i}$ satisfying $P(c)=P$. Obviously, the black hole degeneracy is given by
$$
D^{\rm DL}(a)=\sum_{P=1}^\infty   D^{\rm DL}(a,P)\,.
$$
A graphical confirmation of the ability of the $P(c)$ to identify the peaks in the black hole degeneracy spectrum is given in Fig. \ref{6famrclr}. Here we have represented $D^{DL}(a,P)$ as a function of $a$ for several consecutive values of $P$.

The preceding analysis can be easily carried out for the other countings presented in the paper with similar results. As we have shown the band structure is explained by the $r$-degeneracy. The $m$-degeneracy only plays the role of suppressing some configurations. The main consequence of this is the effective change of the periodicity observed in the black hole degeneracy spectrum.

\section{\label{sect:Conclusions}Conclusions}

We have given a comprehensive account of a number of combinatorial methods that allow us to study in great detail the behavior of the entropy in loop quantum gravity. The power of the techniques that we have employed can be appreciated by using them to study the different counting proposals that appear in the literature. The number theoretic methods based in the solution of diophantine equations provide a very efficient way to study the spectrum of the area operator and characterize the configurations of black holes that determine the entropy. These methods also provide powerful classifying criteria to disentangle the black hole degeneracy spectrum that can be exploited to understand the origin of the entropy microstructure. This detailed information can be codified in an extremely efficient way in suitable generating functions from which it is possible to extract detailed information at will. We illustrate the power of these generating functions by providing complete computations for the smallest black holes in Appendix \ref{App:B}. These computations show in a convincing manner how the interesting microstructure of the entropy comes to life and show that it is a robust feature present in all the different schemes. We give also in appendix \ref{App:C} a new method based in group-theoretic arguments that can be used to obtain, in a unified way, the results concerning the implementation of the projection constraint.

The final goal of the approach that we have described here is to verify if the entropy discretization holds for macroscopic black holes. An important tool in this analysis is the generating function given by \ref{Gpicos} because it isolates the configurations that contribute to the bands that appear in the black hole degeneracy spectrum. Actually the proper identification of this generating function is one of the main new results of the present paper. We want to mention that, although this function is written here for the DL counting, the method described in appendix \ref{App:A} can be used to derive similar expressions for the other countings mentioned in the paper. The main open problem --that we expect to solve with the methods described here-- is the persistence of the observed entropy structure for macroscopic black holes. If this is the case, then LQG would realize in a very non-trivial way the predictions about the effective equally spaced quantization of black hole areas. In our opinion this would lend an important support to the theory.

\acknowledgments

We want to thank Jes\'us Salas for many helpful discussions. The work has been supported by the Spanish MICINN research grants FIS2009-11893, FIS2008-06078-C03-02, FIS2008-01980, ESP2007-66542-C04-01, AYA 2009-14027-C05-01 and  the  Consolider-Ingenio 2010 Program CPAN (CSD2007-00042). IA acknowledges the support provided by a UWM-RGI grant. JDP is also supported by the PHY-0854743 NSF grant and Eberly research funds at Penn State University.

\appendix

\renewcommand \thesubsection{A.\arabic{subsection}}

\section{\label{App:A}Generating functions}
In this appendix we give a pedagogical account of how the different generating functions mentioned  in the paper are derived (a general treatment of generating functions can be found in \cite{Flajolet}). Our construction uses the set of allowed configurations $\mathcal{C}(a)$ consisting of all multisets  $c=\{(k_\alpha^i,N_{k_\alpha^i})\}$ associated with a value of the area $a=\sum_i q_i\sqrt{p_i}$ (as explained in section \ref{subPell}). The black hole degeneracy $D(a)$ can be obtained from $\mathcal{C}(a)$ by incorporating two sources of degeneracy. The first one is related to the possible reorderings ($r$-degeneracy) in each multiset and  is common for all the models. The second corresponds to the additional conditions  that define the   different countings ($m$-degeneracy).

\subsection{Generating function for the $r$-degeneracy}
Given a value of the area $a=\sum_i q_i\sqrt{p_i}$ the $r$-degeneracy is given by
\begin{equation}
D_r(a)=\sum_{c\in\mathcal{C}(a)}
\frac{\big(\sum_i\sum_{\alpha} N_{k^{i}_{\alpha}}\big)!}{\prod_i\prod_\alpha N_{k^i_\alpha}!}\,,\label{Ra}
\end{equation}
where $N_{k^i_\alpha}$ are the solutions to (\ref{diofN}). In order to find the generating function $G_r$ for $D_r(a)$ we split the problem in three steps:
\begin{enumerate}

\item Find the generating function for the number of solutions of each of the diophantine equations (\ref{diofN}), that is, the number
    $$
    \sum_{c\in\mathcal{C}(a)}1\,.
    $$
\item Modify this generating function to introduce the denominators appearing in  the definition (\ref{Ra})    of $D_r(a)$, that is, the number
$$
    \sum_{c\in\mathcal{C}(a)}
\frac{1}{\prod_i\prod_\alpha N_{k^i_\alpha}!}\,.
$$

\item Modify the generating function obtained in the previous step to account also for the numerators in (\ref{Ra}).

\end{enumerate}

The first step is solved by getting a generating function that counts the number of solutions to diophantine equations of the form
\begin{equation}
y^i_1 N_{k^i_1}+y^i_2 N_{k^i_2}+y^i_3 N_{k^i_3}+\cdots =q_i\,,
\label{ecuacionesdiof}
\end{equation}
where the unknowns are the $N_k$ and the $y^i_{\alpha}$ are given by the solutions to the Pell equations.
If we formally consider the function
$$
F_i(x)=\big(x^{0\cdot y^i_1}+x^{1\cdot y^i_1}+x^{2\cdot y^i_1}+\cdots\big)\cdot\big(x^{0\cdot y^i_2}+x^{1\cdot y^i_2}+x^{2\cdot y^i_2}+\cdots\big)\cdots\,,
$$
it is not difficult to see that the coefficient of the $x^{q_i}$ term of the McLaurin expansion of $F_i(x)$ gives precisely the number of solutions of the diophantine equation (\ref{ecuacionesdiof}). Now it is possible to write $F_i(x)$ as
$$
F_i(x)=\frac{1}{1-x^{y^i_1}}\frac{1}{1-x^{y^i_2}}\cdots=\prod_{\alpha=1}^\infty\frac{1}{1-x^{y^i_\alpha}}\,.
$$
A judicious introduction of additional variables allows us to actually compute quantities associated with the solutions of the diophantine equations (\ref{ecuacionesdiof}). For example, if we want to get the solutions themselves, we can introduce a set of variables $\nu_{\alpha}$, $\alpha\in \mathbb{N}$, and take
$$
F_i(x;\nu_1,\nu_2,\ldots)=\frac{1}{1-\nu_1x^{y^i_1}}\frac{1}{1-\nu_2x^{y^i_2}}\cdots=\prod_{\alpha=1}^\infty\frac{1}{1-\nu_\alpha x^{y^i_\alpha}}\,.
$$
The coefficient of $x^{q_i}$ is now a polynomial in the $\nu_\alpha$ such that each monomial of the form $\nu_1^{n_1} \nu_2^{n_2}\cdots$ tells us that there is a solution $N_{k^i_1}=n_1$, $N_{k^i_2}=n_2\ldots$
If we want to obtain just the sum $\sum_\alpha N_{k^i_\alpha}$, we take $\nu_\alpha=\nu$ for all $\alpha\in \mathbb{N}$. In this case, the coefficient of $x^{q_i}$ is a polynomial in the $\nu$ such that the degree $n$ of each monomial signals the existence of a solution to (\ref{ecuacionesdiof}) with $\sum_\alpha N_{k^i_\alpha}=n$. This idea has been used to obtain the generating function for the peaks given by (\ref{Gpicos}). Finally the substitution of the variables $\nu_\alpha$ by the specific numerical value $\nu_\alpha=2$ would provide a generating function such that the coefficient of $x^{q_i}$ would be $\sum_{sols(q_i)} \prod_\alpha 2^{N_{k^i_\alpha}}$,  where the previous sum extends to the solutions of (\ref{ecuacionesdiof}) for a given $q_i$ and is zero if the equation has no solutions.

As we really have to consider sets of decoupled diophantine equations the actual number of solutions is the product of the number of solutions for each of them and hence the generating function becomes
$$
F(x_1,x_2,\ldots)=\prod_{i=1}^\infty F_i(x_i)=\prod_{i=1}^\infty\prod_{\alpha=1}^\infty\frac{1}{1-x_i^{y^i_\alpha}}\,.
$$
Hence, for $a= q_1\sqrt{p_1}+q_2\sqrt{p_2}+\cdots$, the number that we are looking for is
$$
 \sum_{c\in\mathcal{C}(a)}1=[x_1^{q_1} x_2^{q_2}\cdots]F(x_1,x_2,\ldots)\,,
$$
where $[x_1^{q_1} x_2^{q_2}\cdots]F(x_1,x_2,\ldots)$ denotes the coefficient of the $x_1^{q_1} x_2^{q_2}\cdots$ term in the power series expansion of $F(x_1,x_2,\ldots)$ around $(x_1,x_2,\ldots)=(0,0,\ldots)$.

The factors appearing in the denominators of $D_r(a)$ can be incorporated by modifying the previous generating function $F_i$ as follows
$$
H_i(x)=\Big(\frac{1}{0!}x^{0\cdot y^i_1}+\frac{1}{1!}x^{1\cdot y^i_1}+\frac{1}{2!}x^{2\cdot y^i_1}+\cdots\Big)\cdot\Big(\frac{1}{0!}x^{0\cdot y^i_2}+\frac{1}{1!}x^{1\cdot y^i_2}+\frac{1}{2!}x^{2\cdot y^i_2}+\cdots\Big)\cdots=\prod_{\alpha=1}^\infty e^{x^{y^i_\alpha}}\,.
$$
Now, it is immediate to obtain the number we are looking for as
$$
\sum_{c\in\mathcal{C}(a)}
\frac{1}{\prod_i\prod_\alpha N_{k^i_\alpha}!}=[x^{q_1}_1 x^{q_2}_2 \cdots] H(x_1,x_2,\ldots)\,,$$
where
\begin{equation}
H(x_1,x_2,\ldots)=\prod_{i=1}^\infty H_i(x_i)=\prod_{i=1}^\infty \prod_{\alpha=1}^\infty e^{x_i^{y^i_\alpha}}=\exp\left(\sum_{i=1}^\infty \sum_{\alpha=1}^\infty x_i^{y^i_\alpha}\right)\,.
\label{expo}
\end{equation}

Finally, the factor in the numerator of $D_r(a)$ is $N!$, with $N=\sum_i\sum_\alpha N_{k^i_\alpha}$. In order to introduce this factor we notice that the expansion of the exponential appearing in (\ref{expo}) is
$$
\exp\left(\sum_{i=1}^\infty \sum_{\alpha=1}^\infty x_i^{y^i_\alpha}\right)=1+\left(\sum_{i=1}^\infty \sum_{\alpha=1}^\infty x_i^{y^i_\alpha}\right)+\frac{1}{2!}\left(\sum_{i=1}^\infty \sum_{\alpha=1}^\infty x_i^{y^i_\alpha}\right)^2+\cdots
$$
The key point now is to realize the  $N^{\rm th}$-term in the above expansion,
\begin{equation}
 \frac{1}{N!}\left(\sum_{i=1}^\infty \sum_{\alpha=1}^\infty x_i^{y^i_\alpha}\right)^N\,,
\label{term}
\end{equation}
 gathers  the contributions of those elements of $\mathcal{C}(a)$ satisfying precisely $\sum_i\sum_\alpha N_{k^i_\alpha}=N$. This can be easily seen by introducing the single variable $\nu$ as above and checking that each term of the form (\ref{term}) appears now multiplied by $\nu^N$.
 Then, it suffices to modify $H(x_1,x_2,\ldots)$ by multiplying each term in the previous expansion by $N!$ to get
\begin{eqnarray*}
 G_r(x_1,x_2,\ldots)&=&1+\left(\sum_{i=1}^\infty \sum_{\alpha=1}^\infty x_i^{y^i_\alpha}\right)+\left(\sum_{i=1}^\infty \sum_{\alpha=1}^\infty x_i^{y^i_\alpha}\right)^2+\cdots+\left(\sum_{i=1}^\infty \sum_{\alpha=1}^\infty x_i^{y^i_\alpha}\right)^N+\cdots\\
 &=&\left(1-\sum_{i=1}^\infty \sum_{\alpha=1}^\infty x_i^{y^i_\alpha}\right)^{-1}\,.
 \end{eqnarray*}

Summarizing, given an area eigenvalue $a=q_1\sqrt{p_1}+q_2\sqrt{p_2}+\cdots$, its $r$-degeneracy is given by
$$
D_r(a)=[x^{q_1}_1x^{q_2}_2\cdots ] G_r(x_1,x_2,\ldots)\,.
$$

\subsection{Generating functions for the $m$-degeneracy}

A straightforward extension of the argument given in the previous subsection to compute quantities associated with the solutions to (\ref{ecuacionesdiof}) by introducing auxiliary variables allows us to incorporate the projection constraint (or similar conditions) to explain the $m$-degeneracy. It is straightforward, in particular, to introduce the terms for the projection constraints (see subsections \ref{sub413}, \ref{subsectGM} and \ref{sect:SU2})
\begin{eqnarray*}
&&\prod_{i=1}^\infty \prod_{\alpha=1}^\infty (z^{k^i_\alpha}+z^{-k^i_\alpha})^{N_{k^i_\alpha}}\,\hspace{1.7cm} \textrm{for DL}\,,\\
&&\prod_{i=1}^\infty \prod_{\alpha=1}^\infty \left(\frac{z^{k^i_\alpha+1}-z^{-k^i_\alpha-1}}{z-z^{-1}}\right)^{N_{k^i_\alpha}}\,\quad \textrm{for GM}\,,\\
-\frac{(z-z^{-1})^2}{2}&&\prod_{i=1}^\infty \prod_{\alpha=1}^\infty \left(\frac{z^{k^i_\alpha+1}-z^{-k^i_\alpha-1}}{z-z^{-1}}\right)^{N_{k^i_\alpha}}\,\quad \textrm{for ENP}\,.
\end{eqnarray*}
By doing this we get
\begin{eqnarray*}
G^{\rm DL}(z,x_1,x_2,\ldots)&=&\left(1-\sum_{i=1}^\infty\sum_{\alpha=1}^\infty (z^{k^i_\alpha}+z^{-k^i_\alpha}) x_i^{y^i_\alpha}\right)^{-1}\,,\\
G^{\rm GM}(z,x_1,x_2,\ldots)&=&\left(1-\sum_{i=1}^\infty\sum_{\alpha=1}^\infty \frac{z^{k^i_\alpha+1}-z^{-k^i_\alpha-1}}{z-z^{-1}} x_i^{y^i_\alpha}\right)^{-1}\,,\\
G^{\rm ENP}(z,x_1,x_2,\dots)&=&-\frac{(z-z^{-1})^2}{2}\left(\displaystyle 1-\sum_{i=1}^\infty\sum_{m=1}^\infty \Big( \frac{z^{k^i_m+1}-z^{-k^i_m-1}}{z-z^{-1}}\Big) x_i^{y^i_m}\right)^{-1}\,,
\end{eqnarray*}
and, from them, we obtain the black hole degeneracies as
$$D^{\rm C}(a)=[z^0][x^{q_1}_1 x_2^{q_2}\cdots ] G^{\rm C}(z,x_1,x_2,\ldots)\,,$$
where the index $\rm C$ refers to the counting scheme that we are interested in (i.e DL, GM or ENP). Notice that, as we mentioned in section \ref{sect:Detailed}, the ``master'' generating functions $G^{\mathrm{C}}$ allow us to obtain other generating functions that can be used to study the behavior of the degeneracy for every conceivable subfamily of area eigenvalues.

\section{\label{App:B}Explicit computations}
\renewcommand \thesubsection{B.\arabic{subsection}}
We give here a complete computation of the black hole degeneracy spectrum and entropy, corresponding to the first eigenvalues of the area spectrum, according to the DL prescription \cite{Domagala:2004jt}. On one hand, we will provide explicit computations to concretely show how the different methods introduced in the main body of the paper work. On the other, we want to explore the behavior of the entropy for the smallest black holes. This will allow us to see how the shapes of the black hole degeneracy spectrum --and of the entropy-- as a function of the area arise. Important features such as the linear growth of the entropy with the area and the appearance of the periodicity observed in \cite{Corichi:2006bs,Corichi:2006wn} can be already seen at this level. It is also possible to directly study the role of the projection constraint in the definition of the entropy. The main steps of the computation are:
\begin{itemize}
\item[1)] Determination of all the area eigenvalues smaller than a fixed value $a_+$. For practical purposes we will consider areas smaller than 18 (in units of $4\pi\gamma \ell_p^2$).
\item[2)] Computation of the black hole degeneracy spectrum by using generating functions.
\item[3)] Determination of the entropy according to the Domagala-Lewandowski recipe.
\end{itemize}

\bigskip

\noindent We will discuss these points with some detail.

\bigskip

1) The density of area eigenvalues grows very quickly as a function of the area. There are 354 area eigenvalues smaller than 18. Though they can be easily handled by a computer, they are too many to be listed here so in the following we provide a table with only those eigenvalues smaller than 12. Nevertheless we will extend the plots that we give in this appendix up to areas of around 18. For a given upper bound of the area $a_+$, we have to find all the different numbers of the form
\begin{equation}
2\sum_{I=1}\sqrt{j_I(j_I+1)}
\label{sumaapendix}
\end{equation}
obtained by considering positive half-integers $j_I\in\mathbb{N}/2$ and such that they are smaller than $a_+$. In this process, different choices of the $j_I$ may actually give the same value. This source of degeneracy is taken into account in a precise way by our counting methods (the Pell equation, generating functions and so on). At this level we will just care about the area eigenvalues disregarding their degeneracies.

For $a_+=18$ the maximum allowed value of $j_I$ is obtained by solving the inequality $\sqrt{j(j+1)}<9$; this gives $j=17/2$. On the other hand, if each $j_I$ takes the smallest non-zero allowed value of $1/2$, the maximum number of terms in the sum (\ref{sumaapendix}) is $\lfloor 18/\sqrt{3}\rfloor=10$. This means that we can generate all the sought for area eigenvalues by considering the points in the discrete set $(\mathbb{N}\cup\{0\})^{10}$ contained in the ten-dimensional simplex defined by the condition $\sum_{I=1}^{10}j_I\leq17/2$. The $1/10!$ factor reduction with respect to the computation extended to the cubic grid $\{1/2,1,\ldots,17/2\}^{10}$ is important to reduce the computing time.

\bigskip

2) The generating functions that we have introduced in Appendix \ref{App:A} depend, in principle, on an infinite number of variables $x_i$ associated with the square-free numbers $p_i$. For a finite subset of the area spectrum only the square roots of a finite number of them are relevant. This means that we will only have to consider the finite number of variables associated with them. If we take $a<18$, the square roots of the squarefree numbers that appear in the area eigenvalues are listed in Table \ref{tablasquarefree}. The only variables that we need to write explicitly in the generating functions are $x_1$, $x_2$, $x_3$, $x_4$, $x_5$, $x_7$, $x_9$, $x_{10}$, $x_{18}$, $x_{22}$, $x_{27}$, $x_{88}$, $x_{119}$, $x_{156}$, $x_{198}\,.$ We have to consider then:
\begin{eqnarray*}
& & G^{\rm DL}(z,x_1,x_2,x_3,x_4,x_5,x_7,x_9,x_{10},x_{18},x_{22},x_{27},x_{88},x_{119},x_{156},x_{198})\\
& & \begin{array}{lllll}
=\big(1 & - (z^2 + z^{-2}) x_1^2 & - (z^{16} + z^{-16}) x_1^{12} &-(z + z^{-1}) x_2&  - (z^{6} + z^{-6}) x_2^4 \\
       &- (z^8 + z^{-8}) x_3^4 & - (z^4 + z^{-4}) x_4^2&  - (z^7 + z^{-7}) x_5^3 &- (z^9 + z^{-9}) x_7^3 \\
       &- (z^{14} + z^{-14}) x_9^4 &- (z^3 + z^{-3}) x_{10} & - (z^{10} + z^{-10}) x_{18}^2 &- (z^5 +
       z^{-5}) x_{22} \\
       &- (z^{12} + z^{-12}) x_{27}^2  & - (z^{11} + z^{-11}) x_{88}& - (z^{13} + z^{-13}) x_{119} & - (z^{15} +
       z^{-15}) x_{156} - (z^{17} + z^{-17}) x_{198}\big)^{-1}\,,
       \end{array}
\\
& & G^{\rm DL}_*(x_1,x_2,x_3,x_4,x_5,x_7,x_9,x_{10},x_{18},x_{22},x_{27},x_{88},x_{119},x_{156},x_{198})\\
& &\begin{array}{lllllllll}
=\big(1 & - 2 x_1^2 & - 2 x_1^{12} &-2 x_2&  - 2 x_2^4
       &- 2 x_3^4 & - 2 x_4^2&  - 2 x_5^3 &- 2x_7^3 \\
       &- 2x_9^4 &- 2 x_{10} & - 2 x_{18}^2 &- 2 x_{22}
       &- 2x_{27}^2  & - 2 x_{88}& - 2 x_{119} & - 2x_{156} - 2 x_{198}\big)^{-1}\,.
       \end{array}
\end{eqnarray*}
The coefficients $[z^0][x_1^{q_1}\cdots x_{198}^{q_{198}}]G^{\rm DL}$ and $[x_1^{q_1}\cdots x_{198}^{q_{198}}]G^{\rm DL}_*$, respectively, tell us the values of $D^{\rm DL}(a)$ and $D^{\rm DL}_*(a)$ for
$a=q_1\sqrt{2}+\cdots+q_{198}\sqrt{323}$.  For example, for  $a=2\sqrt{2}+3\sqrt{3}+\sqrt{15}$ (i.e. the row corresponding to $n=71$ in the table given at the end of this appendix) we get
\begin{eqnarray*}
D^{\rm DL}(2\sqrt{2}+3\sqrt{3}+\sqrt{15})&=&[z^0][x_1^2x_2^3x_{10}]G^{\rm DL}=120\,,\\
D^{\rm DL}_*(2\sqrt{2}+3\sqrt{3}+\sqrt{15})&=&[x_1^2x_2^3x_{10}]G^{\rm DL}_*=640\,.
\end{eqnarray*}

\bigskip

3) The entropy is obtained by adding up the values of $D^{\rm DL}(a)$ up to a certain value of the area.

\bigskip

\begin{figure}[htbp]
\includegraphics[width=16.5cm]{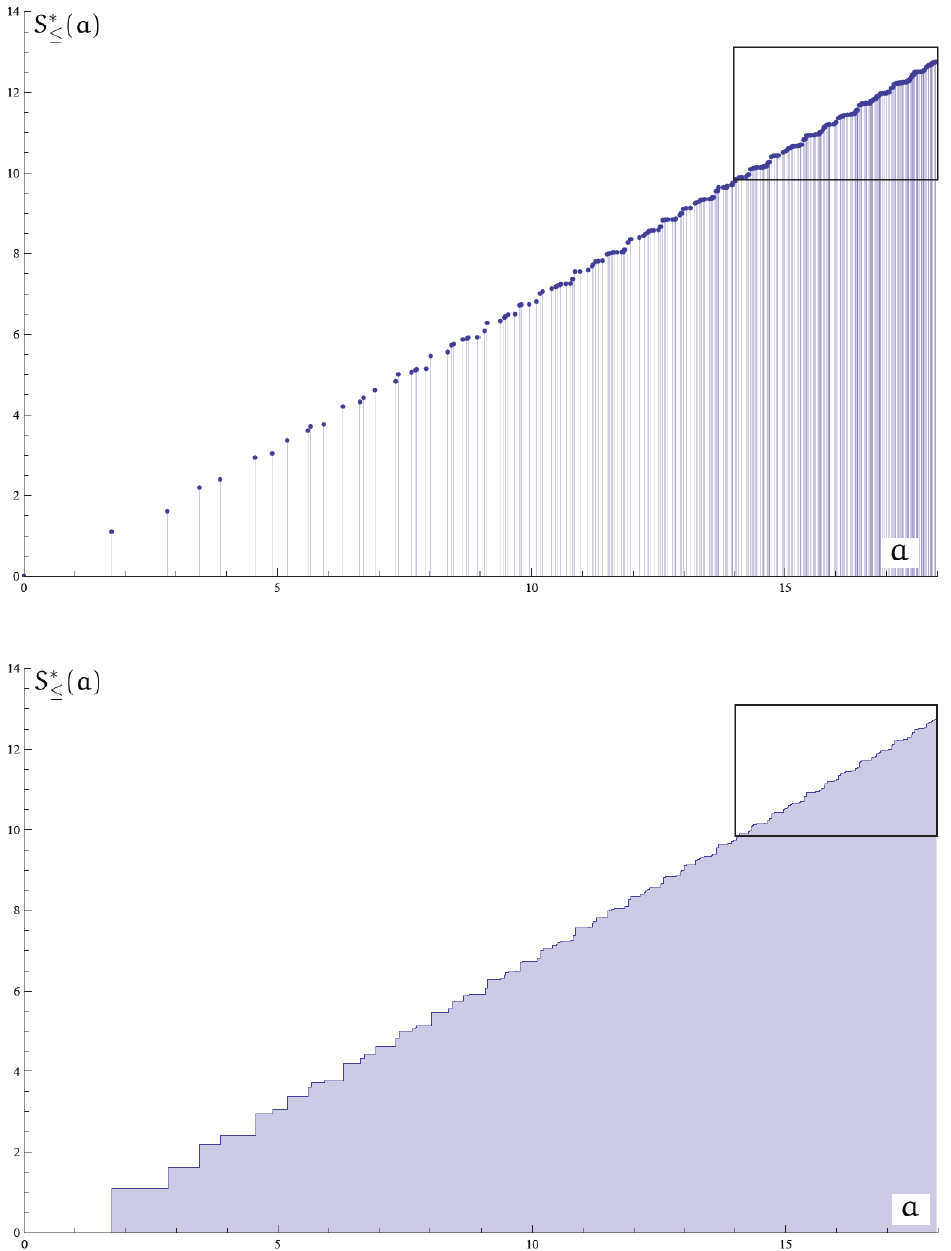}
\caption{Plots of the value of $S_\leq^*(a)$ for the points in the area spectrum  and all the areas smaller than 18 (in units of $4\pi\gamma \ell_p^2$) respectively. A detailed view of the framed parts can be seen in Fig. \ref{Fig:detalle18}.}
\label{Fig:escalera18}
\end{figure}

\begin{figure}[htbp]
\includegraphics[width=16.5cm]{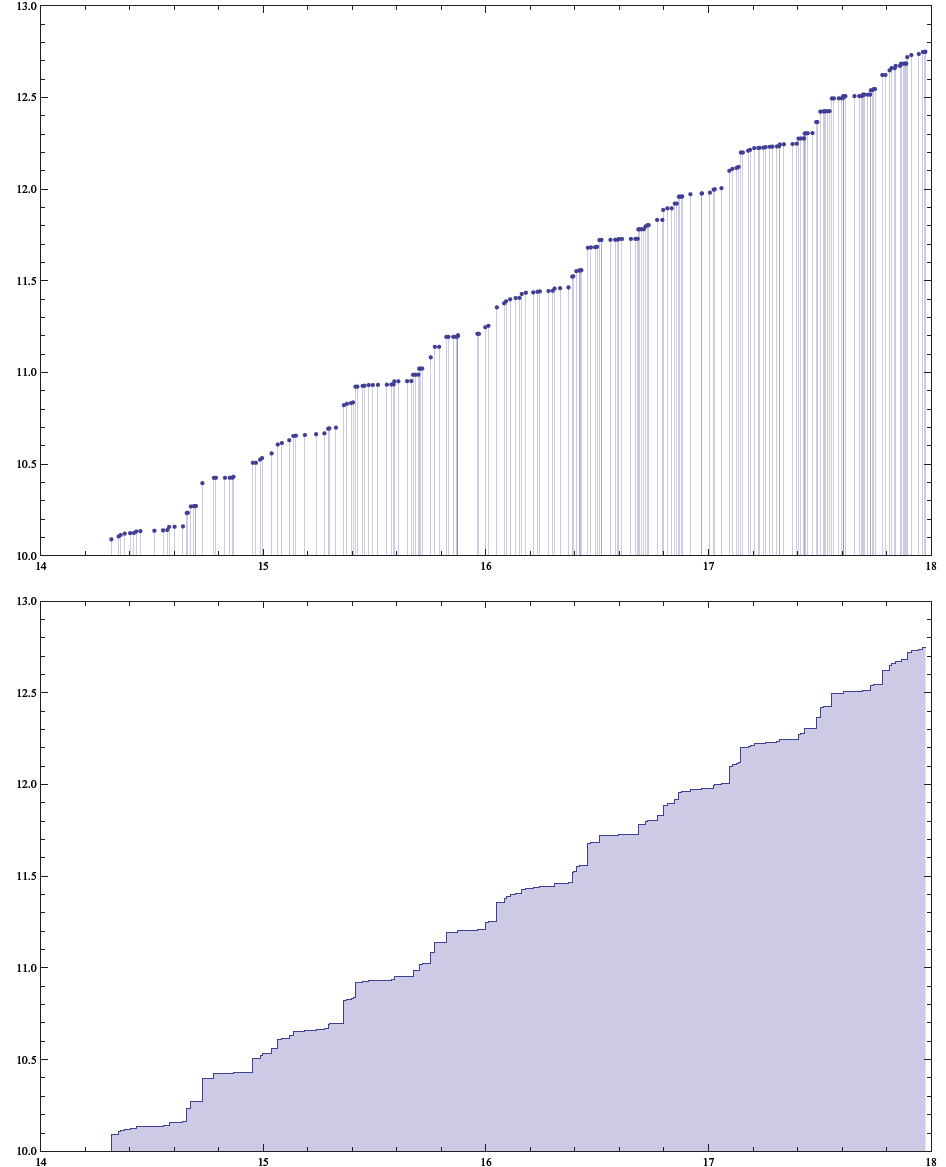}
\caption{Detail of Fig. \ref{Fig:escalera18} for areas $a\in[14,18]$. Notice the steps that appear for the area values considered in this plot.} \label{Fig:detalle18}
\end{figure}

\begin{center}
\begin{figure}[htbp]
\includegraphics[width=16.5cm]{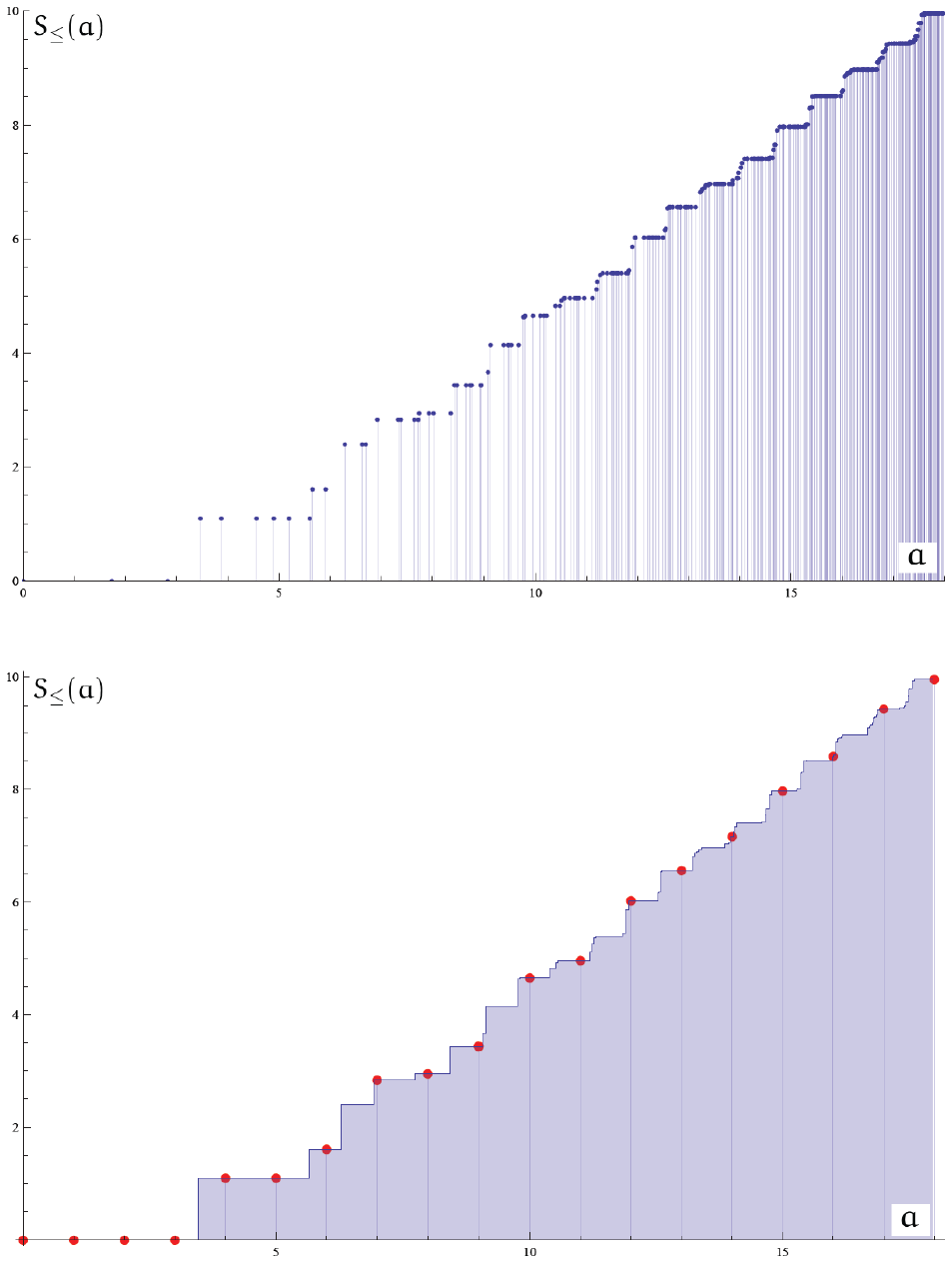}
\caption{Plot of the value of $S_\leq(a)$ for both the points in the area spectrum and all the values of the area smaller than 18 (in units of $4\pi\gamma \ell_p^2$). We also show the entropy values corresponding to prequantized values of the area.} \label{Fig:escalera18proy}
\end{figure}
\end{center}

The table at the end of this appendix shows the explicit values of the black hole degeneracies for the first area eigenvalues (corresponding to areas smaller than 12 in our units) and also their cumulative sum (that in the case of $D^{\rm DL}(a)$ gives the exponential of the entropy). The results of an explicit computation up to areas of the order of 18 are shown in Figs. \ref{Fig:escalera18} to \ref{Fig:escalera18proy}.

We want to make several comments at this point: First, in order to understand some of the features of the entropy and the role of the projection constraint, it is convenient to study the auxiliary entropy $S^*_\leq(a)$ given by (\ref{numint1}) and compare it with the actual entropy $S_\leq(a)$. Figure  \ref{Fig:escalera18} shows both the exact values of $S^*_\leq(a)$ for all the area eigenvalues smaller than $18$ and the values of $S^*_\leq(a)$ as a function defined for all areas  $a\in[0,18]$. It is possible to see that, after a short transient regime, the entropy grows linearly with area and a characteristic staircase structure appears. As can be seen in the detailed plots of Fig. \ref{Fig:detalle18} this is more evident for the largest areas considered, where the plot of the entropy can be effectively approximated by a smooth curve due to the increasing density of the area spectrum. The width of the steps is roughly $0.35$ in our units. This is in satisfactory agreement with the prediction of equation (\ref{0349P}).

\begin{figure}[htbp]
\includegraphics[width=16.5cm]{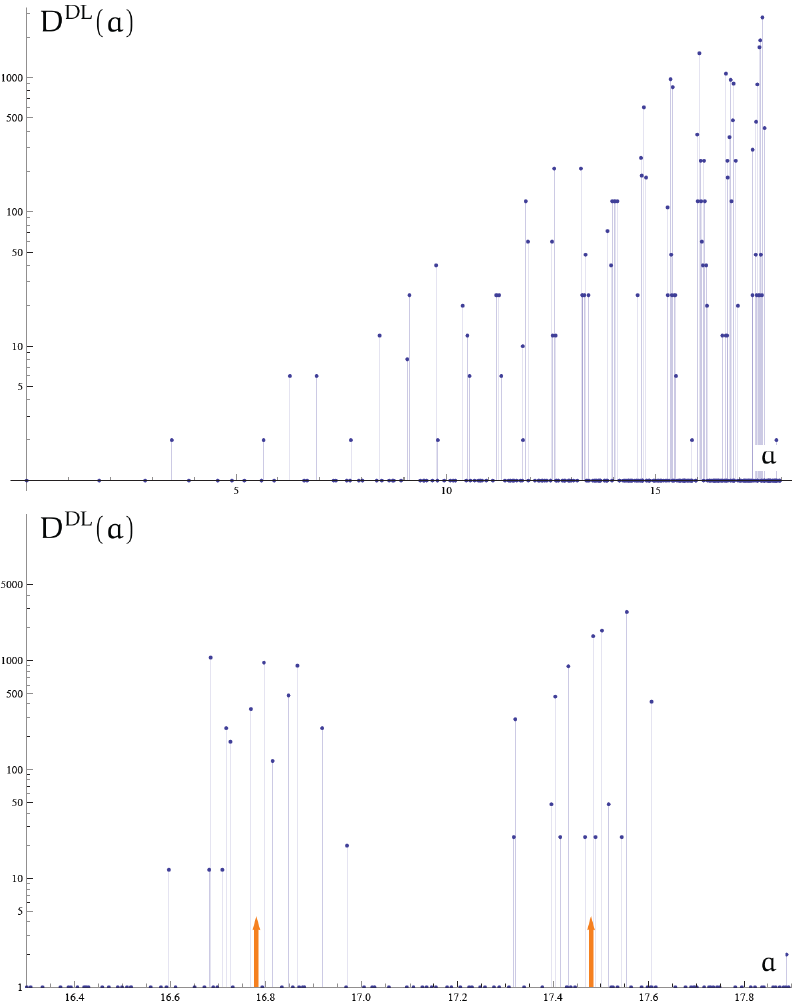}
\caption{Logarithmic plot of $D^{\rm DL}(a)$ for all the points in the area spectrum smaller than 18 (compare this figure with Fig. \ref{Fig:3rootLP}) and a detail of the last two peaks. The arrows mark the position of the bands as predicted by equation (\ref{0349P}). Notice that there are many points in the area spectrum for which $D^{\rm DL}(a)=0$.
} \label{Fig:bandas18proy}
\end{figure}

In order to obtain the physical entropy $S_\leq(a)$ the projection constraint must be incorporated. It is instructive to compare the results obtained in this case with the ones described above. In particular Fig. \ref{Fig:escalera18proy} is the counterpart of  Fig. \ref{Fig:escalera18}. In this case the staircase structure is more evident because the width of the steps doubles (to a value around $0.7$). This doubling is a consequence of the effective suppression of many configurations by the action of the projection constraint as justified in subsection \ref{subsect:peaks}; these configurations are not shown in Fig. \ref{Fig:escalera18proy}. Figure \ref{Fig:escalera18proy} shows also the values of the entropy corresponding to the prequantized values of the areas (that are just integer values in  units of $4\pi\gamma\ell^2_P$). For this subset of areas the growth of the entropy as a function of the area is linear (with the additional logarithmic corrections). The imprint of the staircase structure is this case is the presence of some larger than average jumps (``double jumps'') in the value of the entropy for successive prequantized areas. Finally the last figure (Fig. \ref{Fig:bandas18proy}) shows the black hole degeneracy spectrum with its characteristic peak structure. As it can be seen the peaks are quite pronounced and the distance between them corresponds to the width of the steps in the entropy.

\begin{center}
\begin{tabular}{|c|c|c|c|c|c|}
  \hline
$n$ \T\B & Area eigenvalue $a^{{\scriptscriptstyle \rm LQG}}_n$  \T\B&    $D^{\rm DL}(a^{{\scriptscriptstyle \rm LQG}}_n)$ \T\B & $D^{\rm DL}_*(a^{{\scriptscriptstyle \rm LQG}}_n)$  \T\B&   $\displaystyle \sum_{a'\leq a^{{\scriptscriptstyle \rm LQG}}_n}D^{\rm DL}(a') $ \T\B&  $\displaystyle\sum_{a'\leq a^{{\scriptscriptstyle \rm LQG}}_n}D^{\rm DL}_*(a')$ \T\B \\\hline\hline
0 & $0$ & 1 & 1 & 1 & 1 \\  \hline
1&  $\sqrt{3}$ & 0 & 2 & 1 & 3 \\  \hline
2&  $2\sqrt{2}$ & 0 & 2 & 1 & 5 \\  \hline
3&  $2\sqrt{3}$ & 2 & 4 & 3 & 9 \\  \hline
4&  $\sqrt{15}$ & 0 & 2 & 3 & 11 \\  \hline
5&  $2\sqrt{2}+\sqrt{3}$ & 0 & 8 & 3 & 19 \\  \hline
6&  $2\sqrt{6}$ & 0 & 2 & 3 & 21 \\  \hline
7&  $3\sqrt{3}$ & 0 & 8 & 3 & 29 \\  \hline
8&  $\sqrt{3}+\sqrt{15}$ & 0 & 8 & 3 & 37 \\  \hline
9&  $4\sqrt{2}$ & 2 & 4 & 5 & 41 \\  \hline
10&  $\sqrt{35}$ & 0 & 2 & 5 & 43 \\  \hline
11&  $2\sqrt{2}+2\sqrt{3}$ & 6 & 24 & 11 & 67 \\  \hline
12&  $\sqrt{3}+2\sqrt{6}$ & 0 & 8 & 11 & 75 \\  \hline
13&  $2\sqrt{2}+\sqrt{15}$ & 0 & 8 & 11 &83 \\  \hline
14&  $4\sqrt{3}$ & 6 & 18 & 17 & 101 \\  \hline
15&  $2\sqrt{3}+\sqrt{15}$ & 0 & 24 & 17 & 125 \\  \hline
16&  $4\sqrt{2}+\sqrt{3}$ & 0 & 24 & 17 & 149 \\  \hline
17&  $\sqrt{3}+\sqrt{35}$ & 0 & 8 & 17 & 157 \\  \hline
18&  $2\sqrt{2}+2\sqrt{6}$ & 0 & 8 & 17 & 165 \\  \hline
19&  $2\sqrt{15}$ & 2 & 4 & 19 & 169 \\  \hline
20&  $3\sqrt{7}$ & 0 & 2 & 19 & 171 \\  \hline
21&  $2\sqrt{2}+3\sqrt{3}$ & 0 & 64 & 19 & 235 \\  \hline
22&  $2\sqrt{3}+2\sqrt{6}$ & 0 & 24 & 19 & 259 \\  \hline
23&  $2\sqrt{2}+\sqrt{3}+\sqrt{15}$ & 12 & 48 & 31 & 307 \\  \hline
24&  $6\sqrt{2}$ & 0 & 8 & 31 & 315 \\  \hline
25&  $5\sqrt{3}$ & 0 & 40 & 31 & 355 \\  \hline
26&  $2\sqrt{2}+\sqrt{35}$ & 0 & 8 & 31 & 363 \\  \hline
27&  $2\sqrt{6}+\sqrt{15}$ & 0 & 8 & 31 & 371 \\  \hline
28&  $4\sqrt{5}$ & 0 & 2 & 31 & 373 \\  \hline
29&  $3\sqrt{3}+\sqrt{15}$ & 8 & 64 & 39 & 437 \\   \hline
30&  $4\sqrt{2}+2\sqrt{3}$ & 24 & 96 & 63 & 533 \\   \hline
31&  $2\sqrt{3}+\sqrt{35}$ & 0 & 24 & 63 & 557 \\  \hline
32&  $2\sqrt{2}+\sqrt{3}+2\sqrt{6}$ & 0 & 48 & 63 & 605 \\  \hline
33&  $\sqrt{3}+2\sqrt{15}$ & 0 & 24 & 63 & 629 \\  \hline
34&  $4\sqrt{2}+\sqrt{15}$ & 0 & 24 & 63 & 653 \\  \hline
35&  $\sqrt{3}+3\sqrt{7}$ & 0 & 8 & 63 & 661 \\  \hline
36&  $2\sqrt{2}+4\sqrt{3}$ & 40 & 168 & 103 & 829 \\  \hline
37&  $\sqrt{15}+\sqrt{35}$ & 0 & 8 & 103 & 837 \\  \hline
\end{tabular}
\end{center}

\begin{center}

\begin{tabular}{|c|c|c|c|c|c|}
 \hline
$n$ \T\B& Area eigenvalue $a^{{\scriptscriptstyle \rm LQG}}_n$ \T\B& $D^{\rm DL}(a^{{\scriptscriptstyle \rm LQG}}_n)$ \T\B& $D^{\rm DL}_*(a^{{\scriptscriptstyle \rm LQG}}_n)$ \T\B& $\displaystyle \sum_{a'\leq a^{{\scriptscriptstyle \rm LQG}}_n}D^{\rm DL}(a') $ \T\B& $\displaystyle\sum_{a'\leq a^{{\scriptscriptstyle \rm LQG}}_n}D^{\rm DL}_*(a')$ \T\B \\\hline\hline
38&  $4\sqrt{6}$ & 2 & 4 & 105 & 841 \\  \hline
39&  $3\sqrt{11}$ & 0 & 2 & 105 & 843 \\\hline
40&  $3\sqrt{3}+2\sqrt{6}$ & 0 & 64 & 105 & 907 \\\hline
41&  $2\sqrt{2}+2\sqrt{3}+\sqrt{15}$ & 0 & 192 & 105 &1099 \\\hline
42&  $6\sqrt{2}+\sqrt{3}$ & 0 & 64 & 105 & 1163 \\\hline
43&  $6\sqrt{3}$ & 20 & 88 & 125 & 1251 \\\hline
44&  $2\sqrt{2}+\sqrt{3}+\sqrt{35}$ & 0 & 48 & 125 & 1299 \\\hline
45&  $\sqrt{3}+2\sqrt{6}+\sqrt{15}$ & 12 & 48 & 137 & 1347 \\\hline
46&  $4\sqrt{2}+2\sqrt{6}$ & 6 & 24 & 143 & 1371 \\\hline
47&  $2\sqrt{2}+2\sqrt{15}$ & 0 & 24 & 143 & 1395 \\\hline
48&  $\sqrt{3}+4\sqrt{5}$ & 0 & 8 & 143 & 1403 \\\hline
49&  $2\sqrt{2}+3\sqrt{7}$ & 0 & 8 & 143 & 1411 \\\hline
50&  $4\sqrt{3}+\sqrt{15}$ & 0 & 168 & 143 & 1579 \\\hline
51&  $2\sqrt{6}+\sqrt{35}$ & 0 & 8 & 143 & 1587 \\\hline
52&  $4\sqrt{2}+3\sqrt{3}$ & 0 & 320 & 143 & 1907 \\\hline
53&  $2\sqrt{30}$ & 0 & 2 & 143 & 1909 \\\hline
54&  $3\sqrt{3}+\sqrt{35}$ & 0 & 64 & 143 & 1973 \\\hline
55&  $2\sqrt{2}+2\sqrt{3}+2\sqrt{6}$ & 24 & 192 & 167 & 2165 \\\hline
56 & $2\sqrt{3}+2\sqrt{15}$ & 24 & 96 & 191 & 2261 \\\hline
57&  $4\sqrt{2}+\sqrt{3}+\sqrt{15}$ & 24 & 192 & 215 & 2453 \\\hline
58&  $8\sqrt{2}$ & 6 & 16 & 221 & 2469 \\\hline
59&  $2\sqrt{3}+3\sqrt{7}$ & 0 & 24 & 221 & 2493 \\\hline
60&  $2\sqrt{2}+5\sqrt{3}$ & 0 & 432 & 221 & 2925 \\\hline
61&  $\sqrt{3}+\sqrt{15}+\sqrt{35}$ & 0 & 48 & 221 & 2973 \\\hline
62&  $\sqrt{3}+4\sqrt{6}$ & 0 & 24 & 221 & 2997 \\\hline
63&  $4\sqrt{2}+\sqrt{35}$ & 0 & 24 & 221 & 3021 \\\hline
64&  $2\sqrt{2}+2\sqrt{6}+\sqrt{15}$ & 0 & 48 & 221 & 3069 \\\hline
65& $3\sqrt{15}$ & 0 & 8 & 221 & 3077 \\\hline
66&  $\sqrt{3}+3\sqrt{11}$ & 0 & 8 & 221 & 3085 \\\hline
67&  $2\sqrt{2}+4\sqrt{5}$ & 0 & 8 & 221 & 3093 \\\hline
68&  $3\sqrt{7}+\sqrt{15}$ & 0 & 8 & 221 & 3101 \\\hline
69& $4\sqrt{3}+2\sqrt{6}$ & 10 & 168 & 231 & 3269 \\\hline
70&  $2\sqrt{35}$ & 2 & 4 & 233 & 3273 \\\hline
71&  $2\sqrt{2}+3\sqrt{3}+\sqrt{15}$ & 120 & 640 & 353 & 3913 \\\hline
72&  $6\sqrt{2}+2\sqrt{3}$ & 60 & 320 & 413 & 4233 \\\hline
73&  $\sqrt{143}$ & 0 & 2 & 413 & 4235 \\\hline
  \hline
\end{tabular}
\end{center}

\section{\label{App:C}A group theoretic treatment of the projection constraint}
\renewcommand \thesubsection{C.\arabic{subsection}}

As pointed out in subsection \ref{subsectGM}, the problem of finding the $m$-degeneracy for the GM counting is equivalent to determining the number of irreducible representations --taking into account multiplicities-- that appear in the tensor product $\bigotimes_{I=1}^N[k_I/2]$. The reason is that each of the irreducible representations that appear in the decomposition of the tensor product as a direct sum has, precisely, one basic state with zero total third spin component.

In order to solve this problem we propose a solution based on techniques developed in
the context of conformal field theories \cite{DiFrancesco:1997nk,Agullo:2009zt}. The
starting point is to write the tensor product of two $SU(2)$ irreducible representations in the form
\begin{eqnarray*}\label{tensorprod}
\left[\frac{k_1}{2}\right]\otimes\left[\frac{k_2}{2}\right]=\bigoplus_{k_3=0}^\infty
\mathcal{N}^{k_3}_{k_1k_2}\left[\frac{k_3}{2}\right]\ ,
\end{eqnarray*}
in terms of the fusion numbers $\mathcal{N}^{k_3}_{k_1k_2}$. By taking into account that the tensor product and direct sum of irreducible representations have a direct translation into the behavior of the characters, and in particular:
\begin{enumerate}
\item  The  algebra
of the  characters of the  $SU(2)$ irreducible representations satisfies
$$\chi_{k_1}\cdot \chi_{k_2}=\sum_{k_3} \mathcal{N}^{k_3}_{k_1k_2} \chi_{k_3} \ ,$$
\item and  the characters of irreducible representations  are ortonormal with respect to the $SU(2)$-scalar product induced by the (normalized) Haar measure, i.e.
$$
\langle \chi_{k_1}, \chi_{k_2}\rangle_{SU(2)} =\int_{S^3} \bar{\chi}_{k_1} \chi_{k_2} \,\mathrm{d}\mu_{S^3}=\delta(k_1,k_2)\,,
$$
\end{enumerate}
we can easily obtain the number of irreducible representations in the composition of  $N_k$ spin-$k/2$ associated with a configuration $c=\{(k,N_k)\}$ as
\begin{eqnarray}
d^{\rm GM}_m(c)&=&\sum_{k'=0}^{\infty}
\langle \prod_{k}  \chi_{k}^{N_k}\,,\, \chi_{k'}\rangle_{SU(2)}\nonumber \\
&=&\frac{2}{\pi}\sum_{k'=0}^{\infty}   \int_0^{\pi}
 \left(  \prod_{k}\frac{\sin^{N_k}{(k +1)
\theta}}{\sin^{N_k}{\theta}} \right) \frac{\sin{(k' +1)
\theta}}{\sin{\theta}}\, \sin^2{\theta} \mathrm{d} \theta
\label{mcaracteres} \\
&=&\frac{1}{\pi}\sum_{k'=0}^{\infty}   \int_0^{2\pi}
 \left(  \prod_{k}\frac{\sin^{N_k}{(k +1)
\theta}}{\sin^{N_k}{\theta}} \right) \frac{\sin{(k' +1)
\theta}}{\sin{\theta}}\, \sin^2{\theta} \mathrm{d} \theta\nonumber \,.\end{eqnarray}
Notice that the group coordinate $\theta$ is naturally defined in $\theta\in[0,\pi]$ but, due to the parity properties of the integrand in (\ref{mcaracteres}), it is possible to extend the integration in $\theta$ to $[0,2\pi]$. This will prove useful to solve the present problem in yet another different way. The key idea is to realize that we are actually composing $SU(2)$ representations and asking how many states with vanishing \textit{total} third spin component appear in such composition. This is done by computing, with the help of the characters of the representations, the multiplicity of the $0$-irreducible representation of $U(1)$ in the decomposition of the tensor product of the $SU(2)$ representations involved. The characters $\eta_k$, $k\in \mathbb{Z}$,  of the $U(1)$ irreducible
representations are orthonormal with respect to the standard scalar
product in the circle
\begin{eqnarray*}\langle \eta_{k_1},\eta_{k_2}\rangle_{U(1)}=\int_{S^1} \bar{\eta}_{k_1}\eta_{k_2} \,\mathrm{d}\mu_{S^1} = \int_0^{2 \pi}
 e^{-i k_1 \theta} e^{i k_2 \theta}\,\frac{\mathrm{d}\theta}{2 \pi}=\delta(k_1,k_2) \,. \end{eqnarray*}
We can obtain the number that we are looking for just by projecting the product of characters of the $SU(2)$ representations onto the character $\eta_{k=0}$ of the $U(1)$ irreducible representation
\begin{eqnarray}\label{mdeg1}
d^{\rm GM}_m(c)=
\langle \eta_0\,,\,\prod_k \chi_k^{N_k}\rangle_{U(1)}=\frac{1}{2 \pi} \int_0^{2 \pi}
\prod_{k} \left( \frac{\sin{(k +1) \theta}}{\sin{\theta}}
\right)^{N_k}\, d\theta \, .\end{eqnarray}
The last expression coincides with the one derived in subsection \ref{subsectGM} by using generating functions and shows an interesting interplay between the counting of $SU(2)$ and $U(1)$ labels.

Similar considerations allow us to obtain the formulas corresponding to the ENP and DL countings. In the first case we have to find the multiplicity of the singlet $SU(2)$ irreducible representation  in the composition of the representations appearing in a given configuration. This can be trivially obtained by projecting over the character $\chi_0$. In this way we get
\begin{eqnarray}
\label{enp_c}
d^{\rm ENP}_m(c)=\langle \chi_0\,,\,\prod_k \chi_k^{N_k}\rangle_{SU(2)}= \int_0^{2 \pi}
\sin^2{\theta} \prod_{k} \left( \frac{\sin{(k +1)
\theta}}{\sin{\theta}} \right)^{N_k} \,\frac{d \theta}{\pi}\nonumber \ .\end{eqnarray}
Finally, the result for the DL counting can also be obtained by using \textit{reducible} $U(1)$ representations with characters  $\tilde{\eta}_k= \eta_k+\eta_{-k}$ and projecting over $\eta_0$,
\begin{eqnarray}
d^{\rm DL}_m(c)=\langle \eta_0\,,\,\prod_k \tilde{\eta}_{k}^{N_k}\rangle_{U(1)}=\frac{1}{2\pi}
\int_0^{2 \pi}  \prod_k (2 \cos{k \theta})^{N_k}\,\mathrm{d}\theta.\label{dlmdeg}
\end{eqnarray}


\begin{thebibliography}{43}%
\makeatletter
\providecommand \@ifxundefined [1]{%
 \@ifx{#1\undefined}
}%
\providecommand \@ifnum [1]{%
 \ifnum #1\expandafter \@firstoftwo
 \else \expandafter \@secondoftwo
 \fi
}%
\providecommand \@ifx [1]{%
 \ifx #1\expandafter \@firstoftwo
 \else \expandafter \@secondoftwo
 \fi
}%
\providecommand \natexlab [1]{#1}%
\providecommand \enquote  [1]{``#1''}%
\providecommand \bibnamefont  [1]{#1}%
\providecommand \bibfnamefont [1]{#1}%
\providecommand \citenamefont [1]{#1}%
\providecommand \href@noop [0]{\@secondoftwo}%
\providecommand \href [0]{\begingroup \@sanitize@url \@href}%
\providecommand \@href[1]{\@@startlink{#1}\@@href}%
\providecommand \@@href[1]{\endgroup#1\@@endlink}%
\providecommand \@sanitize@url [0]{\catcode `\\12\catcode `\$12\catcode
  `\&12\catcode `\#12\catcode `\^12\catcode `\_12\catcode `\%12\relax}%
\providecommand \@@startlink[1]{}%
\providecommand \@@endlink[0]{}%
\providecommand \url  [0]{\begingroup\@sanitize@url \@url }%
\providecommand \@url [1]{\endgroup\@href {#1}{\urlprefix }}%
\providecommand \urlprefix  [0]{URL }%
\providecommand \Eprint [0]{\href }%
\@ifxundefined \urlstyle {%
  \providecommand \doi  [0]{\begingroup \@sanitize@url \@doi}%
  \providecommand \@doi [1]{\endgroup \@@startlink {\doibase
  #1}doi:\discretionary {}{}{}#1\@@endlink }%
}{%
  \providecommand \doi  [0]{doi:\discretionary{}{}{}\begingroup
  \urlstyle{rm}\Url }%
}%
\providecommand \doibase [0]{http://dx.doi.org/}%
\providecommand \Doi [0]{\begingroup \@sanitize@url \@Doi }%
\providecommand \@Doi  [1]{\endgroup\@@startlink{\doibase#1}\@@Doi}%
\providecommand \@@Doi [1]{#1\@@endlink}%
\providecommand \selectlanguage [0]{\@gobble}%
\providecommand \bibinfo  [0]{\@secondoftwo}%
\providecommand \bibfield  [0]{\@secondoftwo}%
\providecommand \translation [1]{[#1]}%
\providecommand \BibitemOpen [0]{}%
\providecommand \bibitemStop [0]{}%
\providecommand \bibitemNoStop [0]{.\EOS\space}%
\providecommand \EOS [0]{\spacefactor3000\relax}%
\providecommand \BibitemShut  [1]{\csname bibitem#1\endcsname}%
\bibitem [{\citenamefont {Thiemann}(2007)}]{Thiemann:2007zz}%
  \BibitemOpen
  \bibfield  {author} {\bibinfo {author} {\bibfnamefont {T.}~\bibnamefont
  {Thiemann}},\ }\href@noop {} {\emph {\bibinfo {title} {Modern Canonical
  Quantum General Relativity}}}\ (\bibinfo  {publisher} {Cambridge University
  Press},\ \bibinfo {year} {2007})\BibitemShut {NoStop}%
\bibitem [{\citenamefont {Rovelli}(2004)}]{Rovelli:2004tv}%
  \BibitemOpen
  \bibfield  {author} {\bibinfo {author} {\bibfnamefont {C.}~\bibnamefont
  {Rovelli}},\ }\href@noop {} {\emph {\bibinfo {title} {Quantum Gravity}}}\
  (\bibinfo  {publisher} {Cambridge University Press},\ \bibinfo {year}
  {2004})\BibitemShut {NoStop}%
\bibitem [{\citenamefont {Ashtekar}\ and\ \citenamefont
  {Lewandowski}(2004)}]{Ashtekar:2004eh}%
  \BibitemOpen
  \bibfield  {author} {\bibinfo {author} {\bibfnamefont {A.}~\bibnamefont
  {Ashtekar}}\ and\ \bibinfo {author} {\bibfnamefont {J.}~\bibnamefont
  {Lewandowski}},\ }\Doi {10.1088/0264-9381/21/15/R01} {\bibfield  {journal}
  {\bibinfo  {journal} {Class. Quant. Grav.},\ }\textbf {\bibinfo {volume}
  {21}},\ \bibinfo {pages} {R53} (\bibinfo {year} {2004})},\ \Eprint
  {http://arxiv.org/abs/gr-qc/0404018} {arXiv:gr-qc/0404018} \BibitemShut
  {NoStop}%
\bibitem [{\citenamefont {Ashtekar}\ \emph {et~al.}(1998)\citenamefont
  {Ashtekar}, \citenamefont {Baez}, \citenamefont {Corichi},\ and\
  \citenamefont {Krasnov}}]{Ashtekar:1997yu}%
  \BibitemOpen
  \bibfield  {author} {\bibinfo {author} {\bibfnamefont {A.}~\bibnamefont
  {Ashtekar}}, \bibinfo {author} {\bibfnamefont {J.~C.}~\bibnamefont {Baez}},
  \bibinfo {author} {\bibfnamefont {A.}~\bibnamefont {Corichi}}, \ and\
  \bibinfo {author} {\bibfnamefont {K.}~\bibnamefont {Krasnov}},\ }\Doi
  {10.1103/PhysRevLett.80.904} {\bibfield  {journal} {\bibinfo  {journal}
  {Phys. Rev. Lett.},\ }\textbf {\bibinfo {volume} {80}},\ \bibinfo {pages}
  {904} (\bibinfo {year} {1998})},\ \Eprint
  {http://arxiv.org/abs/gr-qc/9710007} {arXiv:gr-qc/9710007} \BibitemShut
  {NoStop}%
\bibitem [{\citenamefont {Ashtekar}\ \emph
  {et~al.}(2000){\natexlab{a}}\citenamefont {Ashtekar}, \citenamefont {Baez},\
  and\ \citenamefont {Krasnov}}]{Ashtekar:2000eq}%
  \BibitemOpen
  \bibfield  {author} {\bibinfo {author} {\bibfnamefont {A.}~\bibnamefont
  {Ashtekar}}, \bibinfo {author} {\bibfnamefont {J.~C.}\ \bibnamefont {Baez}},
  \ and\ \bibinfo {author} {\bibfnamefont {K.}~\bibnamefont {Krasnov}},\
  }\href@noop {} {\bibfield  {journal} {\bibinfo  {journal} {Adv. Theor. Math.
  Phys.},\ }\textbf {\bibinfo {volume} {4}},\ \bibinfo {pages} {1} (\bibinfo
  {year} {2000}{\natexlab{a}})},\ \Eprint {http://arxiv.org/abs/gr-qc/0005126}
  {arXiv:gr-qc/0005126} \BibitemShut {NoStop}%
\bibitem [{\citenamefont {Ashtekar}\ and\ \citenamefont
  {Krishnan}(2004)}]{Ashtekar:2004cn}%
  \BibitemOpen
  \bibfield  {author} {\bibinfo {author} {\bibfnamefont {A.}~\bibnamefont
  {Ashtekar}}\ and\ \bibinfo {author} {\bibfnamefont {B.}~\bibnamefont
  {Krishnan}},\ }\href@noop {} {\bibfield  {journal} {\bibinfo  {journal}
  {Living Rev. Rel.},\ }\textbf {\bibinfo {volume} {7}},\ \bibinfo {pages} {10}
  (\bibinfo {year} {2004})},\ \Eprint {http://arxiv.org/abs/gr-qc/0407042}
  {arXiv:gr-qc/0407042} \BibitemShut {NoStop}%
\bibitem [{\citenamefont {Jacobson}(2007)}]{Jacobson:2007uj}%
  \BibitemOpen
  \bibfield  {author} {\bibinfo {author} {\bibfnamefont {T.}~\bibnamefont
  {Jacobson}},\ }\Doi {10.1088/0264-9381/24/18/N02} {\bibfield  {journal}
  {\bibinfo  {journal} {Class. Quant. Grav.},\ }\textbf {\bibinfo {volume}
  {24}},\ \bibinfo {pages} {4875} (\bibinfo {year} {2007})},\ \Eprint
  {http://arxiv.org/abs/0707.4026} {arXiv:0707.4026 [gr-qc]} \BibitemShut
  {NoStop}%
\bibitem [{\citenamefont {Carlip}(2000)}]{Carlip:2000nv}%
  \BibitemOpen
  \bibfield  {author} {\bibinfo {author} {\bibfnamefont {S.}~\bibnamefont
  {Carlip}},\ }\Doi {10.1088/0264-9381/17/20/302} {\bibfield  {journal}
  {\bibinfo  {journal} {Class. Quant. Grav.},\ }\textbf {\bibinfo {volume}
  {17}},\ \bibinfo {pages} {4175} (\bibinfo {year} {2000})},\ \Eprint
  {http://arxiv.org/abs/gr-qc/0005017} {arXiv:gr-qc/0005017} \BibitemShut
  {NoStop}%
\bibitem [{\citenamefont {Corichi}\ \emph
  {et~al.}(2007){\natexlab{a}}\citenamefont {Corichi}, \citenamefont {Borja},\
  and\ \citenamefont {Diaz-Polo}}]{Corichi:2006bs}%
  \BibitemOpen
  \bibfield  {author} {\bibinfo {author} {\bibfnamefont {A.}~\bibnamefont
  {Corichi}}, \bibinfo {author} {\bibfnamefont {E.~F.}\ \bibnamefont {Borja}},
  \ and\ \bibinfo {author} {\bibfnamefont {J.}~\bibnamefont {Diaz-Polo}},\
  }\Doi {10.1088/0264-9381/24/1/013} {\bibfield  {journal} {\bibinfo  {journal}
  {Class. Quant. Grav.},\ }\textbf {\bibinfo {volume} {24}},\ \bibinfo {pages}
  {243} (\bibinfo {year} {2007}{\natexlab{a}})},\ \Eprint
  {http://arxiv.org/abs/gr-qc/0605014} {arXiv:gr-qc/0605014} \BibitemShut
  {NoStop}%
\bibitem [{\citenamefont {Corichi}\ \emph
  {et~al.}(2007){\natexlab{b}}\citenamefont {Corichi}, \citenamefont {Borja},\
  and\ \citenamefont {Diaz-Polo}}]{Corichi:2006wn}%
  \BibitemOpen
  \bibfield  {author} {\bibinfo {author} {\bibfnamefont {A.}~\bibnamefont
  {Corichi}}, \bibinfo {author} {\bibfnamefont {E.~F.}\ \bibnamefont {Borja}},
  \ and\ \bibinfo {author} {\bibfnamefont {J.}~\bibnamefont {Diaz-Polo}},\
  }\Doi {10.1103/PhysRevLett.98.181301} {\bibfield  {journal} {\bibinfo
  {journal} {Phys. Rev. Lett.},\ }\textbf {\bibinfo {volume} {98}},\ \bibinfo
  {pages} {181301} (\bibinfo {year} {2007}{\natexlab{b}})},\ \Eprint
  {http://arxiv.org/abs/gr-qc/0609122} {arXiv:gr-qc/0609122} \BibitemShut
  {NoStop}%
\bibitem [{\citenamefont {Corichi}\ \emph
  {et~al.}(2007){\natexlab{c}}\citenamefont {Corichi}, \citenamefont {Borja},\
  and\ \citenamefont {Diaz-Polo}}]{Corichi:2007zz}%
  \BibitemOpen
  \bibfield  {author} {\bibinfo {author} {\bibfnamefont {A.}~\bibnamefont
  {Corichi}}, \bibinfo {author} {\bibfnamefont {E.~F.}\ \bibnamefont {Borja}},
  \ and\ \bibinfo {author} {\bibfnamefont {J.}~\bibnamefont {Diaz-Polo}},\
  }\Doi {10.1088/1742-6596/68/1/012031} {\bibfield  {journal} {\bibinfo
  {journal} {J. Phys. Conf. Ser.},\ }\textbf {\bibinfo {volume} {68}},\
  \bibinfo {pages} {012031} (\bibinfo {year} {2007}{\natexlab{c}})},\ \Eprint
  {http://arxiv.org/abs/gr-qc/0703116} {arXiv:gr-qc/0703116} \BibitemShut
  {NoStop}%
\bibitem [{\citenamefont {Bekenstein}(1974)}]{Bekenstein:1974jk}%
  \BibitemOpen
  \bibfield  {author} {\bibinfo {author} {\bibfnamefont {J.~D.}\ \bibnamefont
  {Bekenstein}},\ }\href@noop {} {\bibfield  {journal} {\bibinfo  {journal}
  {Lett. Nuovo Cim.},\ }\textbf {\bibinfo {volume} {11}},\ \bibinfo {pages}
  {467} (\bibinfo {year} {1974})}\BibitemShut {NoStop}%
\bibitem [{\citenamefont {Bekenstein}\ and\ \citenamefont
  {Mukhanov}(1995)}]{Bekenstein:1995ju}%
  \BibitemOpen
  \bibfield  {author} {\bibinfo {author} {\bibfnamefont {J.~D.}\ \bibnamefont
  {Bekenstein}}\ and\ \bibinfo {author} {\bibfnamefont {V.~F.}\ \bibnamefont
  {Mukhanov}},\ }\Doi {10.1016/0370-2693(95)01148-J} {\bibfield  {journal}
  {\bibinfo  {journal} {Phys. Lett.},\ }\textbf {\bibinfo {volume} {B360}},\
  \bibinfo {pages} {7} (\bibinfo {year} {1995})},\ \Eprint
  {http://arxiv.org/abs/gr-qc/9505012} {arXiv:gr-qc/9505012} \BibitemShut
  {NoStop}%
\bibitem [{\citenamefont {Agullo}\ \emph
  {et~al.}(2008){\natexlab{a}}\citenamefont {Agullo}, \citenamefont
  {Barbero~G.}, \citenamefont {Borja}, \citenamefont {Diaz-Polo},\ and\
  \citenamefont {Villase\~nor}}]{Agullo:2008yv}%
  \BibitemOpen
  \bibfield  {author} {\bibinfo {author} {\bibfnamefont {I.}~\bibnamefont
  {Agullo}}, \bibinfo {author} {\bibfnamefont {J.~F.}\ \bibnamefont
  {Barbero~G.}}, \bibinfo {author} {\bibfnamefont {E.~F.}\ \bibnamefont
  {Borja}}, \bibinfo {author} {\bibfnamefont {J.}~\bibnamefont {Diaz-Polo}}, \
  and\ \bibinfo {author} {\bibfnamefont {E.~J.~S.}\ \bibnamefont
  {Villase\~nor}},\ }\Doi {10.1103/PhysRevLett.100.211301} {\bibfield
  {journal} {\bibinfo  {journal} {Phys. Rev. Lett.},\ }\textbf {\bibinfo
  {volume} {100}},\ \bibinfo {pages} {211301} (\bibinfo {year}
  {2008}{\natexlab{a}})},\ \Eprint {http://arxiv.org/abs/0802.4077}
  {arXiv:0802.4077 [gr-qc]} \BibitemShut {NoStop}%
\bibitem [{\citenamefont {Sahlmann}(2008)}]{Sahlmann:2007jt}%
  \BibitemOpen
  \bibfield  {author} {\bibinfo {author} {\bibfnamefont {H.}~\bibnamefont
  {Sahlmann}},\ }\Doi {10.1088/0264-9381/25/5/055004} {\bibfield  {journal}
  {\bibinfo  {journal} {Class. Quant. Grav.},\ }\textbf {\bibinfo {volume}
  {25}},\ \bibinfo {pages} {055004} (\bibinfo {year} {2008})},\ \Eprint
  {http://arxiv.org/abs/0709.0076} {arXiv:0709.0076 [gr-qc]} \BibitemShut
  {NoStop}%
\bibitem [{\citenamefont {Sahlmann}(2007)}]{Sahlmann:2007zp}%
  \BibitemOpen
  \bibfield  {author} {\bibinfo {author} {\bibfnamefont {H.}~\bibnamefont
  {Sahlmann}},\ }\Doi {10.1103/PhysRevD.76.104050} {\bibfield  {journal}
  {\bibinfo  {journal} {Phys. Rev.},\ }\textbf {\bibinfo {volume} {D76}},\
  \bibinfo {pages} {104050} (\bibinfo {year} {2007})},\ \Eprint
  {http://arxiv.org/abs/0709.2433} {arXiv:0709.2433 [gr-qc]} \BibitemShut
  {NoStop}%
\bibitem [{\citenamefont {Barbero~G.}\ and\ \citenamefont
  {Villase\~nor}(2008)}]{BarberoG.:2008ue}%
  \BibitemOpen
  \bibfield  {author} {\bibinfo {author} {\bibfnamefont {J.~F.}\ \bibnamefont
  {Barbero~G.}}\ and\ \bibinfo {author} {\bibfnamefont {E.~J.~S.}\ \bibnamefont
  {Villase\~nor}},\ }\Doi {10.1103/PhysRevD.77.121502} {\bibfield  {journal}
  {\bibinfo  {journal} {Phys. Rev.},\ }\textbf {\bibinfo {volume} {D77}},\
  \bibinfo {pages} {121502} (\bibinfo {year} {2008})},\ \Eprint
  {http://arxiv.org/abs/0804.4784} {arXiv:0804.4784 [gr-qc]} \BibitemShut
  {NoStop}%
\bibitem [{\citenamefont {Barbero~G.}\ and\ \citenamefont
  {Villase\~nor}(2009)}]{G.:2008mj}%
  \BibitemOpen
  \bibfield  {author} {\bibinfo {author} {\bibfnamefont {J.~F.}\ \bibnamefont
  {Barbero~G.}}\ and\ \bibinfo {author} {\bibfnamefont {E.~J.~S.}\ \bibnamefont
  {Villase\~nor}},\ }\Doi {10.1088/0264-9381/26/3/035017} {\bibfield  {journal}
  {\bibinfo  {journal} {Class. Quant. Grav.},\ }\textbf {\bibinfo {volume}
  {26}},\ \bibinfo {pages} {035017} (\bibinfo {year} {2009})},\ \Eprint
  {http://arxiv.org/abs/0810.1599} {arXiv:0810.1599 [gr-qc]} \BibitemShut
  {NoStop}%
\bibitem [{\citenamefont {Meissner}(2004)}]{Meissner:2004ju}%
  \BibitemOpen
  \bibfield  {author} {\bibinfo {author} {\bibfnamefont {K.~A.}\ \bibnamefont
  {Meissner}},\ }\Doi {10.1088/0264-9381/21/22/015} {\bibfield  {journal}
  {\bibinfo  {journal} {Class. Quant. Grav.},\ }\textbf {\bibinfo {volume}
  {21}},\ \bibinfo {pages} {5245} (\bibinfo {year} {2004})},\ \Eprint
  {http://arxiv.org/abs/gr-qc/0407052} {arXiv:gr-qc/0407052} \BibitemShut
  {NoStop}%
\bibitem [{\citenamefont {Agullo}\ \emph
  {et~al.}(2008){\natexlab{b}}\citenamefont {Agullo}, \citenamefont {Borja},\
  and\ \citenamefont {Diaz-Polo}}]{Agullo:2008eg}%
  \BibitemOpen
  \bibfield  {author} {\bibinfo {author} {\bibfnamefont {I.}~\bibnamefont
  {Agullo}}, \bibinfo {author} {\bibfnamefont {E.~F.}\ \bibnamefont {Borja}}, \
  and\ \bibinfo {author} {\bibfnamefont {J.}~\bibnamefont {Diaz-Polo}},\ }\Doi
  {10.1103/PhysRevD.77.104024} {\bibfield  {journal} {\bibinfo  {journal}
  {Phys. Rev.},\ }\textbf {\bibinfo {volume} {D77}},\ \bibinfo {pages} {104024}
  (\bibinfo {year} {2008}{\natexlab{b}})},\ \Eprint
  {http://arxiv.org/abs/0802.3188} {arXiv:0802.3188 [gr-qc]} \BibitemShut
  {NoStop}%
\bibitem [{\citenamefont {Domagala}\ and\ \citenamefont
  {Lewandowski}(2004)}]{Domagala:2004jt}%
  \BibitemOpen
  \bibfield  {author} {\bibinfo {author} {\bibfnamefont {M.}~\bibnamefont
  {Domagala}}\ and\ \bibinfo {author} {\bibfnamefont {J.}~\bibnamefont
  {Lewandowski}},\ }\Doi {10.1088/0264-9381/21/22/014} {\bibfield  {journal}
  {\bibinfo  {journal} {Class. Quant. Grav.},\ }\textbf {\bibinfo {volume}
  {21}},\ \bibinfo {pages} {5233} (\bibinfo {year} {2004})},\ \Eprint
  {http://arxiv.org/abs/gr-qc/0407051} {arXiv:gr-qc/0407051} \BibitemShut
  {NoStop}%
\bibitem [{\citenamefont {Ghosh}\ and\ \citenamefont
  {Mitra}(2006)}]{Ghosh:2006ph}%
  \BibitemOpen
  \bibfield  {author} {\bibinfo {author} {\bibfnamefont {A.}~\bibnamefont
  {Ghosh}}\ and\ \bibinfo {author} {\bibfnamefont {P.}~\bibnamefont {Mitra}},\
  }\Doi {10.1103/PhysRevD.74.064026} {\bibfield  {journal} {\bibinfo  {journal}
  {Phys. Rev.},\ }\textbf {\bibinfo {volume} {D74}},\ \bibinfo {pages} {064026}
  (\bibinfo {year} {2006})},\ \Eprint {http://arxiv.org/abs/hep-th/0605125}
  {arXiv:hep-th/0605125} \BibitemShut {NoStop}%
\bibitem [{\citenamefont {Kaul}\ and\ \citenamefont
  {Majumdar}(1998)}]{Kaul:1998xv}%
  \BibitemOpen
  \bibfield  {author} {\bibinfo {author} {\bibfnamefont {R.~K.}\ \bibnamefont
  {Kaul}}\ and\ \bibinfo {author} {\bibfnamefont {P.}~\bibnamefont
  {Majumdar}},\ }\Doi {10.1016/S0370-2693(98)01030-2} {\bibfield  {journal}
  {\bibinfo  {journal} {Phys. Lett.},\ }\textbf {\bibinfo {volume} {B439}},\
  \bibinfo {pages} {267} (\bibinfo {year} {1998})},\ \Eprint
  {http://arxiv.org/abs/gr-qc/9801080} {arXiv:gr-qc/9801080} \BibitemShut
  {NoStop}%
\bibitem [{\citenamefont {Engle}\ \emph {et~al.}(2009)\citenamefont {Engle},
  \citenamefont {Perez},\ and\ \citenamefont {Noui}}]{Engle:2009vc}%
  \BibitemOpen
  \bibfield  {author} {\bibinfo {author} {\bibfnamefont {J.}~\bibnamefont
  {Engle}}, \bibinfo {author} {\bibfnamefont {A.}~\bibnamefont {Perez}}, \ and\
  \bibinfo {author} {\bibfnamefont {K.}~\bibnamefont {Noui}},\ }
  \Doi {10.1103/PhysRevLett.105.031302} {\bibfield  {journal}
  {\bibinfo  {journal} {Phys. Rev. Lett.},\ }\textbf {\bibinfo {volume} {105}},\
  \bibinfo {pages} {031302} (\bibinfo {year} {2010})},\ \Eprint
  {http://arxiv.org/abs/gr-qc/0905.3168} {arXiv:0905.3168 [gr-qc]} \BibitemShut
  {NoStop}%
\bibitem [{\citenamefont {Agullo}\ \emph
  {et~al.}(2009){\natexlab{a}}\citenamefont {Agullo}, \citenamefont {Borja},\
  and\ \citenamefont {Diaz-Polo}}]{Agullo:2009zt}%
  \BibitemOpen
  \bibfield  {author} {\bibinfo {author} {\bibfnamefont {I.}~\bibnamefont
  {Agullo}}, \bibinfo {author} {\bibfnamefont {E.~F.}\ \bibnamefont {Borja}}, \
  and\ \bibinfo {author} {\bibfnamefont {J.}~\bibnamefont {Diaz-Polo}},\ }\Doi
  {10.1088/1475-7516/2009/07/016} {\bibfield  {journal} {\bibinfo  {journal}
  {JCAP},\ }\textbf {\bibinfo {volume} {0907}},\ \bibinfo {pages} {016}
  (\bibinfo {year} {2009}{\natexlab{a}})},\ \Eprint
  {http://arxiv.org/abs/0903.1667} {arXiv:0903.1667 [hep-th]} \BibitemShut
  {NoStop}%
\bibitem [{\citenamefont {Corichi}(2009)}]{Corichi:2009wn}%
  \BibitemOpen
  \bibfield  {author} {\bibinfo {author} {\bibfnamefont {A.}~\bibnamefont
  {Corichi}},\ }\href@noop {} {\bibfield  {journal} {\bibinfo  {journal} {Adv.
  Sci. Lett.},\ }\textbf {\bibinfo {volume} {2}},\ \bibinfo {pages} {236}
  (\bibinfo {year} {2009})},\ \Eprint {http://arxiv.org/abs/0901.1302}
  {arXiv:0901.1302 [gr-qc]} \BibitemShut {NoStop}%
\bibitem [{\citenamefont {Horowitz}(2000)}]{Hor}%
  \BibitemOpen
  \bibfield  {author} {\bibinfo {author} {\bibfnamefont {G.}~\bibnamefont
  {Horowitz}},\ }\href@noop {} {\enquote {\bibinfo {title} {{Quantum Gravity at
  the Turn of the Millennium}},}\ } (\bibinfo {year} {2000}),\ \Eprint
  {http://arxiv.org/abs/gr-qc/0011089} {arXiv:gr-qc/0011089} \BibitemShut
  {NoStop}%
\bibitem [{\citenamefont {Smolin}(1995)}]{Smolin:1995vq}%
  \BibitemOpen
  \bibfield  {author} {\bibinfo {author} {\bibfnamefont {L.}~\bibnamefont
  {Smolin}},\ }\Doi {10.1063/1.531251} {\bibfield  {journal} {\bibinfo
  {journal} {J. Math. Phys.},\ }\textbf {\bibinfo {volume} {36}},\ \bibinfo
  {pages} {6417} (\bibinfo {year} {1995})},\ \Eprint
  {http://arxiv.org/abs/gr-qc/9505028} {arXiv:gr-qc/9505028} \BibitemShut
  {NoStop}%
\bibitem [{\citenamefont {Krasnov}(1997)}]{Krasnov:1996tb}%
  \BibitemOpen
  \bibfield  {author} {\bibinfo {author} {\bibfnamefont {K.~V.}\ \bibnamefont
  {Krasnov}},\ }\Doi {10.1103/PhysRevD.55.3505} {\bibfield  {journal} {\bibinfo
   {journal} {Phys. Rev.},\ }\textbf {\bibinfo {volume} {D55}},\ \bibinfo
  {pages} {3505} (\bibinfo {year} {1997})},\ \Eprint
  {http://arxiv.org/abs/gr-qc/9603025} {arXiv:gr-qc/9603025} \BibitemShut
  {NoStop}%
\bibitem [{\citenamefont {Krasnov}(1998)}]{Krasnov:1996wc}%
  \BibitemOpen
  \bibfield  {author} {\bibinfo {author} {\bibfnamefont {K.~V.}\ \bibnamefont
  {Krasnov}},\ }\Doi {10.1023/A:1018820916342} {\bibfield  {journal} {\bibinfo
  {journal} {Gen. Rel. Grav.},\ }\textbf {\bibinfo {volume} {30}},\ \bibinfo
  {pages} {53} (\bibinfo {year} {1998})},\ \Eprint
  {http://arxiv.org/abs/gr-qc/9605047} {arXiv:gr-qc/9605047} \BibitemShut
  {NoStop}%
\bibitem [{\citenamefont {Rovelli}(1996){\natexlab{a}}}]{Rovelli:1996dv}%
  \BibitemOpen
  \bibfield  {author} {\bibinfo {author} {\bibfnamefont {C.}~\bibnamefont
  {Rovelli}},\ }\Doi {10.1103/PhysRevLett.77.3288} {\bibfield  {journal}
  {\bibinfo  {journal} {Phys. Rev. Lett.},\ }\textbf {\bibinfo {volume} {77}},\
  \bibinfo {pages} {3288} (\bibinfo {year} {1996}{\natexlab{a}})},\ \Eprint
  {http://arxiv.org/abs/gr-qc/9603063} {arXiv:gr-qc/9603063} \BibitemShut
  {NoStop}%
\bibitem [{\citenamefont {Rovelli}(1996){\natexlab{b}}}]{Rovelli:1996ti}%
  \BibitemOpen
  \bibfield  {author} {\bibinfo {author} {\bibfnamefont {C.}~\bibnamefont
  {Rovelli}},\ }\href@noop {} {\bibfield  {journal} {\bibinfo  {journal} {Helv.
  Phys. Acta},\ }\textbf {\bibinfo {volume} {69}},\ \bibinfo {pages} {582}
  (\bibinfo {year} {1996}{\natexlab{b}})},\ \Eprint
  {http://arxiv.org/abs/gr-qc/9608032} {arXiv:gr-qc/9608032} \BibitemShut
  {NoStop}%
\bibitem [{\citenamefont {Kaul}\ and\ \citenamefont
  {Majumdar}(2000)}]{Kaul:2000kf}%
  \BibitemOpen
  \bibfield  {author} {\bibinfo {author} {\bibfnamefont {R.~K.}\ \bibnamefont
  {Kaul}}\ and\ \bibinfo {author} {\bibfnamefont {P.}~\bibnamefont
  {Majumdar}},\ }\Doi {10.1103/PhysRevLett.84.5255} {\bibfield  {journal}
  {\bibinfo  {journal} {Phys. Rev. Lett.},\ }\textbf {\bibinfo {volume} {84}},\
  \bibinfo {pages} {5255} (\bibinfo {year} {2000})},\ \Eprint
  {http://arxiv.org/abs/gr-qc/0002040} {arXiv:gr-qc/0002040} \BibitemShut
  {NoStop}%
\bibitem [{\citenamefont {Engle}\ \emph {et~al.}(2010)\citenamefont {Engle},
  \citenamefont {Noui}, \citenamefont {Perez},\ and\ \citenamefont
  {Pranzetti}}]{Engle:2010kt}%
  \BibitemOpen
  \bibfield  {author} {\bibinfo {author} {\bibfnamefont {J.}~\bibnamefont
  {Engle}}, \bibinfo {author} {\bibfnamefont {K.}~\bibnamefont {Noui}},
  \bibinfo {author} {\bibfnamefont {A.}~\bibnamefont {Perez}}, \ and\ \bibinfo
  {author} {\bibfnamefont {D.}~\bibnamefont {Pranzetti}},\ }
  \Doi {10.1103/PhysRevD.82.044050} {\bibfield  {journal} {\bibinfo
   {journal} {Phys. Rev.},\ }\textbf {\bibinfo {volume} {D82}},\ \bibinfo
  {pages} {044050} (\bibinfo {year} {2010})},\ \Eprint
  {http://arxiv.org/abs/1006.0634} {arXiv:1006.0634 [gr-qc]} \BibitemShut
  {NoStop}%
\bibitem [{\citenamefont {Krasnov}\ and\ \citenamefont
  {Rovelli}(2009)}]{Krasnov:2009pd}%
  \BibitemOpen
  \bibfield  {author} {\bibinfo {author} {\bibfnamefont {K.}~\bibnamefont
  {Krasnov}}\ and\ \bibinfo {author} {\bibfnamefont {C.}~\bibnamefont
  {Rovelli}},\ }\Doi {10.1088/0264-9381/26/24/245009} {\bibfield  {journal}
  {\bibinfo  {journal} {Class. Quant. Grav.},\ }\textbf {\bibinfo {volume}
  {26}},\ \bibinfo {pages} {245009} (\bibinfo {year} {2009})},\ \Eprint
  {http://arxiv.org/abs/0905.4916} {0905.4956 [gr-qc]} \BibitemShut
  {NoStop}%
\bibitem [{\citenamefont {Ashtekar}\ \emph
  {et~al.}(2000){\natexlab{b}}\citenamefont {Ashtekar}, \citenamefont
  {Corichi},\ and\ \citenamefont {Krasnov}}]{Ashtekar:1999wa}%
  \BibitemOpen
  \bibfield  {author} {\bibinfo {author} {\bibfnamefont {A.}~\bibnamefont
  {Ashtekar}}, \bibinfo {author} {\bibfnamefont {A.}~\bibnamefont {Corichi}}, \
  and\ \bibinfo {author} {\bibfnamefont {K.}~\bibnamefont {Krasnov}},\
  }\href@noop {} {\bibfield  {journal} {\bibinfo  {journal} {Adv. Theor. Math.
  Phys.},\ }\textbf {\bibinfo {volume} {3}},\ \bibinfo {pages} {419} (\bibinfo
  {year} {2000}{\natexlab{b}})},\ \Eprint {http://arxiv.org/abs/gr-qc/9905089}
  {arXiv:gr-qc/9905089} \BibitemShut {NoStop}%
\bibitem [{\citenamefont {Barbero~G.}\ \emph {et~al.}(2009)\citenamefont
  {Barbero~G.}, \citenamefont {Lewandowski},\ and\ \citenamefont
  {Villase\~nor}}]{FernandoBarbero:2009ai}%
  \BibitemOpen
  \bibfield  {author} {\bibinfo {author} {\bibfnamefont {J.~F.}\ \bibnamefont
  {Barbero~G.}}, \bibinfo {author} {\bibfnamefont {J.}~\bibnamefont
  {Lewandowski}}, \ and\ \bibinfo {author} {\bibfnamefont {E.~J.~S.}\
  \bibnamefont {Villase\~nor}},\ }\Doi {10.1103/PhysRevD.80.044016} {\bibfield
  {journal} {\bibinfo  {journal} {Phys. Rev.},\ }\textbf {\bibinfo {volume}
  {D80}},\ \bibinfo {pages} {044016} (\bibinfo {year} {2009})},\ \Eprint
  {http://arxiv.org/abs/0905.3465} {arXiv:0905.3465 [gr-qc]} \BibitemShut
  {NoStop}%
\bibitem [{\citenamefont {Griffiths}(1965)}]{G}%
  \BibitemOpen
  \bibfield  {author} {\bibinfo {author} {\bibfnamefont {R.~B.}\ \bibnamefont
  {Griffiths}},\ }\href@noop {} {\bibfield  {journal} {\bibinfo  {journal} {J.
  Math. Phys.},\ }\textbf {\bibinfo {volume} {6}},\ \bibinfo {pages} {1447}
  (\bibinfo {year} {1965})}\BibitemShut {NoStop}%
\bibitem [{\citenamefont {Krasnov}(1999)}]{Krasnov:1997yt}%
  \BibitemOpen
  \bibfield  {author} {\bibinfo {author} {\bibfnamefont {K.~V.}\ \bibnamefont
  {Krasnov}},\ }\Doi {10.1088/0264-9381/16/2/018} {\bibfield  {journal} {\bibinfo
   {journal} {Class. Quant. Grav.},\ }\textbf {\bibinfo {volume} {16}},\ \bibinfo
  {pages} {563} (\bibinfo {year} {1999})},\ \Eprint
  {http://arxiv.org/abs/gr-qc/9710006} {arXiv:gr-qc/9710006} \BibitemShut
  {NoStop}%
\bibitem [{\citenamefont {Burton}(2002)}]{Burton}%
  \BibitemOpen
  \bibfield  {author} {\bibinfo {author} {\bibfnamefont {D.~M.}\ \bibnamefont
  {Burton}},\ }\href@noop {} {\emph {\bibinfo {title} {Elementary Number
  Theory}}}\ (\bibinfo  {publisher} {McGraw-Hil},\ \bibinfo {year}
  {2002})\BibitemShut {NoStop}%
\bibitem [{\citenamefont {De~Raedt}\ \emph {et~al.}(2001)\citenamefont
  {De~Raedt}, \citenamefont {Michielsen}, \citenamefont {De~Raedt},\ and\
  \citenamefont {Miyashita}}]{DeRaedt}%
  \BibitemOpen
  \bibfield  {author} {\bibinfo {author} {\bibfnamefont {H.}~\bibnamefont
  {De~Raedt}}, \bibinfo {author} {\bibfnamefont {K.}~\bibnamefont
  {Michielsen}}, \bibinfo {author} {\bibfnamefont {K.}~\bibnamefont
  {De~Raedt}}, \ and\ \bibinfo {author} {\bibfnamefont {S.}~\bibnamefont
  {Miyashita}},\ }\href@noop {} {\bibfield  {journal} {\bibinfo  {journal}
  {Phys. Lett.},\ }\textbf {\bibinfo {volume} {A290}},\ \bibinfo {pages} {227}
  (\bibinfo {year} {2001})}\BibitemShut {NoStop}%
\bibitem [{\citenamefont {Di~Francesco}\ \emph {et~al.}(1997)\citenamefont
  {Di~Francesco}, \citenamefont {Mathieu},\ and\ \citenamefont
  {Senechal}}]{DiFrancesco:1997nk}%
  \BibitemOpen
  \bibfield  {author} {\bibinfo {author} {\bibfnamefont {P.}~\bibnamefont
  {Di~Francesco}}, \bibinfo {author} {\bibfnamefont {P.}~\bibnamefont
  {Mathieu}}, \ and\ \bibinfo {author} {\bibfnamefont {D.}~\bibnamefont
  {Senechal}},\ }\href@noop {} {\emph {\bibinfo {title} {Conformal Field
  Theory}}}\ (\bibinfo  {publisher} {Springer},\ \bibinfo {year}
  {1997})\BibitemShut {NoStop}%
\bibitem [{\citenamefont {B{\"o}ttcher}\ and\ \citenamefont
  {Grudsky}(2005)}]{Bottcher}%
  \BibitemOpen
  \bibfield  {author} {\bibinfo {author} {\bibfnamefont {A.}~\bibnamefont
  {B{\"o}ttcher}}\ and\ \bibinfo {author} {\bibfnamefont {S.~M.}\ \bibnamefont
  {Grudsky}},\ }\href@noop {} {\emph {\bibinfo {title} {Spectral Properties of
  Banded Toeplitz Matrices}}}\ (\bibinfo  {publisher} {SIAM, Philadelphia},\
  \bibinfo {year} {2005})\BibitemShut {NoStop}%
\bibitem [{\citenamefont {Agullo}\ \emph
  {et~al.}(2009){\natexlab{b}}\citenamefont {Agullo}, \citenamefont
  {Barbero~G.}, \citenamefont {Borja}, \citenamefont {Diaz-Polo},\ and\
  \citenamefont {Villase\~nor}}]{Agullo:2009eq}%
  \BibitemOpen
  \bibfield  {author} {\bibinfo {author} {\bibfnamefont {I.}~\bibnamefont
  {Agullo}}, \bibinfo {author} {\bibfnamefont {J.~F.}\ \bibnamefont
  {Barbero~G.}}, \bibinfo {author} {\bibfnamefont {E.~F.}\ \bibnamefont
  {Borja}}, \bibinfo {author} {\bibfnamefont {J.}~\bibnamefont {Diaz-Polo}}, \
  and\ \bibinfo {author} {\bibfnamefont {E.~J.~S.}\ \bibnamefont
  {Villase\~nor}},\ }\Doi {10.1103/PhysRevD.80.084006} {\bibfield  {journal}
  {\bibinfo  {journal} {Phys. Rev.},\ }\textbf {\bibinfo {volume} {D80}},\
  \bibinfo {pages} {084006} (\bibinfo {year} {2009}{\natexlab{b}})},\ \Eprint
  {http://arxiv.org/abs/0906.4529} {arXiv:0906.4529 [gr-qc]} \BibitemShut
  {NoStop}%
\bibitem [{\citenamefont {Flajolet}\ and\ \citenamefont
  {Sedgewick}(2009)}]{Flajolet}%
  \BibitemOpen
  \bibfield  {author} {\bibinfo {author} {\bibfnamefont {P.}~\bibnamefont
  {Flajolet}}\ and\ \bibinfo {author} {\bibfnamefont {R.}~\bibnamefont
  {Sedgewick}},\ }\href@noop {} {\emph {\bibinfo {title} {Analytic
  Combinatorics}}}\ (\bibinfo  {publisher} {Cambridge University Press},\
  \bibinfo {year} {2009})\BibitemShut {NoStop}%
\end{thebibliography}


\end{document}